\documentclass[conference]{IEEEtran}
\IEEEoverridecommandlockouts
\usepackage{multirow}
\usepackage{enumitem}
\usepackage{float}
\usepackage{subfigure}
\usepackage{enumitem}
\usepackage{epstopdf}
\usepackage{colortbl}
\usepackage{cite}
\usepackage{amsmath,amssymb,amsfonts}
\usepackage{algorithmic}
\usepackage[ruled,vlined]{algorithm2e}
\usepackage{graphicx}
\usepackage{textcomp}
\usepackage{xcolor}
\newcommand{\tabincell}[2]{\begin{tabular}{@{}#1@{}}#2\end{tabular}}
\def\BibTeX{{\rm B\kern-.05em{\sc i\kern-.025em b}\kern-.08em
    T\kern-.1667em\lower.7ex\hbox{E}\kern-.125emX}}
\begin{document}

\title{Improve3C: Data Cleaning on Consistency and Completeness with Currency}
\author{\IEEEauthorblockN{Xiaoou Ding\IEEEauthorrefmark{1}, Hongzhi Wang\IEEEauthorrefmark{2}, Jiaxuan Su\IEEEauthorrefmark{3}, Jianzhong Li\IEEEauthorrefmark{2}, Hong Gao\IEEEauthorrefmark{2}}
\IEEEauthorblockA{School of Computer Science and Technology \\Harbin Institute of Technology,
Harbin, Heilongjiang, China 150001\\
Email: \IEEEauthorrefmark{1}dingxiaoou\_hit@163.com, \IEEEauthorrefmark{2}\{wangzh, lijzh, honggao\}@hit.edu.cn, \IEEEauthorrefmark{3}itx351@gmail.com}
}
%\author{\IEEEauthorblockN{Xiaoou Ding Shell\IEEEauthorre
%fmark{1}, Homer Simpson\IEEEauthorrefmark{2}, James K
%irk\IEEEauthorrefmark{3}, Montgomery Scott\IEEEautho
%rrefmark{4} and Eldon Tyrell\IEEEauthorrefmark{5}}
%email address}}
\maketitle
\begin{abstract}
Data quality plays a key role in big data management today. With the explosive growth of data from a variety of sources, the quality of data is faced with multiple problems. Motivated by this, we study the multiple data quality improvement on completeness, consistency and currency in this paper. For the proposed problem, we introduce a 4-step framework, named \textsf{Improve3C}, for detection and quality improvement on incomplete and inconsistent data without timestamps. We compute and achieve a relative currency order among records derived from given currency constraints, according to which inconsistent and incomplete data can be repaired effectively considering the temporal impact. For both effectiveness and efficiency consideration, we carry out inconsistent repair ahead of incomplete repair. Currency-related consistency distance is defined to measure the similarity between dirty records and clean ones more accurately. In addition, currency orders are treated as an important feature in the training process of incompleteness repair. The solution algorithms are introduced in detail with examples. A thorough experiment on one real-life data and a synthetic one verifies that the proposed method can improve the performance of dirty data cleaning with multiple quality problems which are hard to be cleaned by the existing approaches effectively.
\end{abstract}

\begin{IEEEkeywords}
component, formatting, style, styling, insert
\end{IEEEkeywords}

\section{Introduction}
Data quality plays the key role in data-centric applications~\cite{Sidi2012Data}. The quality problems in data are often quite serious and trouble data transaction steps (\emph{e.g.}, acquisition, copy, querying).
Specifically, currency, consistency and completeness (3C for short) are three important issues in data quality~\cite{Fan2013Data}. For example, various information systems store data with different formats or semantic. It may lead to costly consistency problems in multi-source data integration. In addition
with imperfect integrity standard of information systems, the records in database may have missing values. Worse still, the low frequency in data update makes it out-of-date to some degree when the timestamps are missing or incomplete under the loose and imprecise copy functions of data among sources.
These three problems result in the low reliability of data, which adds to the confusion and misunderstanding in data applications.
The low-quality data sets may result in negative impact on many fields.

Researchers have gone a long way in data quality and data cleaning, particularly in consistency and completeness. It is acknowledged that consistency and completeness are likely to affect each other during repairing, rather than completely isolated \cite{Cappiello2008Time,Fan2013Data}. We find that currency issues also seriously impact the repair of inconsistent and incomplete values. These mixed data problem are challenged to be both detected and repaired, as illustrated in the following example.

\newtheorem{example}{Example}
\begin{example}
Table 1 shows a part of personal career information collected from the talent pool of different companies, describing two individuals (entities), Mike and Helen. Each record has 9 attributes. \textsf{Level} is an industry-recognized career rank, while \textsf{Title} is the post of the employee. \textsf{City} describes the place where the company is located, and \textsf{Address} records the commercial districts the \textsf{Company} belongs to. \textsf{Email} reports the current professional email.

Specifically, ``\emph{ME, RE, RA}'' and ``\emph{MR}'' represents major engineer, research engineer, assistant researcher, and major researcher, respectively. ``\emph{Zhongguancun}, \emph{Xuhui}'', etc. are well-known landmarks of different cities in China. Abbreviations are used in \textsf{Email}. As the data came from multi-sources, and the timestamp is missing. Inconsistent and incomplete problems also exist in attributes.
\end{example}
\begin{table*}[t]
\centering
%\big
\caption{Personal career information for Mike and Helen}
\label{tab2}
\scalebox{1}{
\begin{tabular}{cr|ccccccccc|}
\cline{3-11}
%\toprule
 &
& \textbf{Name} & \textbf{Level} & \textbf{Title} & \textbf{Company} &\textbf{Address} & \textbf{City} & \textbf{Salary} & \textbf{Email} & \textbf{Group} \\
%\noalign{\smallskip}
\cline{3-11}
%\noalign{\smallskip}
%\midrule
$\mathtt{E}_1$: & $r_1$:& Mike & P2 & E & Baidu & Zhongguancun & Beijing & 13k & M@Bai & Java \\
%\hline
& $r_2$:& \cellcolor[rgb]{1.00,0.8,0.8}{Mike} & \cellcolor[rgb]{1.00,0.8,0.8}{P2} & \cellcolor[rgb]{1.00,0.8,0.8}{E} & \cellcolor[rgb]{1.00,0.8,0.8}{Baidu} & \cellcolor[rgb]{1.00,0.5,0.5}{Tongzhou} & \cellcolor[rgb]{1.00,0.8,0.8}{Beijing} & \cellcolor[rgb]{1.00,0.8,0.8}{13k} & \cellcolor[rgb]{1.00,0.8,0.8}{M@Bai} & \cellcolor[rgb]{1.00,0.8,0.8}{Map} \\
 & $r_3$:& Mike & P2 & ME & Baidu & Zhongguancun & Beijing & 15k & M@Bai & Map \\
%\hline
%\hline
 & $r_4$:& Mike & P3 & ME & Baidu & Zhongguancun & Beijing & 20k & M@Bai & Map \\
%\hline
 & $r_5$:& \cellcolor[rgb]{1.00,0.8,0.8}{Mike} & \cellcolor[rgb]{1.00,0.8,0.8}{P4} & \cellcolor[rgb]{1.00,0.8,0.8}{E} & \cellcolor[rgb]{1.00,0.8,0.8}{Alibaba} & \cellcolor[rgb]{1.00,0.5,0.5}{Zhongguancun} & \cellcolor[rgb]{1.00,0.5,0.5}{Beijing} & \cellcolor[rgb]{1.00,0.8,0.8}{22k} & \cellcolor[rgb]{1.00,0.5,0.5}{M@Bai} & \cellcolor[rgb]{1.00,0.8,0.8}{Tmall} \\
%\hline
 & $r_6$:& Mike & P4 & E & Alibaba & XiXi & Hangzhou & 22k & M@ali & Financial  \\
 & $r_7$: & Mike & P4 & RE & Alibaba & XiXi & Hangzhou & 23k & M@ali & Financial \\
\cline{3-11}
\cline{3-11}
$\mathtt{E}_2$: & $r_8$:& Helen & P2 & RA & Tencent & Binhai & Shenzhen & 15k & H@QQ & Game \\
 & $r_9$:& \cellcolor[rgb]{1.00,0.8,0.8}{Helen} & \cellcolor[rgb]{1.00,0.8,0.8}{P3} & \cellcolor[rgb]{1.00,0.8,0.8}{R} & \cellcolor[rgb]{1.00,0.8,0.8}{Tencent} & \cellcolor[rgb]{1.00,0.8,0.8}{Binhai} & \cellcolor[rgb]{1.00,0.8,0.8}{Shenzhen} &\cellcolor[rgb]{1.00,0.5,0.5}{ } & \cellcolor[rgb]{1.00,0.8,0.8}{H@QQ} & \cellcolor[rgb]{1.00,0.8,0.8}{Game} \\
 & $r_{10}$:& \cellcolor[rgb]{1.00,0.8,0.8}{Helen} & \cellcolor[rgb]{1.00,0.8,0.8}{P3} & \cellcolor[rgb]{1.00,0.8,0.8}{R} & \cellcolor[rgb]{1.00,0.8,0.8}{Tencent} & \cellcolor[rgb]{1.00,0.5,0.5}{} & \cellcolor[rgb]{1.00,0.5,0.5}{} & \cellcolor[rgb]{1.00,0.8,0.8}{18k} & \cellcolor[rgb]{1.00,0.8,0.8}{H@QQ} & \cellcolor[rgb]{1.00,0.8,0.8}{Financial} \\
 & $r_{11}$: & Helen & P3 & R & Tencent & Xuhui & Shanghai & 20k & H@QQ & Social Network \\
& $r_{12}$:& Helen & P4 & R & Microsoft & Zhongguancun &Beijing  & 22k & H@outlook & Social Computing \\
 & $r_{13}$:& Helen & P4 & MR & Microsoft & Zhongguancun & Beijing & 22k & H@outlook & Social Computing \\
& $r_{14}$:& \cellcolor[rgb]{1.00,0.8,0.8}{Helen} & \cellcolor[rgb]{1.00,0.8,0.8}{P5} & \cellcolor[rgb]{1.00,0.5,0.5}{} & \cellcolor[rgb]{1.00,0.8,0.8}{Microsoft} & \cellcolor[rgb]{1.00,0.8,0.8}{Zhongguancun} & \cellcolor[rgb]{1.00,0.8,0.8}{Beijing} & \cellcolor[rgb]{1.00,0.5,0.5}{} & \cellcolor[rgb]{1.00,0.8,0.8}{H@outlook} & \cellcolor[rgb]{1.00,0.8,0.8}{Social Computing} \\
%$r_{14}$ & $\mathtt{E}_2$& Helen & P5 & MR & Microsoft & Zhongguancun & Beijing & 25k & H@outlook & Social Computing \\
\cline{3-11}
\end{tabular}}
\end{table*}
\begin{table*}[t]
\centering
%\big
\caption{Repair dirty data}
\label{tab2}
\scalebox{1}{
\begin{tabular}{r|l|lc|l|}
\cline{2-5}
%\noalign{\smallskip}
%\toprule
& \textbf{Dirty attributes} & \textbf{After repair} & & \textbf{Explanation} \\
%\noalign{\smallskip}
\cline{2-5}
%\noalign{\smallskip}
%\midrule
$r_2$:& [\textsf{Address}]=``\emph{Tongzhou}'' &  \cellcolor[rgb]{0.8,1,0.8}{($a$).``\emph{Zhongguancun}''} & \cellcolor[rgb]{0.8,1,0.8}{$\checkmark$}& Can be well-repaired by CFD or records similarity.  \\
\cline{2-5}
$r_5$:& [\textsf{Address}], [\textsf{City}], [\textsf{Email}]   &  \cellcolor[rgb]{0.8,1,0.8}{($a$).``\emph{Xixi}'',``\emph{Hangzhou}'',``\emph{M@ali}''} & \cellcolor[rgb]{0.8,1,0.8}{$\checkmark$} & An effective repair from currency-related consistency method.  \\
\cline{2-5}
  &[\textsf{Company}], [\textsf{Group}]  &\cellcolor[rgb]{1,0.8,0.8}{($b$).\emph{Baidu}, ``\emph{ML}''} & \cellcolor[rgb]{1,0.8,0.8}{$\times$} & A poor repair without taking account currency issues. \\
\cline{2-5}
\cline{2-5}
 $r_{9}$:&[\textsf{Salary}] missing &  \cellcolor[rgb]{0.8,1,0.8}{($a$).``\emph{15K}''} &
\cellcolor[rgb]{0.8,1,0.8}{$\checkmark$}& A proper clean value.\\
\cline{2-5}
$r_{10}$:& [\textsf{Address}] missing, [\textsf{City}] missing  &  \cellcolor[rgb]{0.8,1,0.8}{($a$).``\emph{Xuhui}'', ``\emph{Shanghai}''} & \cellcolor[rgb]{0.8,1,0.8}{$\checkmark$}& An effective repair from currency-related completeness methods.\\
\cline{3-5}
 &   &  \cellcolor[rgb]{1,0.8,0.8}{($b$).``\emph{Binhai}'', ``\emph{Shenzhen}''} & \cellcolor[rgb]{1,0.8,0.8}{$\times$}& A poor repair fails to capture the closet current values.\\
\cline{2-5}
$r_{14}$:& [\textsf{Title}] missing  &  \cellcolor[rgb]{0.8,1,0.8}{($a$).``\emph{MR}''} & \cellcolor[rgb]{0.8,1,0.8}{$\checkmark$}& An accurate and current repair\\
\cline{3-5}
&   &  \cellcolor[rgb]{0.8,0.8,1}{($b$).``\emph{R}''} & \cellcolor[rgb]{0.8,0.8,1}{$\otimes$}& The repair is less accurate and current.\\
\cline{2-5}
\end{tabular}}
\end{table*}

As outlined in red in Table 1, dirty values exist in 5 records. An incorrect address happens in $r_2$, since Baidu (Beijing) is not located in Thongzhou district. $r_5$ describes Mike works in Alibaba (Hangzhou). However, it reports the city is Beijing, and he is using a Baidu email at the same time. It leads to a confusion, and we can conclude that inconsistent values exist in [\textsf{Address}], [\textsf{City}] and [\textsf{Email}], or even in [\textsf{Company}] and [\textsf{Group}] of $r_5$. For Helen, $r_{9}$, $r_{10}$ and $r_{14}$ contain missing values. We fail to know when she began working in Shanghai and how much is her current salary.

With existing data repairing methods, we can adopt some optional repair schema in Table 2. The incorrect address in $r_2$ can be repaired to ``\emph{Zhangguancun}'' according to a \emph{CFD}: $(r_i[\textsf{Company}]=``Baidu" \wedge r_i[\textsf{City}]=``Beijing"\longrightarrow r_i[\textsf{Address}]=``Zhongguancun"$). We can give a relative clean value ``15\emph{K}'' to $r_9$'s missing salary referring to its most similar record $r_8$, but things are not simple when repairing other dirty values. The company and group that Mike works in do not coincide with the city and his working email in $r_5$. It is possible to clean $r_5$ with the same values of $r_4$. However, Mike has actually began working in Alibaba at the time of $r_5$, which implies that $r_5$ is more current than $r_4$. Thus, this repair is a poor one without considering the temporal issues. For $r_{10}$ and $r_{14}$ of Helen, the edit distance $\textsc{Dist}(r_{9},r_{10}) = Dist(r_{10},r_{11})$ makes it difficult to distinguish which is closer to $r_{10}$, and it also presents no currency difference among $r_{9},r_{10}$ and $r_{11}$. Similarly, it seems no difference to repair $r_{14}$ with either ``\emph{R}" or ``\emph{MR}" because of the equal $Dist(r_{12},r_{14})$ and $Dist(r_{13},r_{14})$.

From the above, without the guidance of available timestamps,
it is difficult to clean the inconsistent and incomplete values. %Without consideration currency issues among records, the repair accuracy and effectiveness are challenged to perform well.
If cleaning them simply with the values from their most similar records, we are likely to obtain wrongly repaired data.% and even run into serious mistakes when using the data.
Thus, the repairing of data quality problems in currency, consistency and completeness together is in demand.

However, the development of the repairing of mixed quality issues is faced with challenges.
Firstly, with the attributes' changing and evolution with time, the temporal and current features in records influence the repairing accuracy, which becomes the key point in data quality management. Moreover, as some overall fundamental problems are already known as computationally hard \cite{Fan2012Foundations,4Fan2012Determining}, multi-errors data repairing makes this problem even more challenged. Worse still, repairing some errors may cause another kind of errors. Without a sophisticated method, it may be costly to repair dirty data due to the iteratively repairing of the errors caused by data repairing.

As yet, works on cleaning multiple errors in completeness, consistency and currency are still inadequate. On the one hand, currency orders are difficult to determine when timestamps are unavailable. Existing currency repairing methods mostly depend on the definite timestamps, and few works provide feasible algorithms or even models for the data with the absence of valid timestamps. On the other hand, though inconsistency and incompleteness coexist in databases, both issues fails to be solved explicitly.

Motivated by this, we study the repairing approach of incompleteness and inconsistency with currency. Both incompleteness and inconsistency can be solved more effectively with currency information. We use an example to illustrate the benefit of currency in data repairing.

For instance, better repairs are shown in Table 2 as marked in green. We deduce a currency order for $r_5$ that the title of an employee in a company is increasing in the real world. Thus, Mike's title can only change from E to ME when he works in the same company. Similarly, the salary is always monotonically increasing. $r_5$ is expected to be more current than $r_4$. We repair $r_5$'s address, city and email with ``\emph{Xixi}'', ``\emph{Hangzhou}'' and ``\emph{M@ali}". The occurrence of dirty data is possibly because the delay between the database update and changes in the real world. If working emails fail to be well-repaired, both employees and companies will suffer losses.

For the dirty records of Helen, we repair $[\textsf{Address}]=``Xuhui"$, $ [\textsf{City}]=``Shanghai"$ of $r_{10}$ with a CFD: ($r_i[\textsf{Company}] = ``Tencent" \wedge  r_i[\textsf{Group} ] = ``Financial" \longrightarrow r_i[\textsf{City}] = ``Shanghai"$). It reveals that Helen has already changed her work to the financial group in Shanghai at P3. It improves the accuracy of her career information. With a currency order: R $\prec_{\textsf{Title}}$ MR can we know Helen has become a MR at P4, and $r_{13}$ is more current than $r_{12}$. According to anther currency order: $P4 \prec_{\textsf{Level}} P5$, $r_{14}$ is the most current and freshness record now. Its missing title and salary are supposed to be filled with the present of most current values, \emph{i.e.}, ``\emph{P5}'' and ``22\emph{k}'', respectively. It indicates that Helen's salary is no less than 22k at P5 as a MR in her group. These cases indicate the complex conditions in dirty data, and the necessary of the interaction method in data cleaning on 3C. From Table 2, the combination of these three issues makes contributions to improve the accuracy of data cleaning.

\textbf{Contributions}. In this paper, we propose a framework of data repair together with currency, consistency and completeness, named \textsf{Improve3C}. To make sufficient usage of currency information hidden in the database, we propose a currency order computation method with currency constraints, which achieves a reliable time-related replacement when the timestamps of the database is not valid. In this way, we are able to discovery and awaken the internal knowledge from records in databases to maximize the repairing effectiveness. We summarize our contributions in this paper as follows:

(1) We propose a comprehensive data repairing approach for consistency, completeness and currency. To the best of our knowledge, it is the first study on data quality improvement on completeness and consistency of the data sets without reliable timestamps.

(2) We propose a 4-step framework \textsf{Improve3C} of multiple data quality problems detection and quality improvement. A total currency order schema is performed by processing the currency order graph with currency constraints.

(3) Moreover, we propose the currency and consistency \emph{Difference} metric between the dirty data and the standard one to repair the inconsistent attributes together with \emph{CFD}s and currency orders. In addition, we propose the solution for repairing incomplete values with naive Bayesian, where the currency order is considered as a key feature for classification training process.

(4) We conduct a thorough experiment on both real-life and synthetic data. The experimental results verify \textsf{Improve3C} can detect and repair the mixed dirty data effectively. Our framework can improve the performance of the existing methods in low-quality data repairing. Our strategy also achieves high efficiency compared with the treatment of the dimensions independently.

\textbf{Organization}. The rest of the paper is organized as follows: Section 2 discusses the basic definitions and the overview of our method. Section 3 introduces construction and conflict detection on currency graph, and Section 4 discusses algorithms and examples for currency order determination. Section 5 (resp. Section 6) presents inconsistency repairing (resp. incompleteness imputation) process. Experimental study is reported in Section 7. Section 8 reviews the related work, and Section 9 draws the conclusion.
\section{Overview}
\label{Overview}
In this section, we first introduce necessary background and fundamental definitions in Section \ref{3.1}, and then propose our method framework \textsf{Improve3C} in Section 2.2.
\subsection{Basic Definitions}
%\subsection{Problem Definition}
\label{3.1}
The \emph{currency constraints} (also named as currency rules) are used to determine the currency of data under the circumstances the timestamps are not available. Definition \ref{definition1} presents the semantic of currency constraints adopted in our method referring to the one proposed in \cite{Fan2012Foundations}. We use \emph{CC}s for short below in this paper.
\newtheorem{definition}{Definition}
\begin{definition}
\label{definition1}
(\emph{Currency constraints}). In the set of currency constraints, $\mathrm{\Phi}=\big\{r_i [\textsf{eID}]=r_j [\textsf{eID}]\wedge \psi\mid i,j\in [1,N]\big \}$, $N$ is the total record number  in dataset $\mathcal{D}$. $r_i$ and $r_j$ are two records in $\mathcal{D}$. $\psi$ represents the predicate in an instance of a \emph{CC}. \textsf{eID} represent ID number to identify the same person. There are mainly three kinds of constraints regarding $\psi$:
\newline \indent $\mathrm{(a)}$ $\psi_1$: $ (r_i [A_\mathrm{k}]=\textrm{Value}[i] \wedge r_j [A_\mathrm{k}]=\textrm{Value}[j] ) \longrightarrow (r_i \prec_{A_\mathrm{k}} r_j) ;$
\newline \indent $\mathrm{(b)}$ $\psi_2$: $ ( r_i [A_\mathrm{k}] \ op \ r_j [A_\mathrm{k}] ) \longrightarrow (r_i \prec_{A_\mathrm{k}} r_j), op \in \{>, <, \geq, \leq,=,\neq\} ;$
\newline \indent $\mathrm{(c)}$ $\psi_3$: $  ( r_i  \prec_{A_\mathrm{k}} r_j ) \longrightarrow (r_i \prec_{A_\mathrm{m}} r_j). $
\newline where $A_k, A_m \in \mathcal{A}$, and $\mathcal{A}$ is the set of attributes in $\mathcal{D}$. Value[$\cdot$] is the value of the attribute. $\prec_{A}$ is the currency order determined on $A$.
\end{definition}
Accordingly, we can draw the currency constraints
adopted in Table 1 as follows:
\newline \indent $\psi_1$: $(r_i[\textsf{Salary}] < r_j[\textsf{Salary}]) \longrightarrow (r_i \prec_{\textsf{Salary}} r_j )$.
\newline \indent $\psi_2$: $(r_i[\textsf{Level}] < r_j[\textsf{Level}]) \longrightarrow (r_i \prec_{\textsf{Level}} r_j )$.
\newline \indent $\psi_3$: $(r_i[\textsf{Company}]= r_j[\textsf{Company}] \wedge r_i[\textsf{Title}]=``E" \wedge r_j[\textsf{Title}]=``ME") \longrightarrow (r_i \prec_{\textsf{Title}} r_j )$.
\newline \indent $\psi_4$: $(r_i[\textsf{Company}]= r_j[\textsf{Company}] \wedge r_i[\textsf{Title}]=``RA" \wedge r_j[\textsf{Title}]=``R") \longrightarrow (r_i \prec_{\textsf{Title}} r_j )$.
\newline \indent $\psi_5$: $(r_i \prec_{\textsf{Title}} r_j) \longrightarrow (r_i \prec_{\textsf{Group}} r_j )$.
\newline \indent The conditional function dependencies (\emph{CFD}s for short) have been developed to detect and resolute inconsistency in a data set or among datasets \cite{3Fan2009Conditional}. Sound researches have been done in inconsistency repairing \cite{Cong2007Improving,8Fan2014Conflict}. Based on this, we adopt \emph{CFD}s in our framework to improve data consistency as discussed in Definition \ref{definition2}.
\begin{definition}
\label{definition2}
(\emph{Conditional functional dependencies}). On a relation schema $\mathcal{R}$, $\mathrm{\Sigma}$ is the set of all the \emph{CFD}s. A \emph{CFD} is defined as $\varphi: \mathcal{R}(A_l \rightarrow A_r, t_\mathrm{p})$, where $A_l$ (resp. $A_r$) is denoted as the antecedent (resp. consequent) of $\varphi$, \emph{i.e.}, LHS($\varphi$), (resp. RHS($\varphi$)). $A_l, A_r \subseteq \mathcal{A}$, where
\newline \indent $\mathrm{(a)}$ $A_l \rightarrow A_r$ is a standard FD, and
\newline \indent $\mathrm{(b)}$ $t_\mathrm{p}$ is a tableau that  either $t_\mathrm{p}[A]$ is a constant value from the attribute value domain $dom(A)$ or an unnamed variable ``\_'' which draws values from $dom(A)$.
\end{definition}
Accordingly, below are some of the \emph{CFD}s the records in Table 1 should satisfy.
\newline \indent $\varphi_1$: $(r_i[\textsf{Address}]=``\_")\longrightarrow (r_i[\textsf{City}]=``\_")$.
\newline \indent $\varphi_2$: $(r_i[\textsf{Company}]=``\_") \longrightarrow (r_i[\textsf{Email}]$ $= ``@\_")$.
\newline \indent $\varphi_3$: $(r_i[\textsf{Company}] =
``Tencent"\wedge r_i[\textsf{Group}]=``Games")\longrightarrow (r_i[\textsf{City}]=``Shenzhen")$.
\newline \indent $\varphi_4$: $(r_i[\textsf{Company}]=``Alibaba"\wedge r_i[\textsf{Group}]=``Tmall")\longrightarrow (r_i[\textsf{City}]=``Hangzhou")$.
\newline \indent $\varphi_5$: $(r_i[\textsf{Company}]=``Baidu"\wedge r_i[\textsf{Group}]=``Map")\longrightarrow (r_i[\textsf{City}]=``Beijing")$.
\newline \indent Further, we introduce the low-quality data with mixed problems. As mentioned above, we focus on three vital quality problems on completeness, consistency and currency, thus, the low-quality data in our study is defined in Definition \ref{D}. We outline our problem definition of 3C-data-quality repairing in Definition \ref{problem}.
\begin{definition}
\label{D}
(\emph{Low-quality Data $\mathcal{D}$}). The schema $\mathcal{R}=($\textsf{$A_1$},...,\textsf{$A_n$}) has no timestamps. Some missing values exist in $A_i$ ($A_i \in \mathcal{A}$) in $\mathcal{D}$, and at the same time some value pairs violate the consistency (including \emph{CFD}s in Definition 2) measures. $\mathcal{D}$ is a set including massive instances like $\mathcal{R}$.
%as $D=\{I_e|e\in \mathcal{E}\}$.
\end{definition}
\begin{definition}
\label{problem}
(\emph{Problem Definition}). Given a low-quality data $\mathcal{D}$, data quality rules including a set $\Phi$ of \emph{CC}s and a set $\Sigma$ of \emph{CFD}s, and a confidence $\sigma$ for each attributes. Data quality improvement on $\mathcal{D}$ with completeness, consistency and currency is
to detect the dirty data in $\mathcal{D}$ and repair it into a clean one, denoted by $\mathcal{D}_r$, where
\newline \indent (a) $\forall r (r \in \mathcal{D}_r) $ has a reliable currency order value satisfying the set $\Phi$ of \emph{CC}s, denoted by $(\mathcal{D}_r, \mathcal{D}) \models \Phi$.
\newline \indent (b) $\mathcal{D}_r$ is consistent referring to the set $\Sigma$ of \emph{CFD}s, \emph{i.e.}, $(\mathcal{D}_r, \mathcal{D}) \models \Sigma$.
\newline \indent (c) The missing values in $\mathcal{D}$ are repaired with the clean ones whose confidence $> \sigma$ into $\mathcal{D}_r$.
\newline \indent (d) The repair cost $cost(\mathcal{D}_r, \mathcal{D})$ is as small as possible.
\end{definition}
 \begin{figure}[t]
\centering
  \includegraphics[scale=0.26 ]{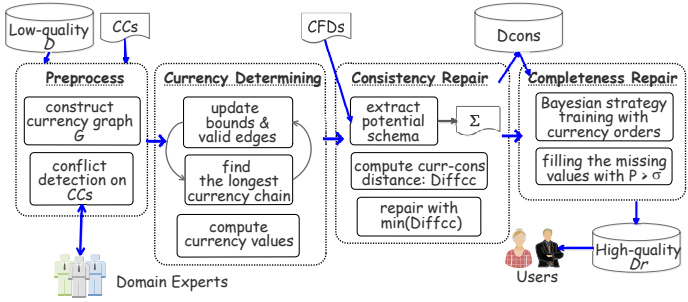}
\caption{Framework overview of Improve3C}
\label{fig:1}       % Give a unique label
\end{figure}
\subsection{Framework}
\label{sec3.2}
We present the proposed 3C data repairing method \textsf{Improve3C} in Figure 1. Since that completeness and consistency are metrics focusing on measuring the quality with features in \emph{values}, while currency describes the temporal order or the volatility of records in the whole data set. We process consistency and completeness repairing in order along the currency order defined in this paper. \textsf{Improve3C} is constructed to serves two purposes: First, each repair operation in \textsf{Improve3C} will not cause any new dirty data which violates one of 3C issues. Second, no dirty data exists on 3C after process \textsf{Improve3C} according to the proposed definitions in this paper. We achieve an overall data repairing on currency, consistency and completeness with the approach \textsf{Improve3C}, which consists four main steps.
\newline \indent (1) We first construct currency graphs for records with the adopted \emph{CC}s, and make conflict detections in the currency graphs. If conflicts exist, the conflicted \emph{CC}s and the involved records will be returned. They are supposed to be fixed by domain experts or revised from business process. This step is introduced in Section \ref{sec3}.
\newline \indent (2) We then determine the currency order of records extracted from \emph{CC}s. We update valid edges and find the longest currency order chain in the currency graph iteratively, and compute currency values to each record. This currency order is obtained as a direct and unambiguous metric among records on currency. Currency order determination is discussed in detail in Section \ref{sec4}.
\newline \indent (3) After that, we repair consistency issues with the global currency orders. We input consistency constraints (\emph{CFD}s in this paper) first, and extract potential consistency schema from the original date set to capture undiscovered consistent tableau. After the consistency schema is determined, we define a metric \emph{Diffcc} to measure the distance between dirty data and clean ones, combining consistency difference with currency orders. We repair the inconsistent data not only according to the consistency schema, but also take into account the currency order, \emph{i.e.}, repair the dirty data with proper values which have the closest current time point. The process is reported in Section \ref{sec5}.
\newline \indent (4) We repair incomplete values with Bayesian strategy in the final step because of its obvious advantages in training both discrete and continuous attributes in relational database. We treat currency orders as a weighted feature and train the complete records to fill in the missing values if the filling probability no less than a confidence measure $\sigma$. Up till now, we achieve high-quality data on 3C. Incompleteness imputation is presented in Section \ref{sec6}.
\newline \indent Specifically, we use \emph{CFD}s as consistency constraints, and other kind of dependencies can be similarly adopted in our framework. We detect and repair consistency problems ahead of completeness in \textsf{Improve3C}, because we are able to repair some missing values (like $r_{10}$ in Table 1) which can be detected by the given \emph{CFD}s. In this case, data completeness achieves a little improvement with consistency solution. The data becomes more complete, beneficial to the accuracy of completeness training model. We can clean the data more effectively for the rest missing values which fails to be captured and fixed by $\Sigma$. Moreover, the repaired part will not give rise to new violation issues on both currency and consistency. On one hand, currency order has been taken into account as an important feature in the training process. The algorithm will provide clean values with nearest currency metrics. On anther hand, the consistency constraints would not let any records escape which have missing and inconsistent values at the same time. With respect to the time costs, the computing time is also decreased in \textsf{Improve3C}.
\section{Conflict Detection in CCs}
\label{sec3}
\begin{figure}[t]
\centering
\subfigure[$\mathtt{E}_1$]{
  \includegraphics[scale=0.23]{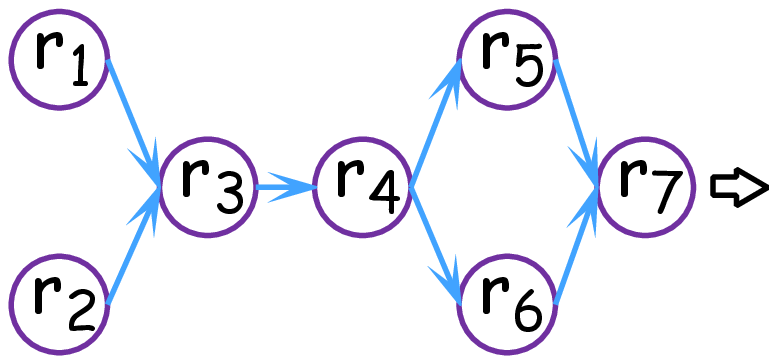}}
\subfigure[$\mathcal{G}_{c1}$ for $\mathtt{E}_1$]{
\includegraphics[scale=0.25]{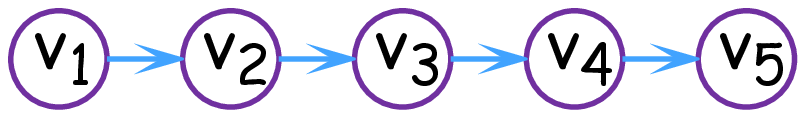}}
\subfigure[$\mathtt{E}_2$]{
  \includegraphics[scale=0.23]{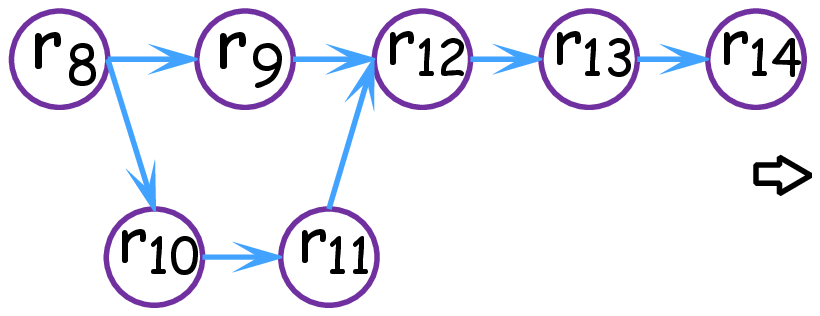}}
\subfigure[$\mathcal{G}_{c2}$ for $\mathtt{E}_2$]{
\includegraphics[scale=0.24]{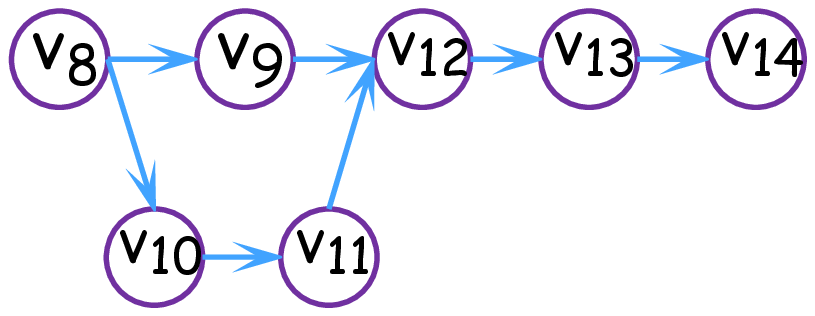}}\caption{Currency graphs for Example 1}
\label{currencygraph}       % Give a unique label
\end{figure}

Conflict resolution of currency constraints is a necessary step in preprocessing for achieving accurate and unambiguous currency order determination. As defined in Definition \ref{CurrencyGraph}, we first construct the directed currency graph $\mathcal{G}_c=(V,E)$ for each entity $\mathtt{E}$ in $\mathcal{D}$, where each $v_i$ in $V$ represents a set of records with the same currency order referring to the same entity. Accordingly, the conflicts on \emph{CC}s can be identified by discovering whether there exists loops in $\mathcal{G}_c$. Conflicts may result from either ambiguous currency constraints or definite currency problems in some records. Without credible external knowledge, these conflicts cannot be resolved. As the conflicts only happen in a small part of data, we detect and return them for artificial process (\emph{e.g.}, repairing by domain experts or assigning crowdsourcing tasks \cite{Li2017Crowdsourced,Zheng2016DOCS}.) The time cost of conflict detection is $O(N)$, where $N$ is the total number of records in $\mathcal{D}$.
\begin{definition}
\label{CurrencyGraph}
(\emph{Currency Graph}). An entity $\mathtt{E}$ has $n$ records in $\mathcal{D}$, denoted by $r(\mathtt{E}) = \{r_1, ... ,r_n\}$. The directed graph $\mathcal{G}_c=(V,E)$ is the currency graph of $\mathtt{E}$,
where $V=\{v[r_i]|r_i \in r(\mathtt{E})\}$ represents the currency order of the records $r_i$ ($i \in [1,n]$) in $r(\mathtt{E})$ concluded by \emph{CC}s. Each $v (v\in V)$
represents a set of records with the same currency order, denoted by $r_j \asymp r_{j+1}, ..., \asymp r_l$. For $v_m$, $v_k$ in $V$, if $v_k$ has higher currency order than $v_m$, \emph{i.e.}, $v_m \prec v_k$, there is an edge $e(m,k)$ from $v_m$ pointing to $v_k$, $(e(m,k) \in E)$, and otherwise $e(k,m)$.
\end{definition}
\begin{example}
According to Definition \ref{CurrencyGraph}, we construct the currency graph for $\mathtt{E}_1$ and $\mathtt{E}_2$ in Example 1 in Figure \ref{currencygraph}. We deduce from the \emph{CC}s in Section 2.1 that $r_1 \asymp r_2$, $r_2 \prec_{\textsf{Title}} r_3$, $r_3 \prec_{\textsf{Salary}} r_4 \asymp r_5$, $r_5 \prec_{\textsf{Level}} r_6$, and $r_6 \prec_{\textsf{Salary}} r_7$ in Figure \ref{currencygraph}(a). $r_1, r_2$ (resp. $r_3, r_4$) is merged to be vertices $v_1$ (resp. $v_3$), as they share the same currency order. Thus, $\mathcal{G}_{c1}= (V_1,E_1) $ is constructed in Figure \ref{currencygraph}(b), where $V_1 = \{v_1, v_2, v_3, v_4, v_5\}$ and $E_1 = \{e(1,2), e(2,3)$ $e(3,4), e(3,5)\}$. Similarly, with $r_8 \prec_{\textsf{Title}} r_9$, $r_9 \asymp r_{10}$, $r_{10} \prec_{\textsf{Level}} r_{11}$, $r_{11} \prec_{\textsf{Salary}} r_{12}$, $r_{12} \prec_{\textsf{Title}} r_{13}$, $\mathcal{G}_{c2} = (V_2,E_2)$ is constructed in Figure \ref{currencygraph}(d), where $E_2 = \{e(1,2), e(2,3), e(3,4), e(4,5)\}$, and $v_2 = \{r_9, r_{10}\}$, $v_3 = \{r_{11}, r_{12}\}$.
\end{example}
\section{Currency Order Determination}
\label{sec4}
\begin{figure}[h]
\centering
\subfigure[$\mathcal{G}_{c1}$]{
  \includegraphics[scale=0.3]{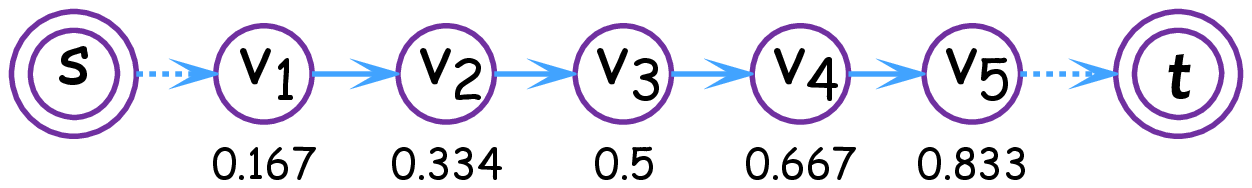}}
\subfigure[$\mathcal{G}_{c2}$]{
\includegraphics[scale=0.3]{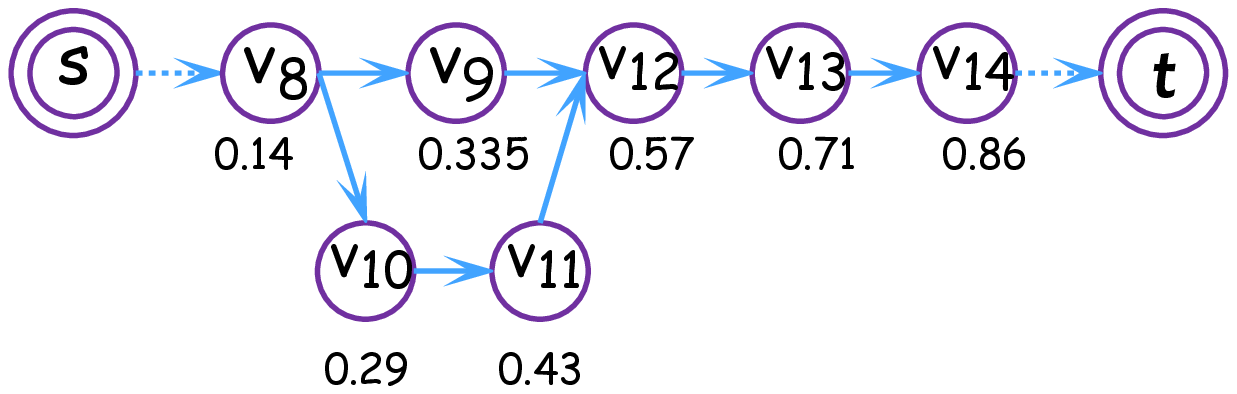}}
\caption{Determine currency values for $\mathcal{G}_{c1}$ and  $\mathcal{G}_{c2}$}
\label{e3}       % Give a unique label
\end{figure}

Since that \emph{CC}s can only describe \emph{partial} orders among values on several target attributes, part of records' currency order still cannot be deduced. Under the circumstances, the data without any currency order reasoning from \emph{CC}s is hard to be evaluated on currency. It motives us to determine data currency on the whole data. We compute and assign currency values to all the vertices in $\mathcal{G}_c$, which achieves an approximate currency order for records.
\newline \indent $\mathcal{G}_c$ becomes a directed acyclic graph after conflict detection. We assign currency order values to all the vertices in $\mathcal{G}_c$ to make all the records comparable on currency. An intuitive approach is to perform topological sorting on $\mathcal{G}_c$ and determine currency order on the sorting results. Unfortunately, the topological sorting result is not always stable \cite{DBLP:books/daglib/0023376}, which could be influenced by the order of graph construction or other external factors. On this occasion, we propose a currency order determination method, which computes currency values more precisely. To some extent, the currency order is a kind of replacement of timestamps when the real timestamps are not available in database. Accordingly, the currency of data is uncovered and the metrics on it assist data quality resolutions on both consistency and completeness.
\newline \indent In currency graphs like $\mathcal{G}_{c2}$ in Example 2, the currency-comparable records of the same entity make up chains, which assists to determine currency values of the graph. We now present the definition of the currency order chains in Definition \ref{Sequence}. Accordingly, the directed edge $e(i,j)$ connects two elements (vertices) $v_i$ and $v_j$ in a currency order chain, where $v_i \prec v_j$, \emph{i.e.}, the records represented by $v_j$ are more current than the ones in $v_i$.
\begin{definition}
\label{Sequence}
(\emph{The Currency Order Chain}) $\mathcal{S}=\{v_1,...,v_m\}$ is a currency order chain of the currency graph $\mathcal{G}_c=(V,E)$, \textbf{iff}.
\newline \indent (a) $\forall$ $v_k \in \mathcal{S}, k \in[1,m)$, there exists an edge $e(k,k+1)$, and $v_k \in V $ and $e(k,{k+1}) \in E$, and
\newline \indent (b) $\forall$ $v_k \in \mathcal{S}, k \in[1,m)$, then $v_k \prec v_{k+1}$.
\end{definition}
When determining currency orders, we are supposed to assign values to the currency order chains in $\mathcal{G}_c$ first. In order to achieve a uniform and accurate determination of currency orders, we propose the currency value computing approach following two steps: (1) We compute and update the currency order bounds of the vertices in $\mathcal{G}_c$, and (2) find the present longest valid chains $\mathcal{S}_{max}$ and value each element in it in ascending order, denoted by CurrValue($v$), ($v \in \mathcal{S}_{max}$).
We recursively repeat the two steps until all the chains have been visited and all the vertices are valued.
\begin{algorithm}[t]
\caption{CurrValue}
\label{CurrValue}
\LinesNumbered
%\begin{scriptsize}
\KwIn{  the currency graph $\mathcal{G}_c=(V,E)$ of the entity $\mathtt{E}$}
\KwOut{  $\mathcal{G}_c$ = (CurrValue($V$), $E$)}
add \textbf{s} and \textbf{t} to $\mathcal{G}_c$, let \textbf{s} points to all 0 in-degree edges and \textbf{t} be pointed from all 0 out-degree edges\;CurrValue(\textbf{s}), \textsf{sup}(\textbf{s}), \textsf{inf}(\textbf{s}) $\leftarrow 0$, CurrValue(\textbf{t}), \textsf{sup}(\textbf{t}), \textsf{inf}(\textbf{t}) $\leftarrow 1$\;
\While {$\exists$ CurrValue($v_i$) has not been determined, $(v_i \in V)$ }
 {\textsf{UpdateValid}($\mathcal{G}_c,\textsf{sup},\textsf{inf}$)\;
$\mathcal{S} \leftarrow$ \textsf{getMaxCandS}($\mathcal{G}_c$), $k \leftarrow |\mathcal{S}|$\;
Value $\leftarrow$ \textsf{inf}($\mathcal{S}[1]$), Inc $\leftarrow$ $\frac{\textsf{sup}(\mathcal{S}[k])-\textsf{inf}(\mathcal{S}[1])}{k+1}$\;
\For {$v \in \mathcal{S}\backslash\{\mathcal{S}[1],\mathcal{S}[k]\}$}
%\STATE Value $\leftarrow$ Value + Inc;
{CurrValue($V$) $\leftarrow$ Value + Inc\;}
}
%\ENDFOR
%\ENDWHILE
return $\mathcal{G}_c$ = (CurrValue($V$), $E$);
%\end{scriptsize}
\end{algorithm}
\newline \indent When finding $\mathcal{S}_{max}$, each CurrValue($v$) is computed depended on the possible minimum and maximum values of $v$, as well as the relative position of $v$ in the involved $\mathcal{S}_{max}$. We adopt the currency order bound to describe these possible min and max values in Definition \ref{Bound}. $\textsf{sup}(v)$ and $\textsf{inf}(v)$ are vital factors for discovering currency order and updating currency values for vertices. The bounds make the value range of CurrValue($v$) as accurate as possible.
\begin{definition}
\label{Bound}
(\emph{The Currency Order Bounds}). When determining currency values, the upper and lower bound of a vertex $v_i$ in $\mathcal{S}=\{v_1,...v_m\}$, ($i \in [1,m)$) is defined as:
\newline \indent (a) The \emph{upper currency order bound} of $v_i$ is $\textsf{sup}(v_i) =$ $\min$\{CurrValue($v[i \cdot]$)\}. $v[i \cdot]$ represents the descendant vertex connecting from $v_i$.
\newline \indent (b) The \emph{lower currency order bound} of $v_i$ is $\textsf{inf}(v_i) =$ $\max$ \{CurrValue($v[\cdot i]$)\}, where $v[\cdot i]$ represents the ancestor vertex connecting to $v_i$.
\end{definition}
The whole computing process is shown in Algorithm \ref{CurrValue}. We first add a global start and terminal node \emph{i.e.}, \textbf{s} and \textbf{t} to the graph to ensure all currency orders are located in the domain $(0,1)$. \textbf{s} points to all 0-in-degree vertices, and its currency value and bounds is set 0. Similarly, \textbf{t} are connected from all 0-out-degree vertices and CurrValue(\textbf{t})=\textsf{sup}(\textbf{t})=\textsf{inf}(\textbf{t})=1. After that, we begin to compute currency values of vertices.
\newline\indent  In lines 3-11, we repeatedly find the longest candidate chain in $\mathcal{G}_c$ and compute currency values of the elements in it (Algorithm 2). In the loop, we update $\mathcal{S}$'s present bounds, and determine the validation of the involved edges (line 4). This function will be outlined in Algorithm \ref{UpdateValid} below.
\newline\indent After that, we find the present longest candidate chain $\mathcal{S}$ in line 5 (Algorithm \ref{getMaxSequence}), where $k=|\mathcal{S}|$ is the length of $\mathcal{S}$,\emph{ i.e.}, the number of elements in $\mathcal{S}$. Next, we assign normalized currency values to each $v$ in $\mathcal{S}$ in lines 7-10. Since that bounds are determined, we use the lower (resp. upper) bound of the first (resp. last) element \textsf{inf}($\mathcal{S}[1]$) (resp. \textsf{sup}($\mathcal{S}[k]$)) in $\mathcal{S}$ to compute currency values of all elements in $\mathcal{S}$. Finally we obtain the valued currency graph of $\mathtt{E}$.
\begin{example}
We now determine currency values in $\mathcal{G}_{c1}$ and $\mathcal{G}_{c2}$. In Figure \ref{e3}(a), $\mathcal{S}_{max} =\{v_1,v_2,v_3,v_4 \}$ is found after insert \textbf{s, t} to the graph. For each vertex in $\mathcal{S}_{max}$, CurrValue($v_i$)= CurrValue($v_1$) + $\frac{\textsf{sup}(v_k)-\textsf{inf}(v_1)}{k+1}, k=4$. For $\mathcal{G}_{c2}$ in Figure \ref{e3}(b), we find $\mathcal{S}_{max} = \{v_8, v_{10}, v_{11}, v_{12}, v_{13}, v_{14}\}$ and compute CurrValue($v_i$) in it to be $\{0.14, 0.29, 0.43, 0.57,$ $0.71, 0.86\}$. After that, only remain $v_9$'s currency value has not been determined. We use $\textsf{sup}(v_{12})$ and $\textsf{inf}(v_{8})$ to obtain CurrValue($v_9$)= 0.335.
\end{example}
Next, we address the two main steps in currency order determination in detail. We introduce bounds and valid edges update process in Section \ref{4.1}, and discuss the longest candidate chain discovery in \ref{4.2}.
\subsection{Updating Bounds and Valid Edges}
\label{4.1}
%Due to the fact that the directed edges in the currency graph $\mathcal{G}_c$ are deduced from the partial currency order described by \emph{CC}s. These edges do form several chains in $\mathcal{G}_c$.
As mentioned above, a chain reveals a length of transitive currency orders deduced from part of currency order described by \emph{CC}s, and different chains may come cross through vertices. Thus, not all edges contribute to find the longest chain of $\mathcal{G}_c$ during each iteration. During the computing course, we are supposed to determine whether a vertex can make up $\mathcal{S}_{max}$ by computing the bounds of it.
\newline \indent The edges selected to form $\mathcal{S}_{max}$ are called \emph{valid} edges in this paper. That is, the candidate $\mathcal{S}_{max}$ exists in the currency order chains forms with valid edges. We update the validation of the present edges with Definition \ref{Valid} during each iteration. Thus, we can effectively find $\mathcal{S}_{max}$ according to these valid edges (discussed in Section \ref{4.2}).
\begin{definition}
\label{Valid}
(\emph{Validation of Edges}).  The edge $e(i,j)$ is a valid edge ($e(i,j)\in E$) under three cases:
\newline \indent (a) If both CurrValue($v_i$) and CurrValue($v_j$) has not determined, $e(i,j)$ is a valid edge \textbf{iff}. $\textsf{sup}(v_i)$ = $\textsf{sup}(v_j)$, and $\textsf{inf}(v_i)$ = $\textsf{inf}(v_j)$.
\newline \indent (b) If CurrValue($v_i$) is determined and CurrValue($v_j$) is not, $e(i,j)$ is a valid edge \textbf{iff}. $\textsf{inf}(v_i)$ = $\textsf{inf}(v_j)$.
\newline \indent (c) If CurrValue($v_j$) is determined and CurrValue($v_i$) is not, $e(i,j)$ is a valid edge \textbf{iff}. $\textsf{sup}(v_i)$ = $\textsf{sup}(v_j)$.
%\textbf{iff}. $\forall$ $v_k \in \{v_i,v_j\}$, both $\textsf{sup}(v_k)$ and $\textsf{inf}(v_k)$ can be directly determined by $e(i,j)$, when the currency value of $v_k$ has not been determined yet.
\end{definition}
Note that if the currency values on both $v_i$ and $v_j$ is determined, $e(i,j)$ is certainly not a valid edge, because $v_i$ and $v_j$ have been already visited in previous iterations. As we have obtained their currency values, $e(i,j)$ will not be valid in the present updating function.
\begin{algorithm}[t]
\caption{UpdateValid}
\LinesNumbered
\label{UpdateValid}
%\begin{scriptsize}
%\scriptsize{}
%\begin{algorithmic}[1]
\KwIn{  the currency graph $\mathcal{G}_c$, \textsf{sup} and \textsf{inf}}
\KwOut{  the updated $\mathcal{G}_c$, \textsf{sup} and \textsf{inf}.}
mark all the $e (e \in E)$ of $\mathcal{G}_c$ as invalid edges\;
\textsf{UpdateOneWay}($\mathcal{G}_c$, \textsf{inf}, $>$)\;
\textsf{UpdateOneWay}($\mathcal{G}_c^{T}$, \textsf{sup}, $<$)\;
\ForEach {$e(i,j) \in E$}{
\If {\textrm{(\textsf{inf}[$v_i$] = \textsf{inf}[$v_j$] $\vee$ CurrValue($v_i$) is not determined) and (\textsf{sup}[$v_i$] = \textsf{sup}[$v_j$] $\vee$ CurrValue($v_j$) is not determined)}}
{label $e(i,j)$ as a valid edge\;}
}
\textbf{Function} \textsf{UpdateOneWay}($\mathcal{G}$, bound $\in$ \{\textsf{sup}, \textsf{inf}\}, $op$ $\in$ \{$<,> $\})\;
\While {$\exists v_i (v_i\in V)$ with 0 in-degree}{
%\STATE {find a vertex $v_i$ with 0 in-degree};
\ForEach {$e(i,j)$}{
\If {$op$(bound[$v_i$], bound[$v_j$])}{
bound[$v_j$] $\leftarrow$ bound[$v_i$]\;}
}
$V \leftarrow V\backslash v_i$\;
}
 \textbf{end Function}\;
restore all $v_i \in V$\;
\Return the updated $\mathcal{G}_c$, \textsf{sup}, \textsf{inf}\;
%\end{algorithmic}
%\end{scriptsize}
\end{algorithm}
\newline \indent Algorithm \ref{UpdateValid} shows the update process of the valid edges and the bounds. We first mark all edges in $\mathcal{G}_c$ as invalid edges. We use the vertex $v_i$ with its determined CurrValue($v_i$) to update the lower bound \textsf{inf}($v[i\cdot]$) of the vertices reachable from $v_i$. Similarly, we update \textsf{sup}($v[i\cdot]$) on the converse graph of $\mathcal{G}_c$ (Line 2-3). Both bounds are updated via a one-way function \textsf{UpdateOneWay}. During the function, (we might as well take \textsf{inf} updating for example), we recursively chose a $v_i$ with 0 in-degree, and enumerate all $v[i\cdot]$. We compare the \textsf{inf} values between $v_i$ and $v[i\cdot]$. If \textsf{inf}($v[i\cdot]$)$ >$ \textsf{inf}($v_i$), \textsf{inf}($v[i\cdot]$) will be updated with \textsf{inf}($v_i$) (Lines 12-13). After all $e(i,\cdot)$ are processed, we (temporarily) removed $v_i$ from $V$ (Line 16). After the function, we enumerate all edges in $E$, and determined whether the edge is a valid one according to Definition \ref{Valid} (Lines 5-6). Finishing validation determination, we recover the vertices deleted in previous iterations and $\mathcal{G}_c$ with updated bounds and labeled valid edges will be returned to Algorithm 1.
\newline \indent Since the structure of $\mathcal{G}_{c1}$ and $\mathcal{G}_{c2}$ is simple, we discuss another case in Example \ref{example4} to present the steps of our method. It is clear and valid to show how the method works on the records with a more complex currency relations.
\begin{example}
\label{example4}
Figure \ref{e4} shows a currency graph $\mathcal{G}_{c3}$, and the present longest chain is $\mathcal{S}_{max(1)}=\{v_1,v_2,v_3,v_4,v_5,$ $v_6,v_7\}$ in Figure \ref{e4}(a), with the present valid edges are marked in blue lines. With the computed CurrValue($v_i$) ($i \in \{1,...,7\}$), we update bounds of the rest vertices, \emph{i.e.}, $v_8,...,v_{12}$, and find next $\mathcal{S}_{max}$ in the rest chains. In Figure \ref{e4}(a), $v_8,v_9,v_{10}$ and $v_{11}$ all reach $v_6$, which is the vertex with the min currency value among descendant vertices of them. According to Definition \ref{Bound}, \textsf{sup}($v_8$, $v_9,v_{10},v_{11}$) = CurrValue($v_6$)=0.75. $v_{12}$ only reaches $v_7$, so \textsf{sup}($v_{12}$) = CurrValue($v_7$)=0.875. Similarly, in the converse graph of $\mathcal{G}_{c3}$, the max\{CurrValue($v[\cdot i]$)\} reachable from $v_8,v_9$ is $v_1$, while $v_{10},v_{11}$ and $v_{12}$ reach $v_2$ in Figure \ref{e4}(b). \textsf{inf}($v_8,v_9$) = CurrValue($v_1$) =0.125, and \textsf{inf}($v_{10},v_{11},v_{12}$)=CurrValue($v_2$)=0.25.
\begin{figure*}[t]
\centering
\subfigure[Update \textsf{sup} for $\mathcal{S}_{max(1)}$]{
  \includegraphics[scale=0.31]{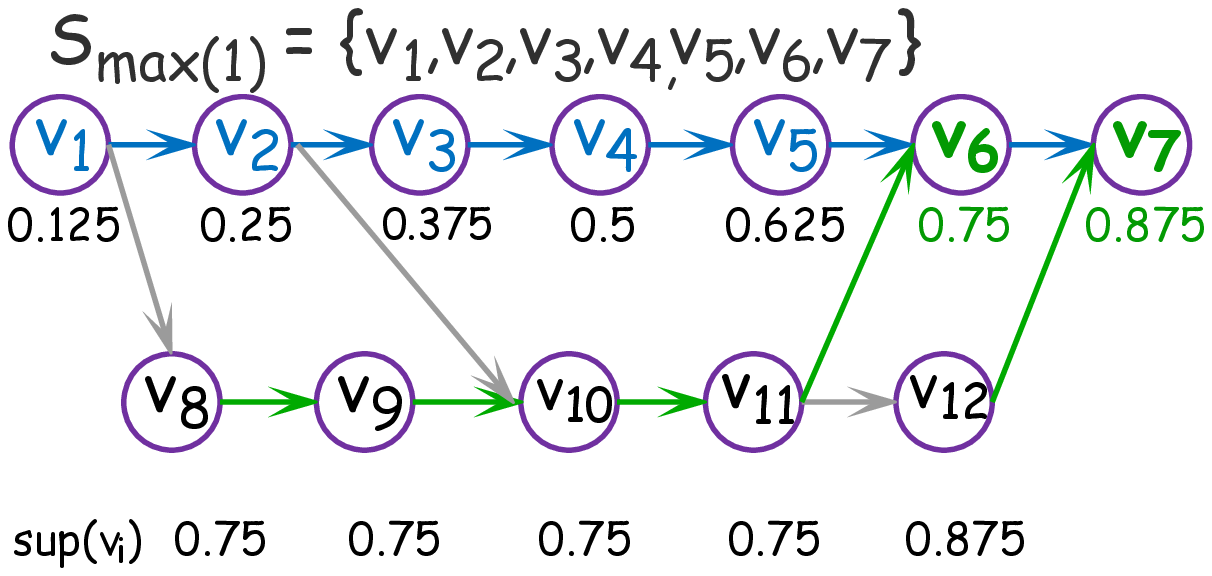}}
\subfigure[Update \textsf{inf} for $\mathcal{S}_{max(1)}$]{
\includegraphics[scale=0.31]{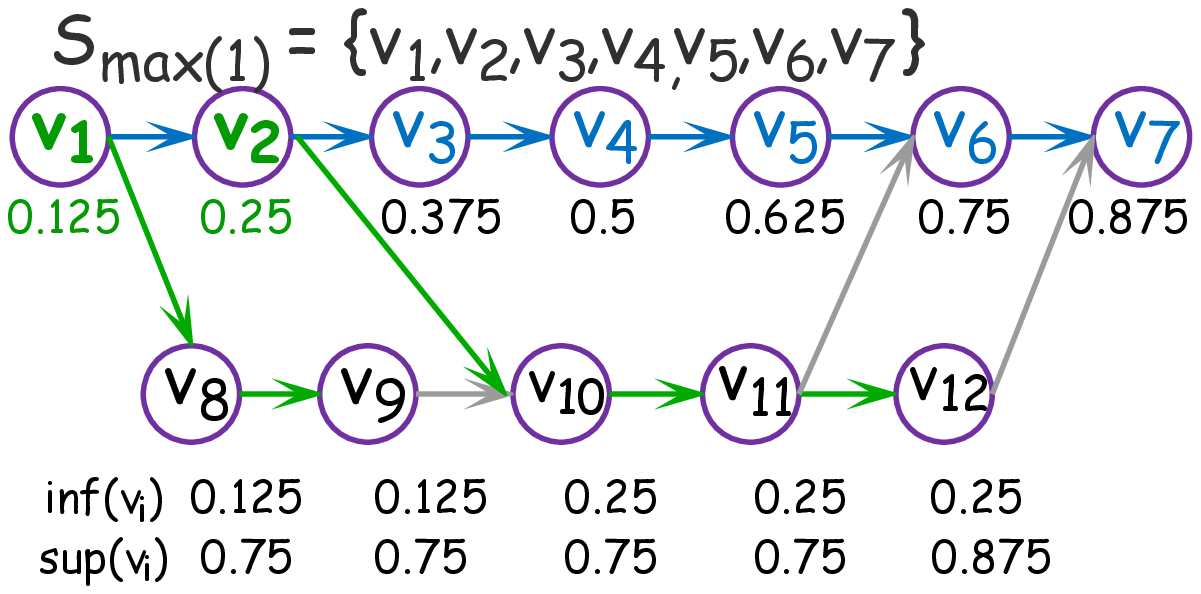}}
\subfigure[Find present valid edges]{
\includegraphics[scale=0.31]{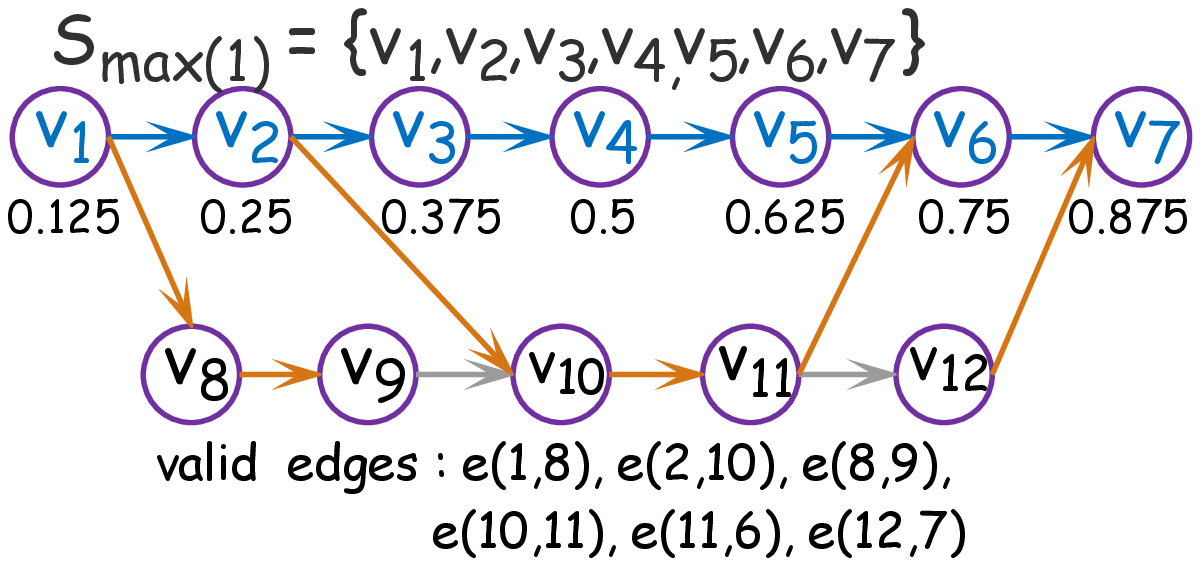}}
\subfigure[Find candidate $\mathcal{S}$]{
  \includegraphics[scale=0.33]{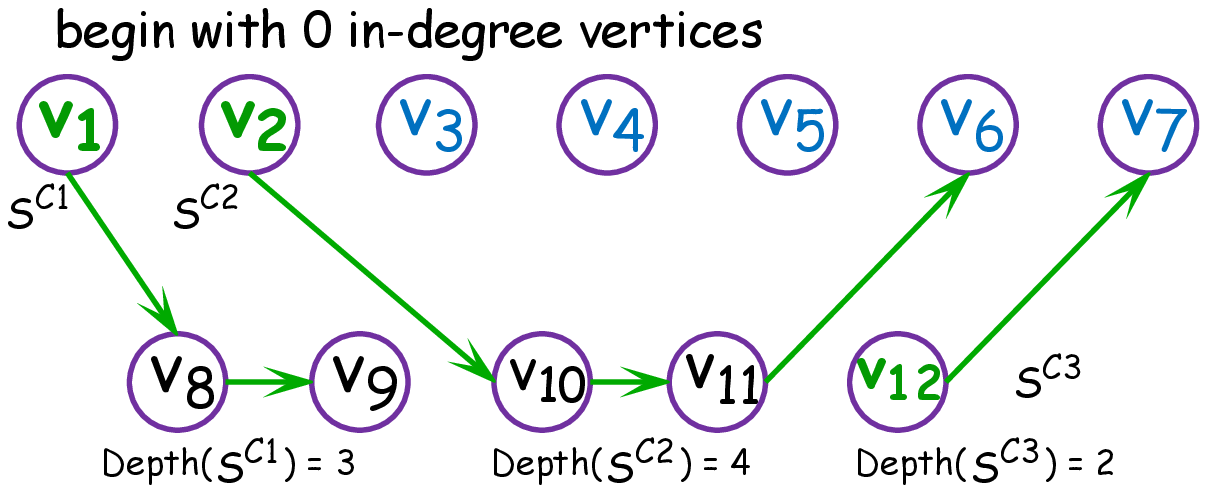}}
\subfigure[Find out $\mathcal{S}_{max(2)}$]{
\includegraphics[scale=0.32]{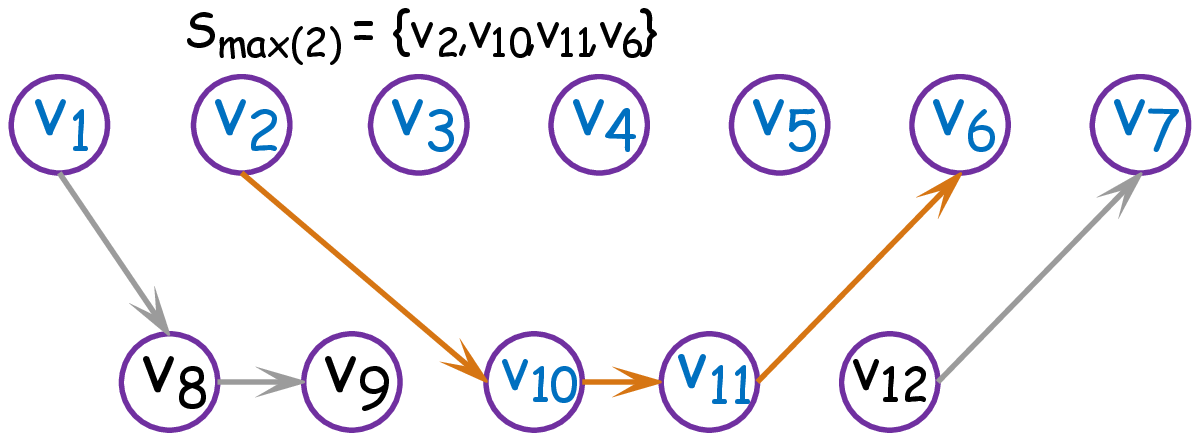}}
\subfigure[Find present valid edges]{
\includegraphics[scale=0.31]{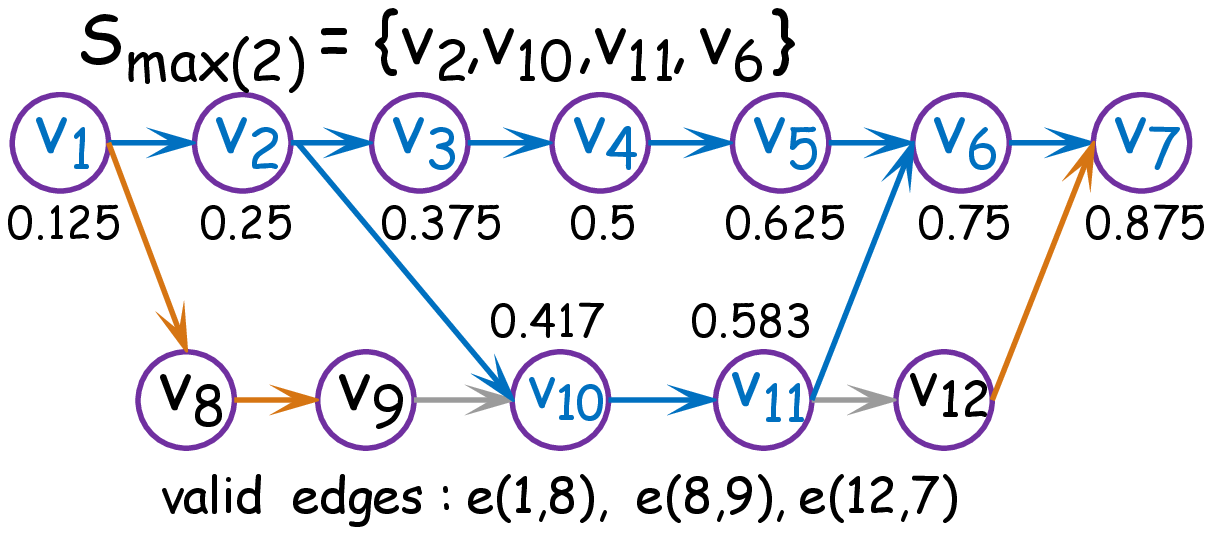}}
\subfigure[Find out $\mathcal{S}_{max(3)}$]{
  \includegraphics[scale=0.31]{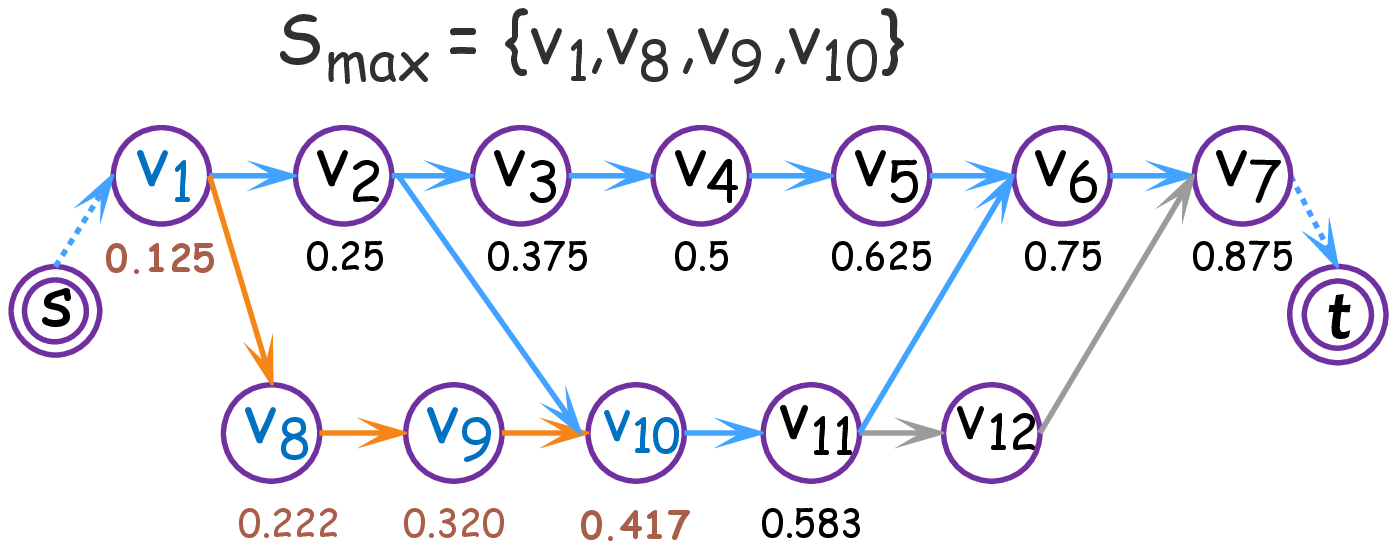}}
\subfigure[Find out $\mathcal{S}_{max(4)}$]{
\includegraphics[scale=0.31]{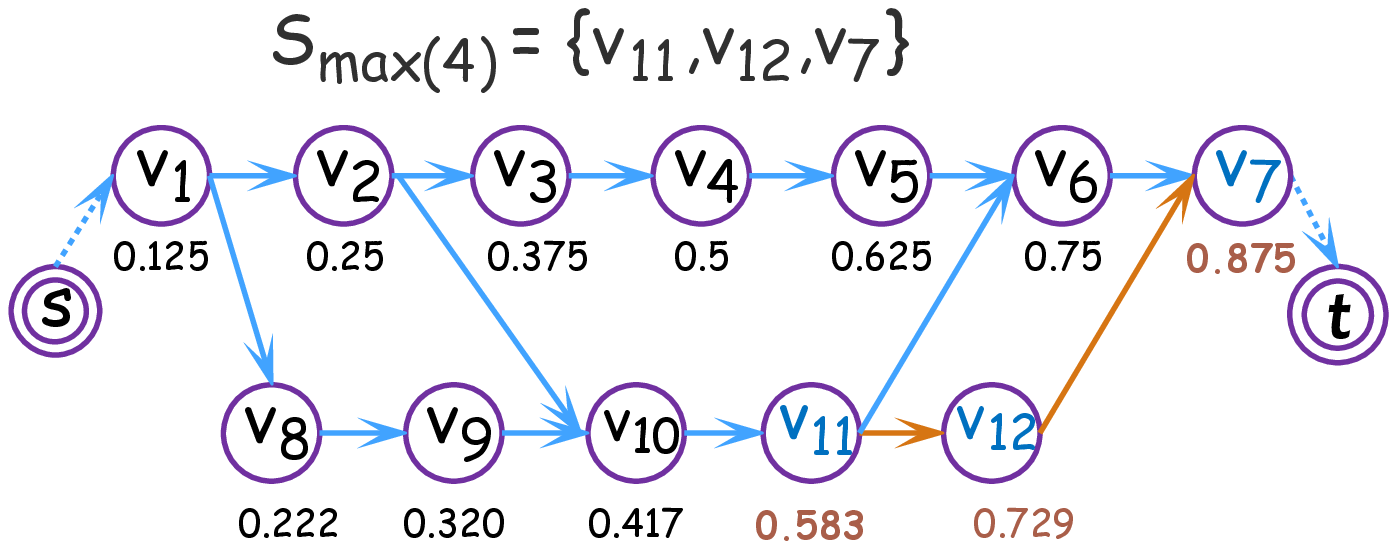}}
\caption{Examples of updating valid edges}
\label{e4}       % Give a unique label
\end{figure*}
\newline \indent As currency values of $v_8,v_9,v_{10},v_{11}$ are not determined, and $v_8, v_9$ (resp. $v_{10}, v_{11}$) has the same \textsf{sup} and \textsf{inf}. $e(8,9)$ and $e(10,11)$ are marked valid according to Definition \ref{Valid}(a). Similarly, $e(1,8)$ and $e(2,10)$ (resp. $e(11,6)$ and $e(12,7)$) are valid referring to Definition \ref{Valid}(a) (resp. Definition \ref{Valid}(c)). The valid edges are marked in orange lines in Figure \ref{e4}(c).
\end{example}
\begin{algorithm}[t]
\caption{getMaxCandS}
\label{getMaxSequence}
%\begin{scriptsize}
%\scriptsize{}
%\begin{algorithmic}[1]
\LinesNumbered
\KwIn{the currency graph $\mathcal{G}_c$}
\KwOut{the longest candidate chain $\mathcal{S}_{\mathrm{max}}$}
Depth $\leftarrow$ 0, pre[ ] $\leftarrow$ Null\;
endDepth $\leftarrow 0$, endPoint $\leftarrow $ Null\;
\While {$\exists v_i (v_i\in V$) with 0 in-degree}{
\If {CurrValue($v_i$) is determined}{
\If {Depth[$v_i$] $>$ endDepth}
{endDepth $\leftarrow$ Depth[$v_i$]\;
endPoint $\leftarrow$ $v_i$\;
}
Depth[$v_i$] $\leftarrow$ 0\;
}
\ForEach {$e(i,k) \in E,$}{
\If {$e(i,k) $ is a valid edge and Depth[$v_i$]+1 $>$ Depth[$v_k$]}{
Depth[$v_k$] $\leftarrow $ Depth[$v_i$]+1\;
pre[$v_k$] $\leftarrow v_i$\;
}
}
delete $v_i$ from $V$\;
}
$\mathcal{S}_{\emph{max}} \leftarrow $ the $\mathcal{S}^{c}$ with endPoint and pre[ ]\;
restore all $v_i \in V$\;
\Return $\mathcal{S}_{\emph{max}}$;
%\end{algorithmic}
%\end{scriptsize}
\end{algorithm}
%\indent \indent \emph{Example 3}.
\subsection{Finding the Longest Candidate Chain}
\label{4.2}
We now introduce how to find the longest candidate chain $\mathcal{S}_{max}$. As the bounds and valid edges are updated (in each iteration), we discover $\mathcal{S}_{max}$ among the vertices connected by valid edges. We first present the definition of candidate chains $\mathcal{S}^c$ in Definition \ref{Candidate}.
\begin{definition}
\label{Candidate}
(\emph{The Candidate Currency Order Chain}). A currency order chain $\mathcal{S}=\{v_1,...,v_m\}$ is a candidate one, denoted by $\mathcal{S}^{c}$, \textbf{iff}.
\newline \indent $\mathrm{(a)}$ CurrValue($v_1$) and CurrValue($v_m$) are known, where $v_1$, $v_m$ is the starting and ended element in $\mathcal{S}$, repectively.
\newline \indent $\mathrm{(b)}$ $ \forall k \in [1,m)$, the directed edge $e(k,{k+1})$ is a valid edge.
\end{definition}
Based on the breadth-first search method, the algorithm \textsf{getMaxCandS} finds the current longest candidate currency order chain among all valid edges. The pseudocode is outlined in Algorithm \ref{getMaxSequence}. We perform topological sorting in lines 3-18 until all vertices in $V$ have been visited. According to Definition \ref{Candidate}, $\mathcal{S}^{c}$ cannot contain such $v_i$ that CurrValue($v_i$) is determined. Thus, when the sorting process arrives line 4, we update the current chain.
For $v_i$ whose CurrValue($v_i$) is not computed, we enumerate all edges beginning from $v_i$, and update each $\mathcal{S}$'s depth with valid edges (lines 12-16). If we reach any invalid edge, we quit the present chain because it cannot form a $\mathcal{S}^{c}$ any longer. We finally restore the edges deleted in pervious computing steps and obtain $\mathcal{S}^{c}_{max}$.
%\indent \indent Algorithm \ref{getMaxSequence} is based on the traditional longest path discovery on directed acyclic graph \cite{?}. The (only) difference between them is \underline{.....}.
%只在已确定时效值点和未确定时效值点进行深度更新，或在两个未确定时效值且由有效边相连的点进行深度更新，以保证选出的链一定是一条“候选链”
\begin{example}
\label{example5}
We continue to introduce finding $\mathcal{S}^c_{max}$ in $\mathcal{G}_{c3}$ from Example 4. As the valid edges have been determined in orange lines in Figure \ref{e4}(d), we find the candidate chains beginning with 0-in-degree valid vertices, \emph{i.e.}, $v_1$, $v_2$ and $v_{12}$, and let them be $\mathcal{S}_1$, $\mathcal{S}_2$ and $\mathcal{S}_3$, respectively. We update the depth of $\mathcal{S}_1$ with the valid $e(1,8)$ and $e(8,9)$. When it reaches $v_{10}$, it is not a $\mathcal{S}^{c}$ any longer, because $e(9,10)$ is not valid. Thus, the depth of $\mathcal{S}^{c1}$ is 3. Similarly, As $e(2,10), e(10,11)$ and $e(11,6)$ are valid, we update Depth($\mathcal{S}^{c2}$) = 4 when it finally reaches $v_6$, while $\mathcal{S}^{c3}$ reaches $v_7$ and Depth($\mathcal{S}^{c3}$) = 2. Thus, the present longest candidate chain is obtained, \emph{i.e.}, $\mathcal{S}^{c}_{max(2)}=\mathcal{S}^{c2}= $\{$v_2,v_{10},v_{11},v_{6}$\} in Figure \ref{e4}(e). We compute CurrValue($v_{10}$) = 0.417 and CurrValue($v_{11}$) = 0.583, according to the currency value of $v_2$ and $v_6$.
\newline \indent The determined vertices are marked in blue in Figure \ref{e4}(f), and we iteratively carry out the above steps. The third longest $\mathcal{S}$ is $\mathcal{S}^{c}_{max(3)} = \{v_1,v_{8},v_{9},v_{10}$\}, thus, CurrValue($v_8$) = 0.222, and CurrValue($v_9$) = 0.320. Finally, we obtain CurrValue($v_{12}$) = 0.729. The currency value determining on the whole graph is finished.
\end{example}
\indent \indent \emph{Complexity}. \textsf{UpdateValid} (Algorithm 2) and \textsf{getMaxCandS} (Algorithm 3) are two steps within the outer loop in Algorithm 1, \textsf{CurrValue}. In Algorithm 2, the \textsf{UpdateOneWay} function costs $O(|V|+|E|)$ time to update bounds of all vertices. It takes $O(|E|)$ to determine valid edges. Thus, Algorithm 2 runs in $O(|V|+|E|)$ time. For Algorithm 3, it takes $O(|V|+|E|)$ in total to find out $\mathcal{S}_{max}$. When computing currency values, the outer loop (lines 3-11) in Algorithm \textsf{CurrValue} costs $O(|V|)$ for the worst. Thus, Algorithm \textsf{CurrValue} takes $O\big(|V|\cdot(|V|+|E|)\big)$ in total.
\newline \indent After the currency orders of records are determined, we further consider repair the inconsistent and incomplete dirty data. In order to achieve no violation on both consistency and completeness after the whole repair, we address inconsistency issues first, and then resolve incompleteness ones.
\section{Inconsistency Repair}
\label{sec5}
As mentioned above, with the attributes evolution among records, currency and consistency issues as well as the interaction between them are both critical to repairing the dirty data violating the constraints (like \emph{CFD}s). To achieve consistency cleaning effectively, we propose an inconsistency repair method with the currency orders obtained above. We first put forward a thought of potential consistency schema extraction in Section \ref{sec5.1}, and then introduce the consistency repair algorithm \textsf{ImpCCons} together with cases study in Section \ref{sec5.2}.
\subsection{Consistency Schema Extraction}
\label{sec5.1}
\emph{CFD}s are used as a general kind of consistency evaluation measure and data quality rule to describe whether the data is clean or not \cite{Fan2012Foundations}. At the meanwhile, the challenges cannot be ignored that high-quality \emph{CFD}s are not easily to be both manually designed and automatically discovered. In this case, some relation schema within attribute values in certain data set may fail to be captured. In the third step of \textsf{Improve3C}, we consider to address a \emph{reliable} relation among enormous records for $\mathcal{D}$ besides \emph{CFD}s in order to detect and repair the violation in data more precisely and sensitively.
\newline \indent For the potential relation on some attributes among records which cannot be process by \emph{CFD}s, we count the total occurrence number $M$ of a schema $\varphi^{+}=(\mathcal{A}_l \rightarrow A_r,t_p$) in $\mathcal{D}$, as the same form with \emph{CFD} in Definition 2. If the ratio between $M$ and the total record number $N$ achieve a given threshold, \emph{i.e}., $\frac{M}{N}\geq \theta^{+}$, we called $\varphi^{+}$ a \emph{reliable schema} in $\mathcal{D}$. Such $\varphi^{+}$ will be added to the consistency constraint set $\Sigma$. The expanded set $\Sigma$ will be applied to guide the repairing of the inconsistent data in $\mathcal{D}$.
\newline \indent This step can be treated as an alternative step to extensively consider the consistency dependencies specifically from certain data set beyond \emph{CFD}s if necessary. Works in \cite{Fan2011Discovering,Papenbrock2015Functional} has been done to discover reasonable functional dependencies, which can guide the setting of $\Sigma$ in our framework. We omit the detailed explanation of the extraction due to the limited space.
\subsection{Algorithm ImpCCons}
\label{sec5.2}
%\underline{And then, we improve both consistency and currency with the proposed \emph{Difference} distance functions.}
We now propose the repairing method considering the effect from both consistency and currency. Intuitively, to repair a dirty record $r$ with a (at least relative) cleaned one, we are supposed to measure the distance (sometimes, the cost) of $r$ with the standard schema. In our method, we first detect the record $r$ violating any $\varphi$ in $\Sigma$, and then compute the consistency-currency distance between $r$ and its neighbor clean records. We also compute the distance between $r$ with the tableau $(t_p)$ of the $\varphi$ violated by $r$. We repair $r$ with the minimum distance.
\newline \indent To address the interaction between consistency and currency, we measure the difference between $r$ with the standard one by the distance of consistency together with currency, \emph{i.e.}, to find the consistent data with the closest currency value. We first present the distance functions on consistency and currency, respectively.
\newline \indent Equation (1) shows the consistency distance between two records denoted by \textsl{cons}$Dist(r_i,r_j)$. $\textsl{Bin}$ is a Boolean function that $\textsl{Bin} (i,j) = 1,$ if $i = j$, and $\textsl{Bin} (i,j) = 0$, otherwise. Equation (2) shows the consistency distance between a record and a $\varphi$ measuring the distance on both LHS($\varphi$) and RHS($\varphi$) referring to Definition \ref{definition2}. $|\textrm{LHS}(\varphi)|$ (resp. $|\textrm{RHS}(\varphi)|$) is the number of attributes in $\mathcal{A}_l$ (resp. $\mathcal{A}_r$). In general, Equation (1),(2) measures the consistency distance as the ratio of the number of violations in the involved attributes. This kind of distance is widely adopted in records distance and similarity measurement \cite{Fan2012Foundations}.
\newline \indent It is sometimes traditionally assumed there are less violations in LHS($\varphi$) than RHS($\varphi$). Repairing methods usually focus on the violations in RHS($\varphi$). However, the violations in LHS($\varphi$) make things even worse and may results in detecting mistakes. Thus, we treat both LHS and RHS equally when computing consistency distance.
\begin{gather}
\label{consDist}
\textsl{cons}{Dist}(r_i,r_j) =  \frac{\sum_{A_k\in \mathcal{A}} \textsl{Bin}(r_i[A_k],r_j[A_k])}{|\mathcal{A}|} \\
\textsl{cons}{Dist}(r_i,\varphi) =  \frac{\sum_{A_k\in \textrm{-HS}} \textsl{Bin}(r_i[A_k],\varphi[A_k])}{|\textrm{LHS}(\varphi)|+|\textrm{RHS}(\varphi)|}
\end{gather}
\indent Equation (\ref{currDist}) measures the currency distance with the difference in currency values. $\Delta \textrm{Curr}(r_i,r_j)=|\textrm{CurrValue}(r_i)-\textrm{CurrValue}(r_j)|$ represents the difference between currency values of $r_i$ and $r_j$ as determined above. $\theta_\omega$ ($\theta_\omega \in (0,1]$) is a threshold which can be set by users or learned from data, describing the max tolerable difference between the currency value of $r_i$ and $r_j$. If $\Delta \textrm{Curr}(r_i,r_j) > \theta_\omega$, which means the currency gap between $r_i$ and $r_j$ are too large to be referred in currency compare, we set $\textsl{curr}{Dist}(r_i,r_j) = \theta_\omega$. \newline \indent Specifically, we set $\textsl{curr}{Dist}(r_i,\varphi) = \theta_\omega$. The currency distance guarantees $r_i$ is closer to its neighbor records in currency order, and has a certain distance with the \emph{CFD} schema whose currency is indefinite to some degree.
\begin{algorithm}[t]
\caption{ImpCCons}
\label{ImproveCons}
%\begin{scriptsize}
%\scriptsize{}
%\begin{algorithmic}[1]
\LinesNumbered
\KwIn{  $\mathcal{D}$ after algorithm 1, $\Sigma$, $\theta^{+}$, $\theta_\omega$, $\alpha$, and $\beta$.}
\KwOut{   the data after consistency repair: $\mathcal{D}_{\textrm{cons}}$}
add $\varphi^{+}$s into $\Sigma$ with $\theta^{+}$\;
\ForEach {$\varphi \in \Sigma$}{
\ForEach {$r_i \in $ Vio($\varphi$)}{
 $\emph{Sch}_{\textsf{repair}}$ $\leftarrow$ Null, $\min$\emph{Diffcc} $\leftarrow +\infty$\;
\ForEach {$r_j \in (\mathcal{D}(\Delta) \cup \varphi(t_p))$}
{\emph{Diffcc} $\leftarrow$ $\alpha \cdot \textsl{cons}{Dist}(r_i,r_j) +\beta \cdot \textsl{cons}{Dist}(r_i,r_j)$\;
$\min$\emph{Diffcc} $\leftarrow$ \emph{Diffcc}, $\emph{Sch}_{\textsf{repair}} \leftarrow r_j$\;
%\STATE update $\emph{Sch}_{\textsf{repair}}$ with ($r_j$, $\min$Diff);
}
update $\emph{Sch}_{\textsf{repair}}$ with ($r_j$, $\min$\emph{Diffcc})\;
}}
\Return $\mathcal{D}_{\textrm{cons}}$\;
%\end{algorithmic}%\end{scriptsize}
\end{algorithm}
\begin{equation}
\label{currDist}
\textsl{curr}{Dist}(r_i,r_j)= \left \{
\begin{array}{cl}
  \Delta \textrm{Curr}(r_i,r_j),  & \Delta \textrm{Curr}(r_i,r_j) < \theta_\omega  \\
  \theta_\omega,     & \textrm{else}
\end{array}
\right.
\end{equation}
\indent Now, we propose the distance metric of records, named \emph{Diffcc} in Definition \ref{Difference} on the both dimensions.
\begin{definition}
\label{Difference}
(\emph{Diffcc}). The currency-consistency difference between two records $r_i$ and $r_j$ is denoted by,
\begin{equation}
  \emph{Diffcc}(r_i,r_j)=\alpha\cdot \textsl{cons}{Dist}(r_i,r_j)+\beta\cdot \textsl{curr}{Dist}(r_i,r_j)
\end{equation}
where $\textsl{cons}{Dist}(r_i,r_j)$ and $\textsl{curr}{Dist}(r_i,r_j)$ are the distance functions defined on consistency and currency, respectively. $\alpha$ and $\beta$ are weight values, and $\alpha,\beta\in(0,1),\alpha+\beta=1$.
\end{definition}
\indent \indent Algorithm \ref{ImproveCons} outlines the
consistency repair process with currency. We first extract potential relation schema $\varphi^{+}$s from $\mathcal{D}$ and add them to $\Sigma$. In the outer loop (Lines 2-11), we detect the satisfaction of each $\varphi$. The records violating a certain $\varphi$ will be marked in the set Vio($\varphi$). In the inner loop (lines 3-9), for each $r_i$ in Vio($\varphi$), we enumerate its neighbor records (selected by $\theta_\omega$) in $\mathcal{D}$ and the schema in $\varphi(t_p)$ to compute the \emph{Diffcc} of $r_i$ with them. We update the present min\emph{Diffcc} and store the corresponding $r_j$ (Line 7). After finishing this loop, we repair $r_i$ with $Sch_{\textsf{repair}}$ according to min\emph{Diffcc}, and obtain a consistent data set $D_{\textrm{cons}}$.
\begin{example}
We now present the repair of $r_5$ in Table 1. We first find out the neighbor records of $r_5$ with $\Delta$Curr($r_i,r_j$) = 0.2. As $r_5$ belongs to $v_4$ in $\mathcal{G}_{c1}$, $r_4$ and $r_6$ are selected, for $\Delta$Curr($r_4,r_5$) = 0.667 - 0.5 = 0.167, $\Delta$Curr($r_5,r_6$) = 0. We then detect $r_5$ violates $\varphi_4$ and $\varphi_5$ as mentioned in Section 3.1 with algorithm \textsf{ImpCCons}. Thus, we compute the \emph{Diffcc} of $r_5$ with $r_4,r_6$, $\varphi_4$ and $\varphi_5$ with $\alpha=0.6,\beta=0.4$. \emph{Diffcc}($r_4,r_5$) is computed
\begin{equation}\label{diff1}
  \emph{Diffcc}(r_4,r_5)=0.6\cdot0.167+(1-0.6)\cdot\frac{5}{9}=0.324
\end{equation}
Similarly, \emph{Diffcc}($r_4,r_5$)=0.177, and \emph{Diffcc}($r_5,\varphi_4$) is,
\begin{equation}\label{diff1}
  \emph{Diffcc}(r_5,\varphi_4)=0.6\cdot0.2+(1-0.6)\cdot\frac{2}{3}=0.387
\end{equation}
And \emph{Diffcc}($r_5,\varphi_5$)=0.267. $r_6$ turns out to be the closest neighbor of $r_5$. We repair the dirty part of $r_5$ as $r_5$[\textsf{Address}]$\rightarrow Xixi$, $r_5$[\textsf{City}]$\rightarrow Hangzhou$ and $r_5$[\textsf{Email}]$\rightarrow M@ali$. Errors in $r_{10}$ can also be captured by Algorithm 4, and we are able to repair it to be $r_{10}$ [\textsf{Address}]$\rightarrow Xuhui$ and $r_{10}$ [\textsf{City}]$\rightarrow Shanghai$ ahead of incompleteness repair step.
\end{example}
\indent \indent \emph{Complexity}. In Algorithm \ref{ImproveCons}, the outer loop in lines 2-10 takes $O(|\Sigma|)$ time to detect the violation on each $\varphi$, where $|\Sigma|$ is the number of consistency constraints. Within the loop, it costs $O\big((N_\Delta)+|\varphi(t_p)|) \cdot N(vio)\big)$ to compute and repair the consistent-violative values on average, where $N(vio)$ represents the number of violative values, $N_\Delta$ is the number of neighbor records (quite smaller than $N$), and $|\varphi(t_p)|$ is the number of $t_p$ in a $\varphi$. To put it together, Algorithm \textsf{ImpCCons} costs $O\big(|\Sigma|\cdot N(vio)\cdot(N_\Delta+|\varphi(t_p)|)\big)$ on average.
\newline \indent During consistency repair, we treat missing values captured by the given $\Sigma$ as a kind of violation of consistency. We are able to repair them by $\Sigma$ in the third step of \textsf{Improve3C}. We do not need to repair those values in completeness repair step. Specially, Algorithm \ref{ImproveCons} performs on the assumption that there is no conflict or ambiguous between the given \emph{CFD}s and \emph{CC}s. Works has been done (like \cite{8Fan2014Conflict}) on conflict resolution with \emph{CFD}s and \emph{CC}s, which has been applied in the preprocess of our method.
\begin{table}
\label{com}
%\begin{small}
\centering
\caption{Analogy between the incomplete repair and Bayesian}
\scalebox{0.9}{
\begin{tabular}{|lcl|} \hline
%\textbf{Notation} & \textbf{Description}\\
%\hline \hline
\tabincell{l}{an incomplete record, \\ $r_{\overline{\textrm{c}}} = \big\{a_1,...,a_{m-1}, \textrm{cV}(r_{\overline{\textrm{c}}})\big\}$}  & $\rightarrow$ & $X = \{a_1,...,a_m\}$\\
\hline
\tabincell{l}{the domain of the missing values, \\ $A^{\overline{c}} = \{z_1,...,z_n\}$} & $\rightarrow$ & $Y = \{y_1,...,y_n\}$\\
\hline
the prior probability, $Pr(A^{\overline{c}})$ & $\rightarrow$ &  $Pr(Y)$\\
\hline
the class-conditional probability, $P(r_{\overline{\textrm{c}}}|A^{\overline{c}})$ & $\rightarrow$ &  $Pr(X|Y)$\\
\hline
the filling posterior probability, $Pr(z_i|r_{\overline{\textrm{c}}})$ & $\rightarrow$ & $Pr(y_i|X)$\\
%\hline
%\tabincell{l}{$P(r_{\overline{\textrm{c}}}|z_i)P(z_i) = $ \\$P(z_i)\big(\prod^{m-1}_{j=1}P(a_j|z_i) + P(\textrm{cV}(r_{\overline{\textrm{c}}})|z_i) \big)$} & $\rightarrow$ & \tabincell{l}{$P(X|y_i)P(y_i) = $ \\$P(y_i)\prod^{m}_{j=1}P(a_j|y_i)$}\\
\hline
\end{tabular}}
%\end{small}
\end{table}
\section{Incompleteness Repair}
\label{sec6}
Repairing missing values is one classical key problem in data completeness solution  \cite{13Wang1996Beyond}. Various methods have been studied in missing value cleaning, such as statistical-based experience-based, learning-oriented, and etc \cite{Batini2009Methodologies}. In the fourth step of \textsf{Improve3C}, we adopt the naive Bayes classification method \cite{DBLP:books/daglib/0087929} which is acknowledged to perform well in data completeness repairing issues. We improve the completeness repair by filling the missing values with a time-related clean value. In general, to capture the temporal evolution in attribute values, we treat CurrValue($r_i$) as an important feature and insert it to the training in naive Bayes.
\newline \indent We first draw the analogy between our completeness repair approach and the general elements in Naive Bayes in Table 3. For an incomplete record $r_{\overline{\textrm{c}}} = \big \{a_1,...,a_{m-1}, \textrm{cV}(r_{\overline{\textrm{c}}})\big \}$, $a_j$ is the value on $A_j$, and we abbreviate currency value of $r_{\overline{\textrm{c}}}$ to be $\textrm{cV}(r_{\overline{\textrm{c}}})$. The missing value of $r_{\overline{\textrm{c}}}$ is on $A^{\overline{c}}$ (\emph{w.l.o.g}., assuming that only one missing value exists in $r_{\overline{\textrm{c}}}$), whose value domain is $z_1,...,z_n$. It makes up the possible value set for the test data like $r_{\overline{\textrm{c}}}$. We adopt the prior probability $Pr(A^{\overline{\textrm{c}}})$ and the class-conditional probability $Pr(r_{\overline{\textrm{c}}}|A^{\overline{\textrm{c}}})$ in Bayes formula in our completeness repair problem.
\newline \indent Accordingly, to classify and repair an incomplete record, the naive Bayes computes the posterior probability for each complete record in Equation (7).
\begin{gather}\label{bayes}
  Pr(z_i|r_{\overline{\textrm{c}}}) = \frac{
   Pr(z_i)\cdot \big(\prod^{m-1}_{j=1} Pr(a_j|z_i) + Pr(cV(r_{\overline{\textrm{c}}})|z_i)\big)}{ Pr(r_{\overline{\textrm{c}}})},\textrm{ and} \nonumber \\
   Pr(z_i)\cdot \prod^{m-1}_{j=1} Pr(a_j|z_i)= Pr(a_1|z_i)Pr(a_2|z_i)...Pr(a_{m-1}|z_i).
\end{gather}
\indent Accordingly, the completeness repairing issue, named \textsf{ImpCCom}, can be solved by the following steps.
\newline \indent {\emph{Step} 1}: Input the data $\mathcal{D}_{\textrm{cons}}$, and the confidence measure $\sigma$. Treat the currency values computed in Section 3.2 as a new attribute, and insert CurrValue($r_i$) to each record.
\newline \indent {\emph{Step} 2}: Detect the records with missing values \emph{i.e.}, $r_{\overline{\textrm{c}}}$s. We treat the set of $r_{\overline{\textrm{c}}}$s as test data.
\newline \indent {\emph{Step} 3}: Construct the training set with complete records in $\mathcal{D}_{\textrm{cons}}$ and preprocess the discrete and continuous data, respectively.
Compute $Pr(z_1|r_{\overline{\textrm{c}}}), Pr(z_2|r_{\overline{\textrm{c}}}), ...,Pr(z_n|r_{\overline{\textrm{c}}})$ with Equation (7).
\newline \indent \emph{Step} 4: Find out $Pr(z_k|r_{\overline{\textrm{c}}})= \max \big\{Pr(z_1|r_{\overline{\textrm{c}}}), Pr(z_2|r_{\overline{\textrm{c}}}), ... ,$ $Pr(z_n|r_{\overline{\textrm{c}}}) \big\}$, and fill the missing value of $r_{\overline{\textrm{c}}}$ with $z_k$ if $Pr(z_k|r_{\overline{\textrm{c}}}) \geq \sigma$.
\newline \indent \emph{Step} 5: Recursively repeat Step 3 and Step 4 until the missing values on all attributes in $\mathcal{A}$ have been solved.
\newline \indent The currency values of records can train the model to repair missing values with the data shares the same (or similar) current order of $r_{\overline{\textrm{c}}}$. Bayesian method is a proper instance in our framework which also performs well in experiments. Other alternative repair methods on completeness can also be adopted in view of the characteristic of the data to be cleaned.
\section{Experimental Study}
\label{sec7}
In this section, we evaluate the experimental performance of the proposed methods. We first introduce the experimental settings in Section \ref{sec7}.1, and discuss the performance of the methods in \ref{sec7}.2.
\begin{table}[t]
\caption{Summary of the data sets}
\label{summary}
\centering
\scalebox{1}{
\begin{tabular}{|ccccc|}
\hline
\textbf{Data} & \#\textbf{Records} & \#\textbf{Entities} & \#\textbf{Tables} & \#\textbf{Key attributes}\\
\hline
\hline
\emph{NBA} & 25,050 & 1560& 4 & 10
\\
\emph{PCI}& 40,000 & 820& 2 & 9
\\
\hline
\end{tabular}}
\end{table}
\subsection{Experimental Settings}
\label{7.1}
% 为验证方法的一般性
\textbf{Experimental Data}. To report the generality of the proposed method, we use one real data and a synthetic one in experiments. Table \ref{summary} summaries the data details.
\newline \indent \textsl{\underline{NBA}}. The NBA player statistics data\footnote{http://databasebasketball.com.} \footnote{http://www.basketball-reference.com} reports over 2800 players' career information in NBA. We select more than 25 thousand records for over 1500 players, where the key attributes we adopted are $\mathcal{A} = $(\textsf{Pid, Name, Age, Nationality, Team, Arena, City, Season, PPG, Scores}). \textsf{Pid} is used to identify different players, and the data describes which \textsf{Team} players belongs to at the corresponding \textsf{Season}. It records the home arena with the city of each team in \textsf{Arena} and \textsf{City}, respectively. \textsf{PPG} presents the averaged points the player achieves pre game in each season, and \textsf{Score} records the total scores of players' career.
\newline \indent We collect data for each player with no less than 5 regular seasons. \textsl{NBA} carries few timestamps. However, values on many attributes (such as \textsf{Scores, PPG}) evolve so frequently with each regular season. We derive a set of \emph{CC}s and \emph{CFD}s, the patterns of which include the following.
\newline \indent $\psi_1$: $\forall r_i, r_j$, $(r_i[\textsf{Pid}] = r_j[\textsf{Pid}]$ and $ r_i[\textsf{Age}] < r_j[\textsf{Age}]) \longrightarrow (r_i \prec_{\textsf{Age}} r_j)$.
\newline \indent $\psi_2$: $\forall r_i, r_j$, $(r_i[\textsf{Pid}] = r_j[\textsf{Pid}]$ and $ r_i[\textsf{Scores}] < r_j[\textsf{Scores}]) \longrightarrow (r_i \prec_{\textsf{Scores}} r_j)$.
\newline \indent $\varphi_1$: $\forall r_i$, $ \big(r_i[\textsf{Season}]=``\_", r_i[\textsf{Team}]=``\_" \big)
\longrightarrow (r_i[\textsf{Arena}]=``\_", r_i[\textsf{City}]=``\_")$.
\newline \indent \textsl{\underline{PCI}}. \underline{P}ersonal \underline{C}areer \underline{I}nformation is a synthetic data adheres to the same schema shown in Example 1, which describes over 800 individuals with 400 thousand records. The constraints we used here have the same patterns with the ones introduced in Section 2.1.
\newline \indent We preprocess the data sets to be clean and use them as the ground truth. To effectively evaluate the methods, we introduce random dirty values \emph{i.e.}, noises to the data under different conditions. \emph{noi}\% are used to describe the noise ratio of the erroneous values to the total number of values.
\begin{figure*}[t]
\centering
\subfigure[NBA, \emph{noi}=10\%, \textsf{P}]{
  \includegraphics[scale=0.2]{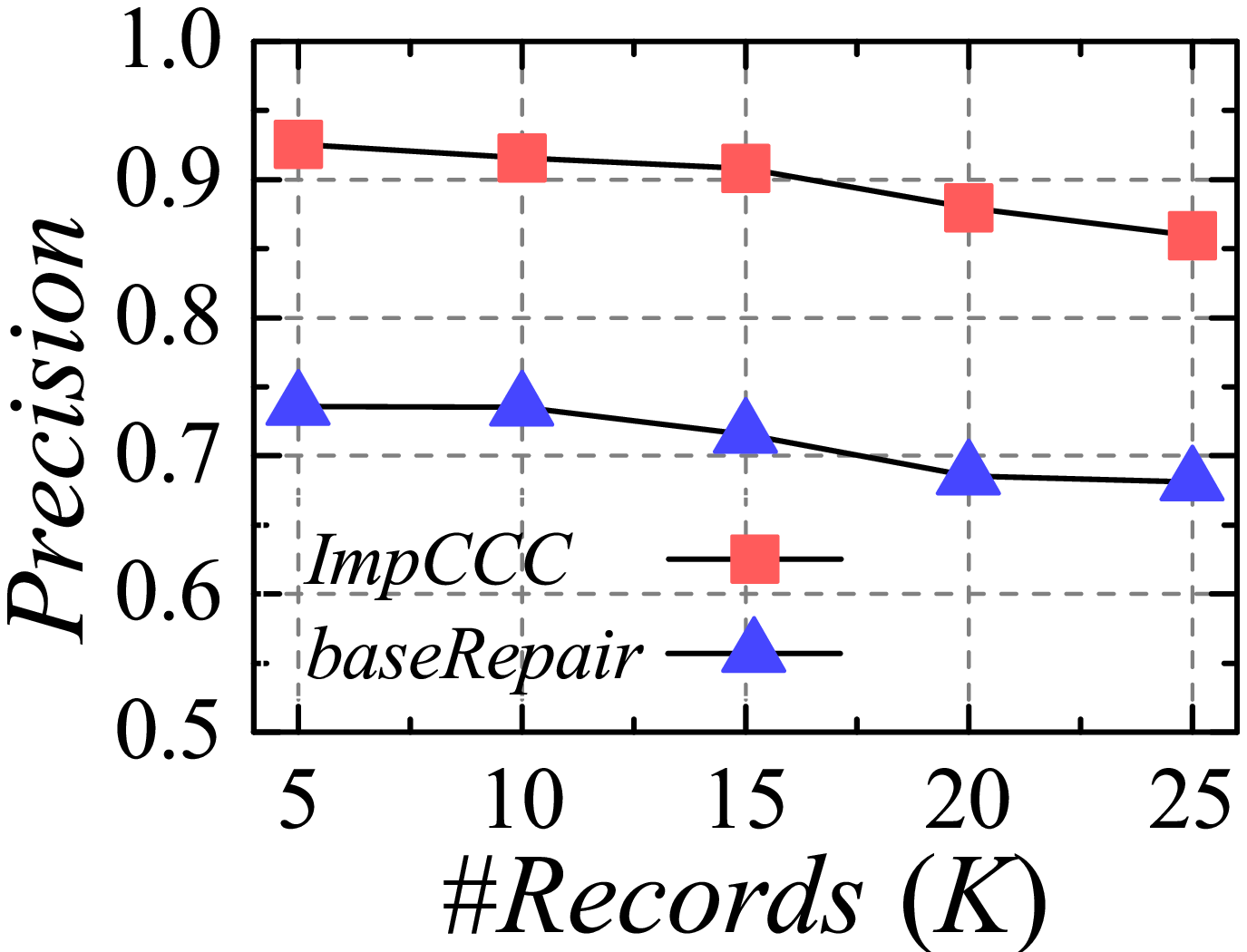}}
\subfigure[NBA, \emph{noi}=10\%, \textsf{R}]{
\includegraphics[scale=0.2]{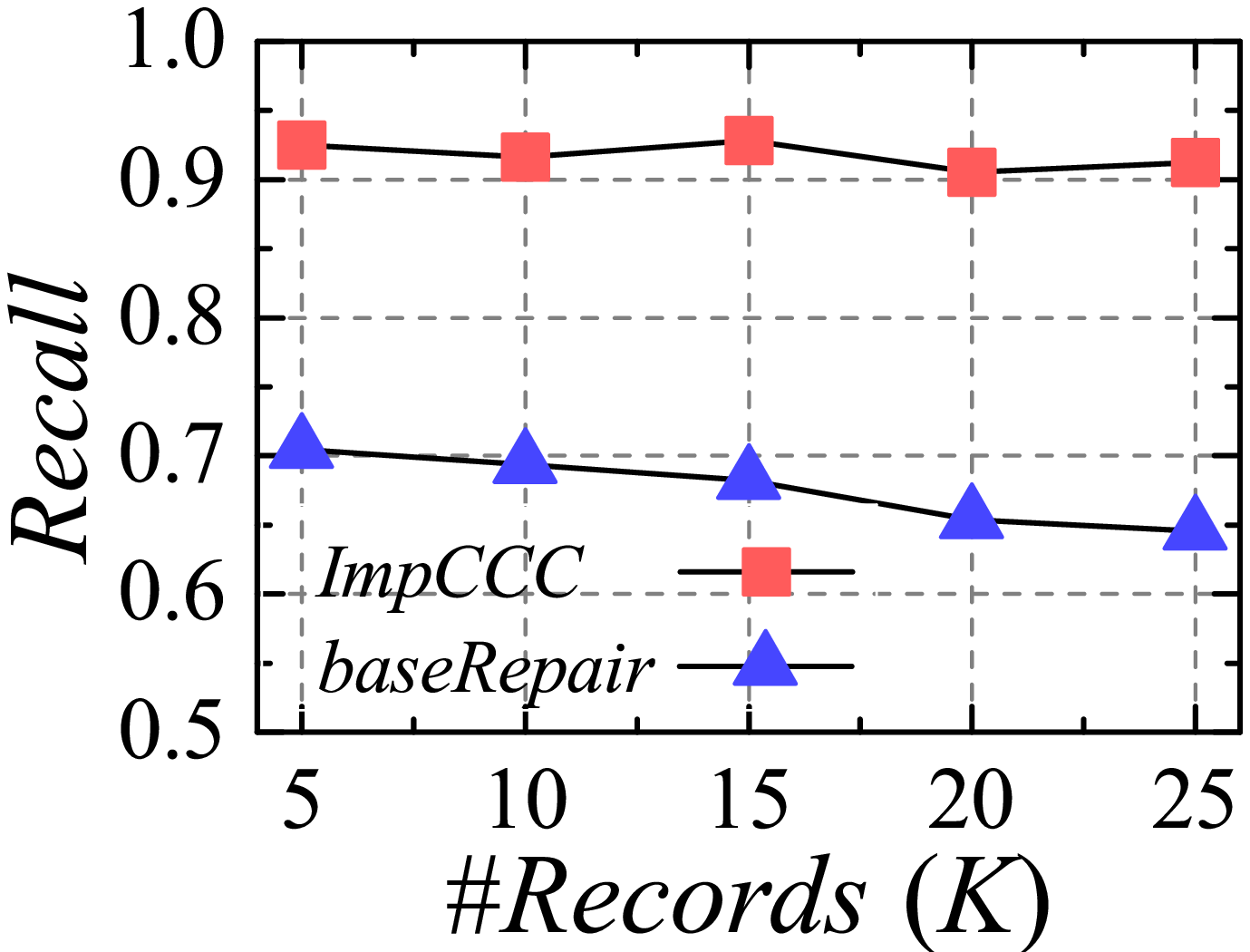}}
\subfigure[PCI, \emph{noi}=10\%, \textsf{P}]{
  \includegraphics[scale=0.2]{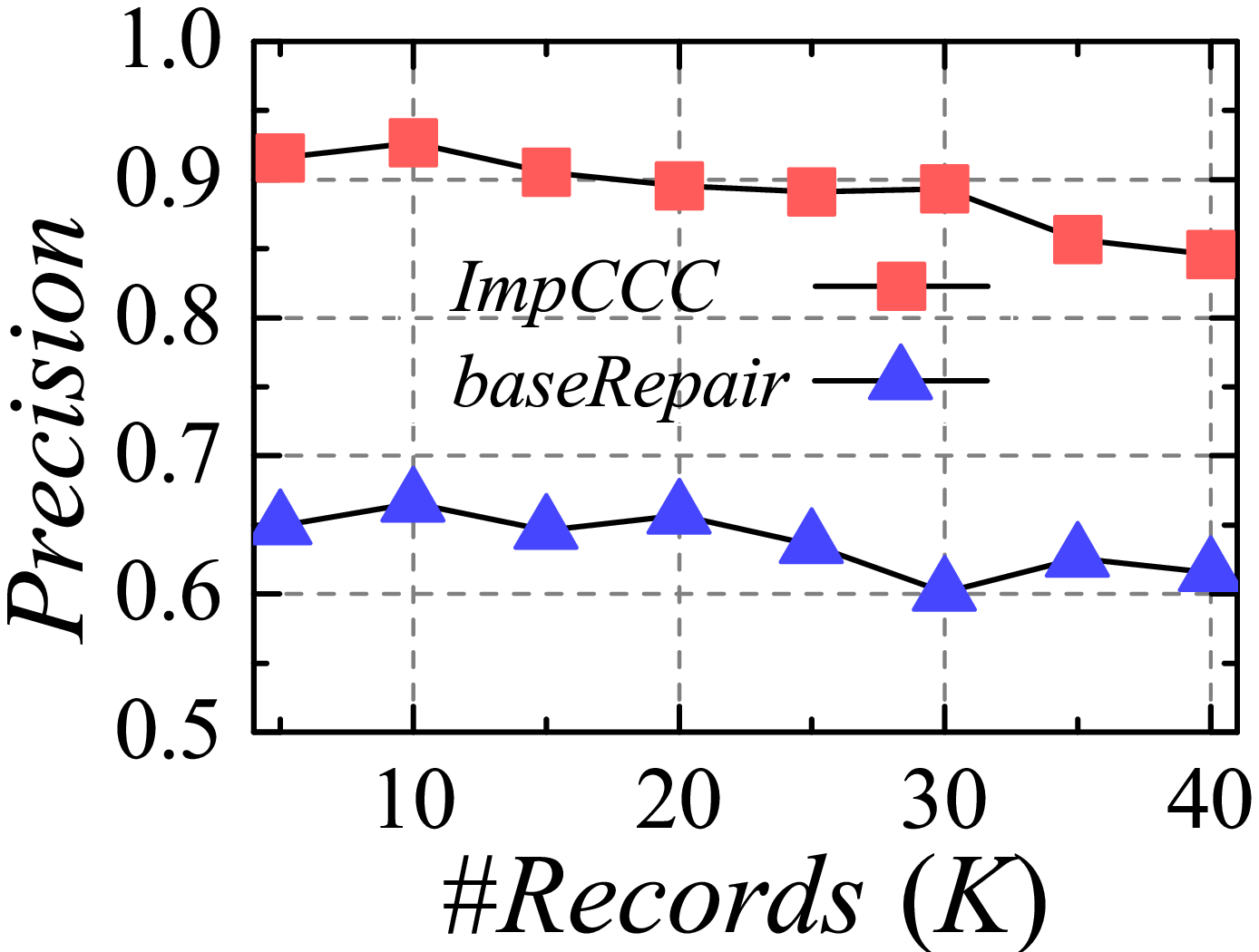}}
\subfigure[PCI, \emph{noi}=10\%, \textsf{R}]{
\includegraphics[scale=0.2]{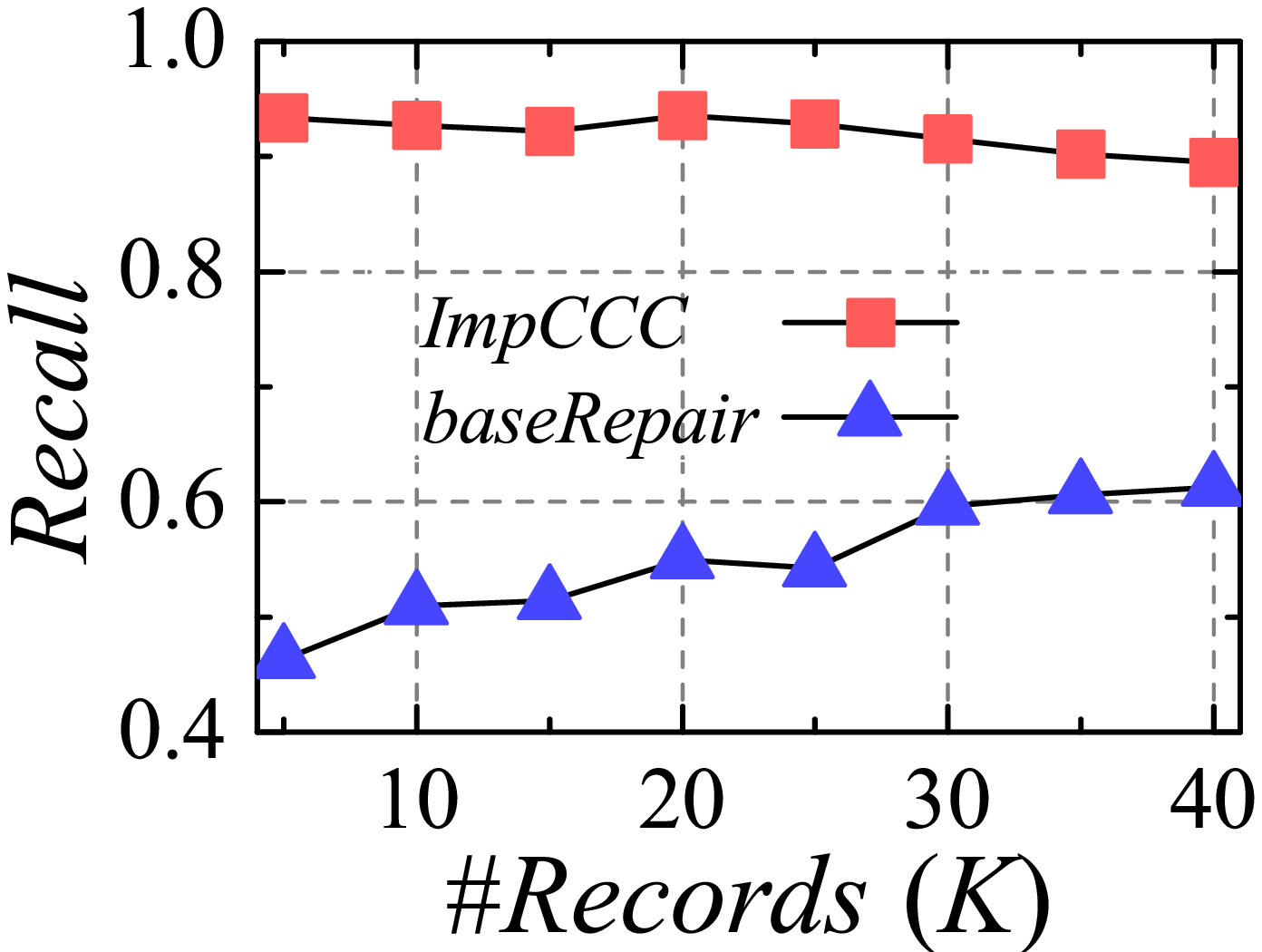}}
\\
\subfigure[NBA, \emph{noi}=10\%, \textsf{P}]{
  \includegraphics[scale=0.2]{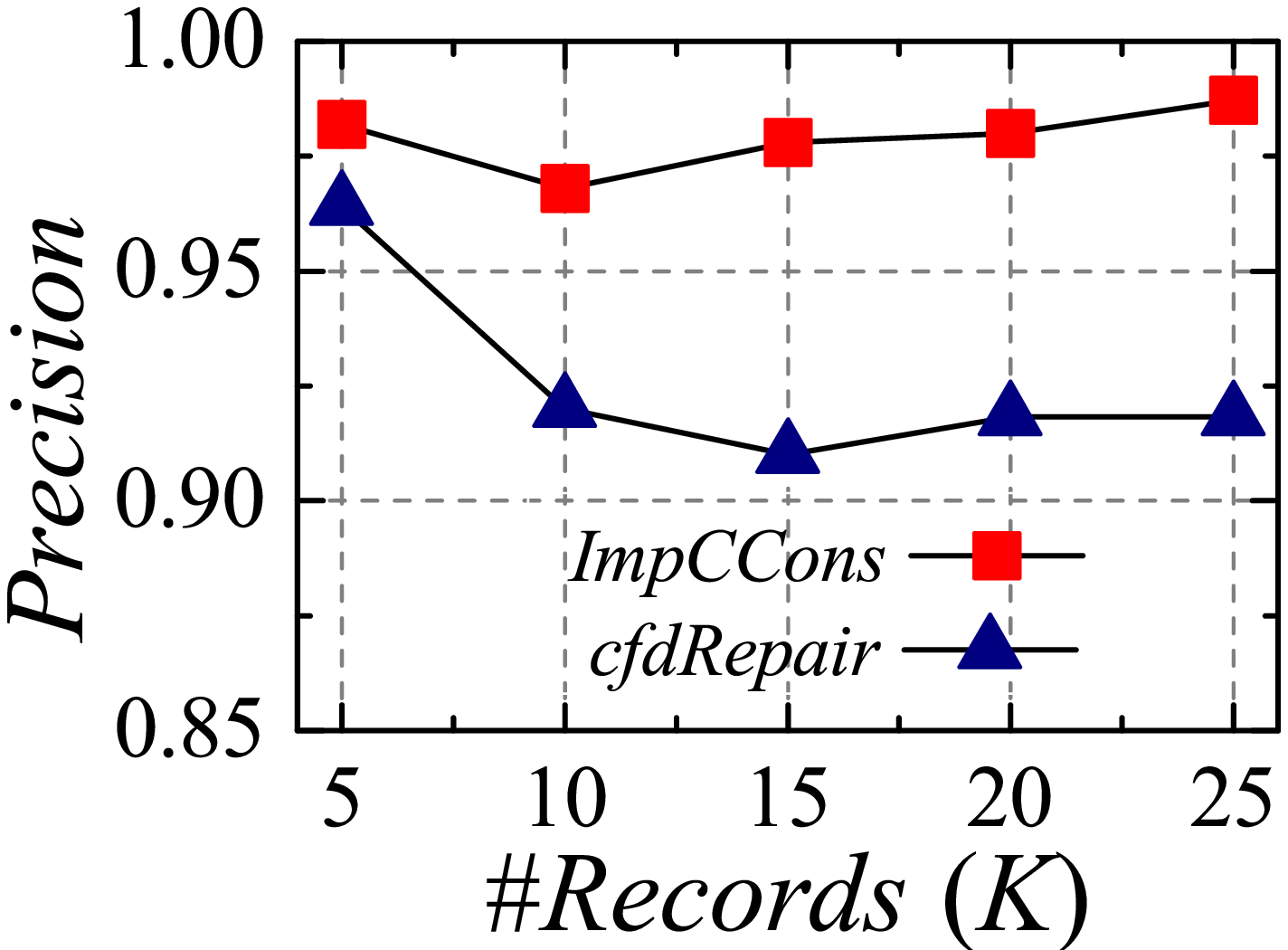}}
\subfigure[NBA, \emph{noi}=10\%, \textsf{R}]{
\includegraphics[scale=0.2]{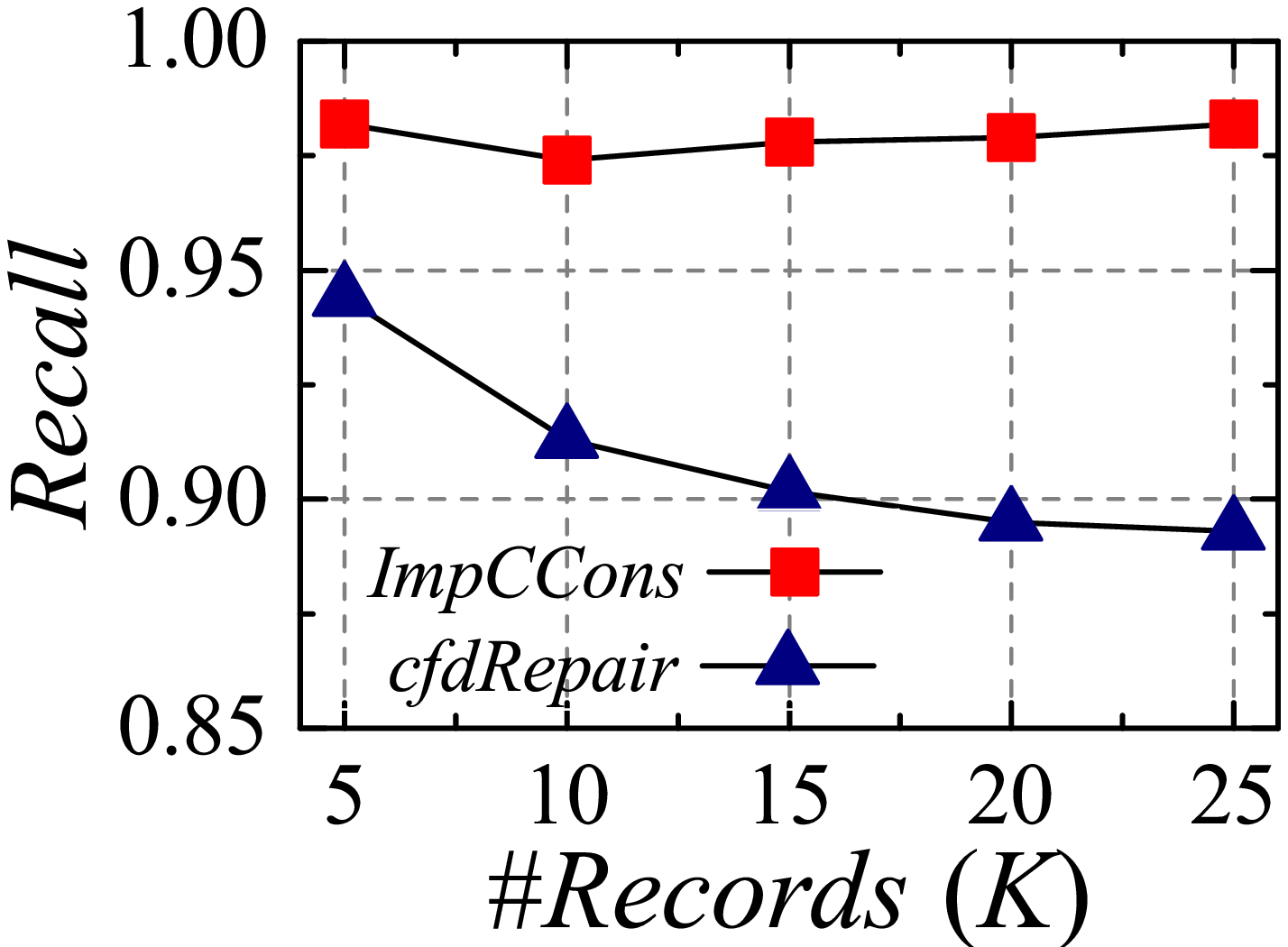}}
\subfigure[PCI, \emph{noi}=10\%, \textsf{P}]{
  \includegraphics[scale=0.2]{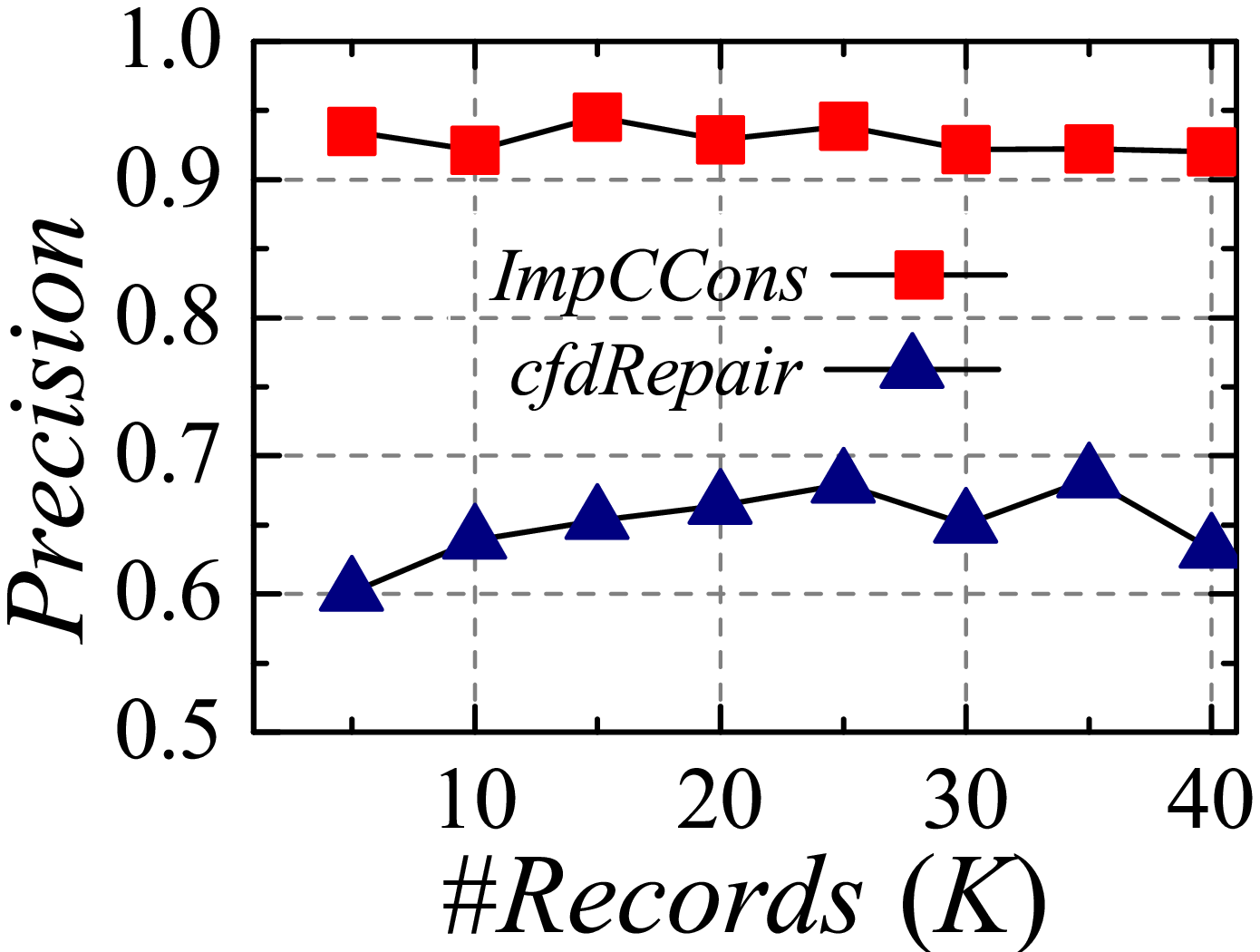}}
\subfigure[PCI, \emph{noi}=10\%, \textsf{R}]{
\includegraphics[scale=0.2]{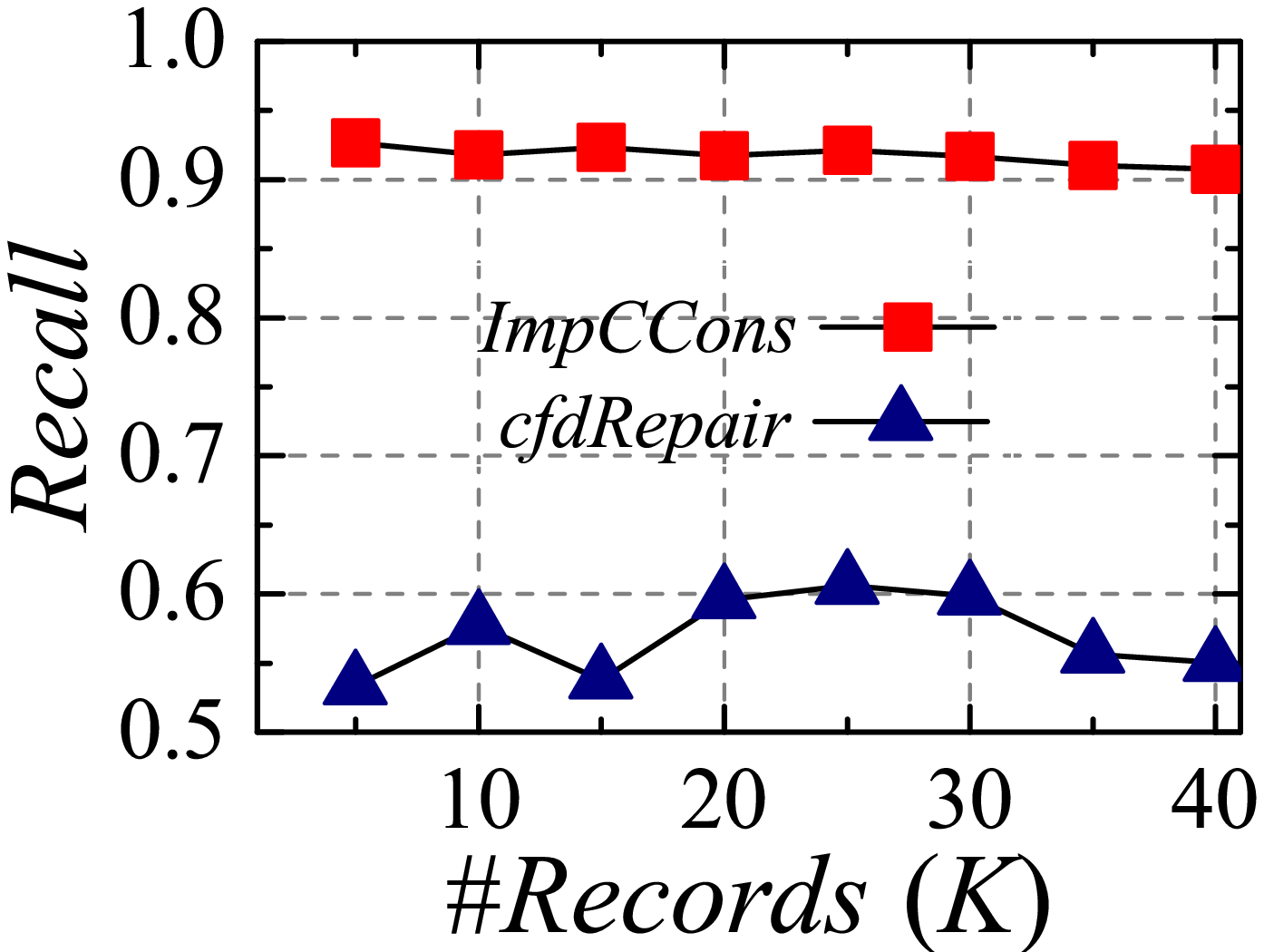}}
\\
\subfigure[NBA, \emph{noi}=10\%, \textsf{P}]{
  \includegraphics[scale=0.2]{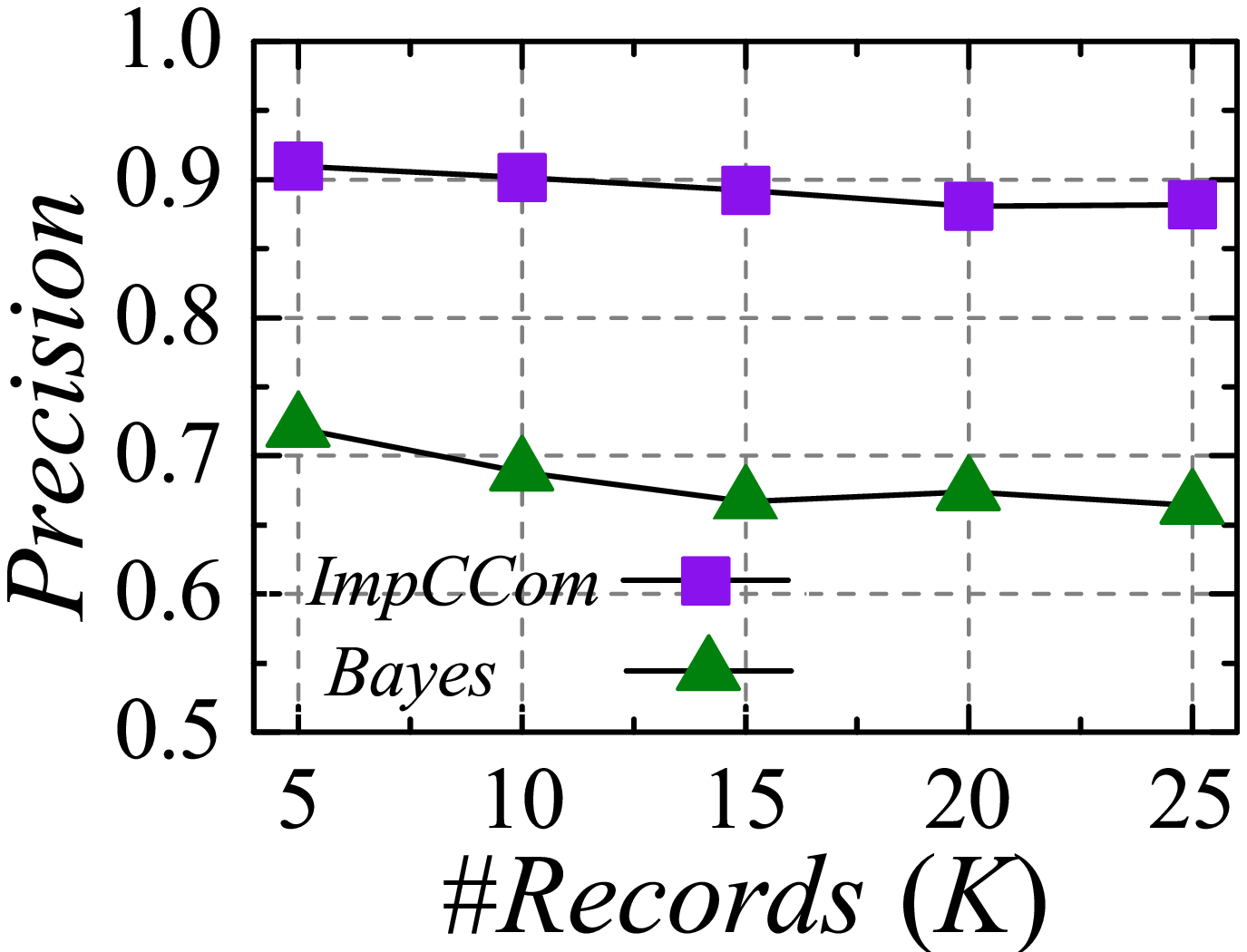}}
\subfigure[NBA, \emph{noi}=10\%, \textsf{R}]{
\includegraphics[scale=0.2]{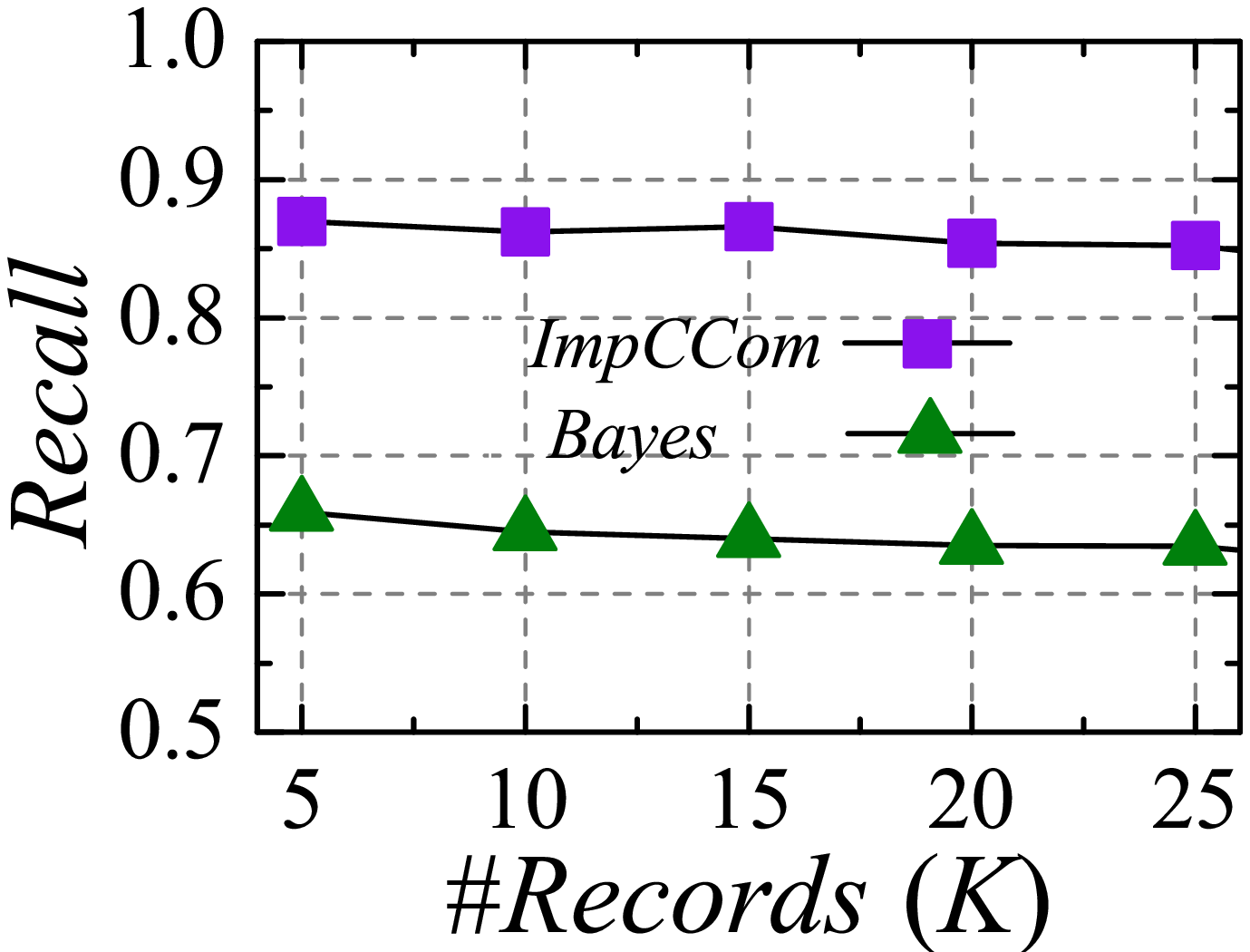}}
\subfigure[PCI, \emph{noi}=10\%, \textsf{P}]{
  \includegraphics[scale=0.2]{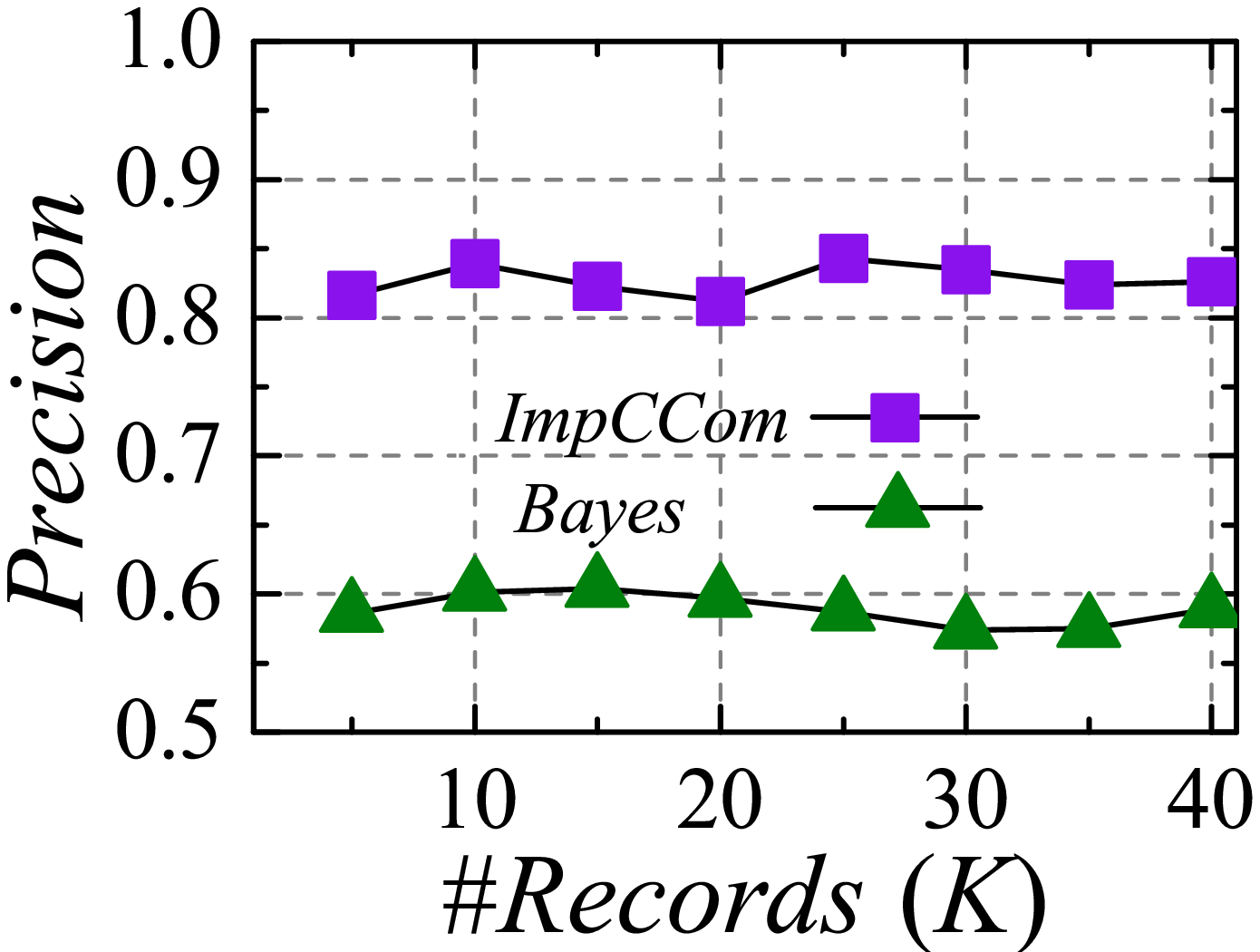}}
\subfigure[PCI, \emph{noi}=10\%, \textsf{R}]{
\includegraphics[scale=0.2]{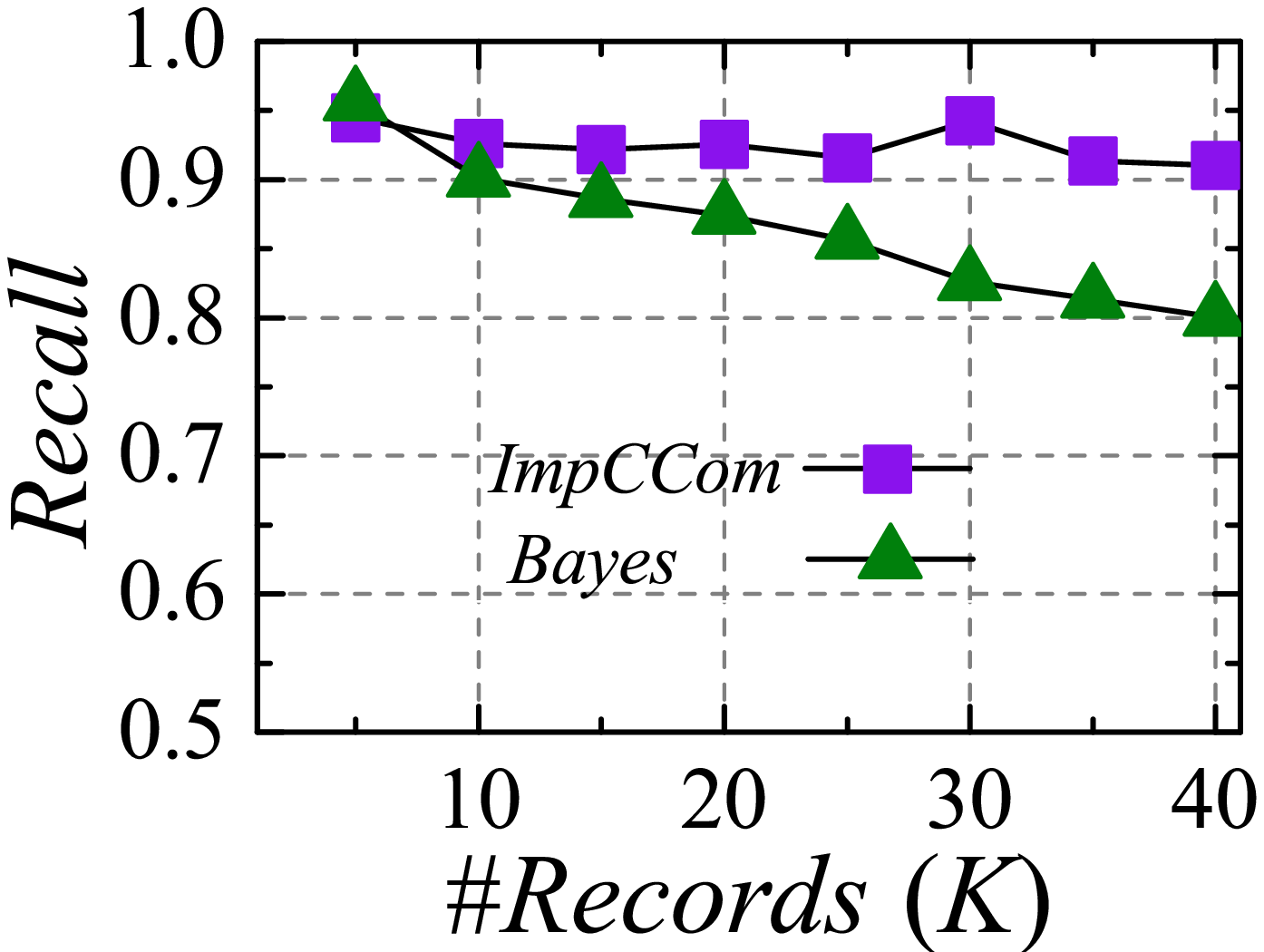}}
\caption{Effectiveness comparison on two data sets}
\label{6}       % Give a unique label
\end{figure*}
%\subsection{Experimental Settings}
%\label{5.2}
\indent \textbf{Implementation}. The experiment ran on a computer with Inter(R) 3.40 GHZ Core i5 CPU and 32GB of RAM.We implemented all the algorithms proposed in Section 3. We also implemented \textsf{ImpCCons} and \textsf{ImpCCom} with the currency order values for performance comparison on consistency and completeness independently. We use two methods to find \emph{CC}s and \emph{CFD}s during the preprocessing. On one hand, we discover according to methods proposed in \cite{Papenbrock2015Functional,Chu2014RuleMiner} to discover \emph{CC}s and \emph{CFD}s. On the other hand, we artificially design some constraints with assistance of credible knowledge base \emph{e.g.}, wikipedia, Baidupedia. These semantic constraints also satisfy the semantic definitions in \cite{Fan2012Foundations} .
\newline \indent \textbf{Baselines}. We implemented several baseline algorithms for comparing the performance of the algorithm \textsf{Improve3C}, \textsf{ImpCCom} and \textsf{ImpCCons} independently with the existing methods.
\newline \indent  (1) \textsf{cfdRepair}. It repairs the dirty data with defined \emph{CFD}s taking no account the temporal detection. It is one common approach for consistency repair \cite{Fan2012Foundations}.
\newline \indent  (2) \textsf{Bayes}. It fills in the missing value with the probability functions on classifications. We adopt Naive Bayes algorithm in \cite{DBLP:books/daglib/0087929}.
\newline \indent  (3) \textsf{baseRepair}. We use \textsf{baseRepair} as the baseline algorithm of repairing all kinds of dirty data. It combines \textsf{cfdRepair} and \textsf{Bayes} methods without taking into account of currency issues.
\newline \indent \textbf{Measure}.
We apply precision (\textsf{P}), recall (\textsf{R}) to measure the effectiveness of algorithms. \textsf{P} is the ratio between the number of values correctly repaired and the total number of repaired values. \textsf{R} is the ratio between the number of values correctly repaired and the total number of dirty values.
\begin{figure*}[t]
\centering
\subfigure[NBA, \#R=15K]{
  \includegraphics[scale=0.2]{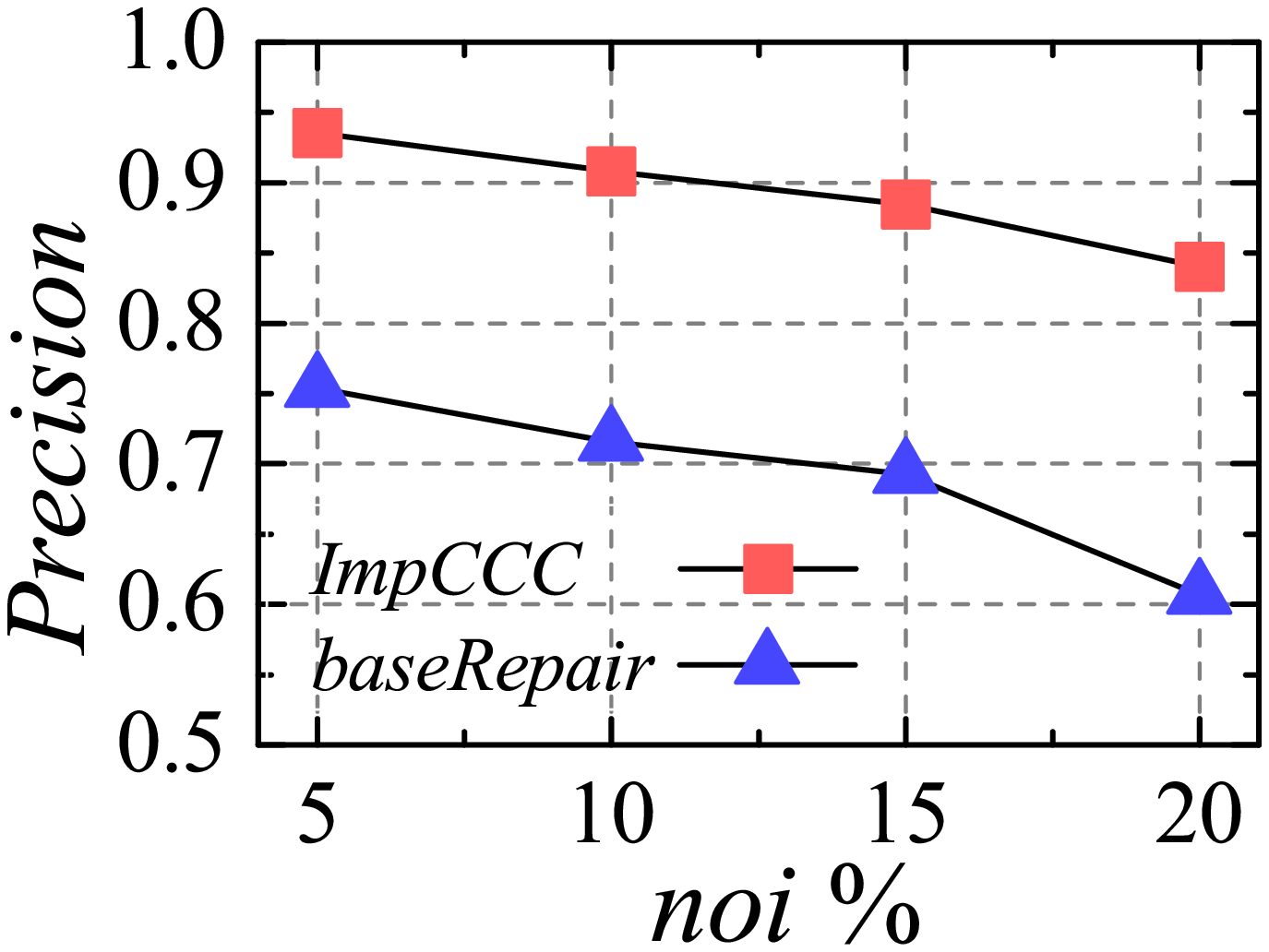}}
\subfigure[NBA, \#R=15K]{
\includegraphics[scale=0.2]{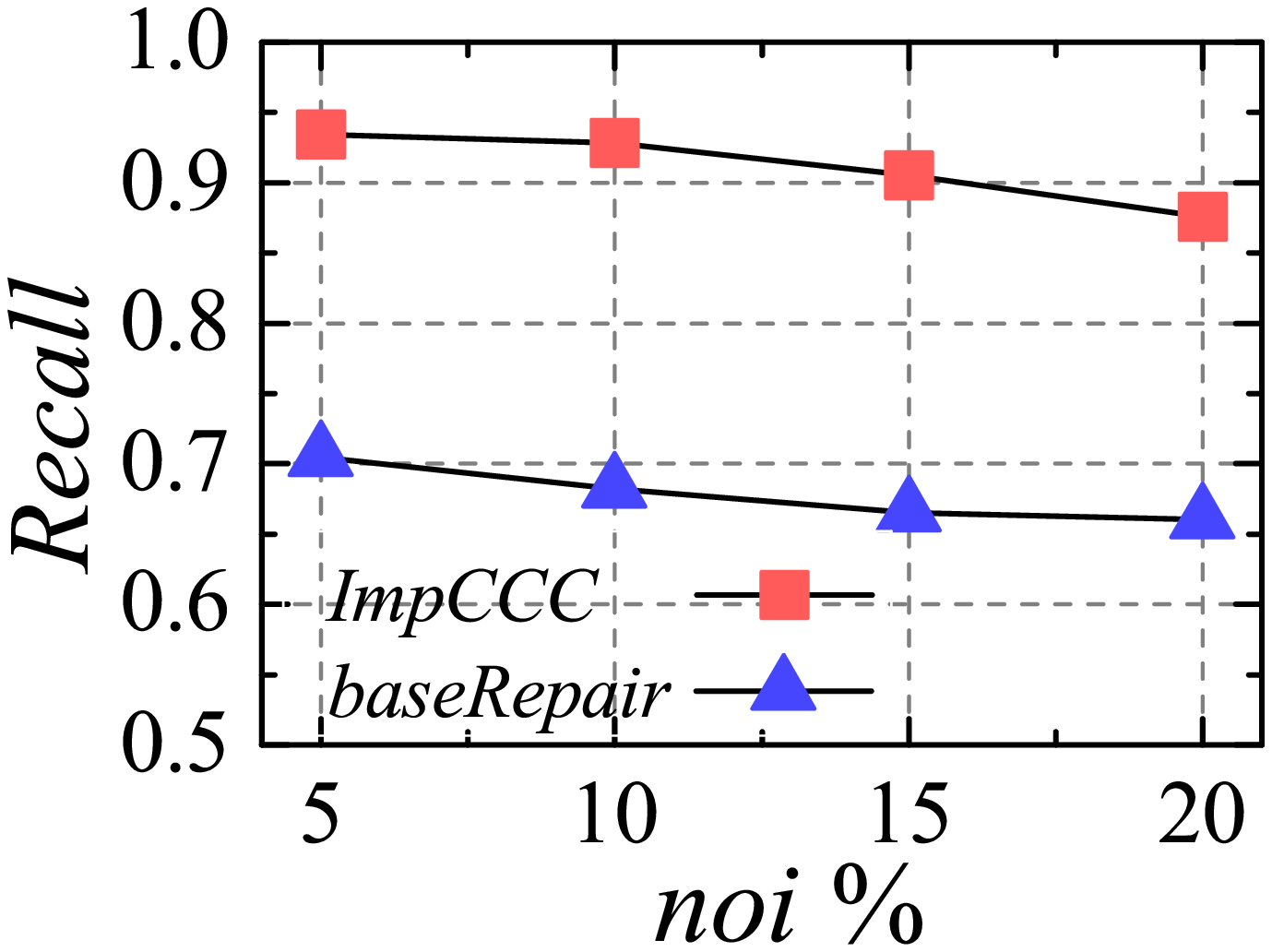}}
\subfigure[PCI, \#R=30K]{
  \includegraphics[scale=0.2]{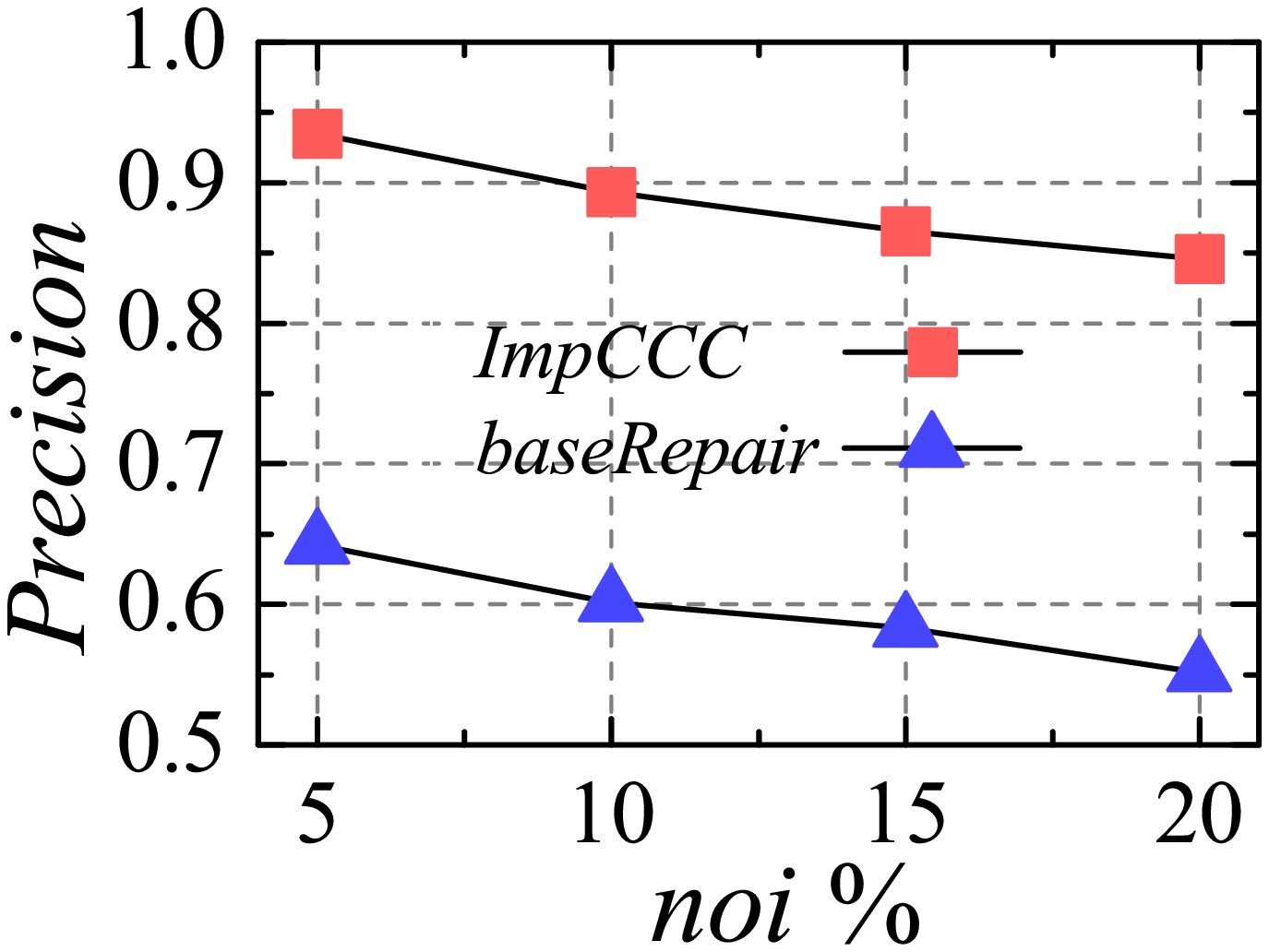}}
\subfigure[PCI, \#R=30K]{
\includegraphics[scale=0.2]{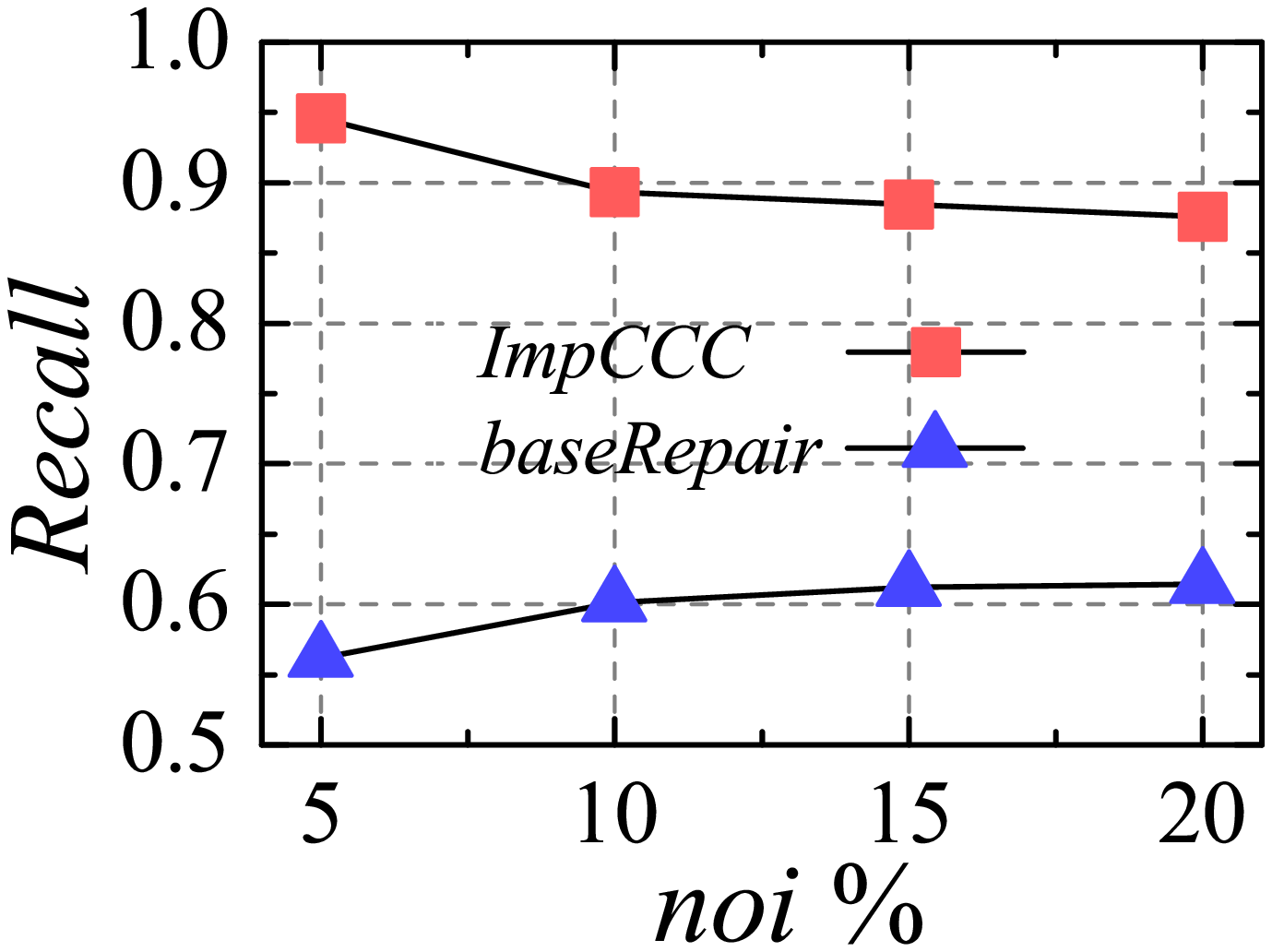}}
\\
\subfigure[NBA, \#R=15K]{
  \includegraphics[scale=0.2]{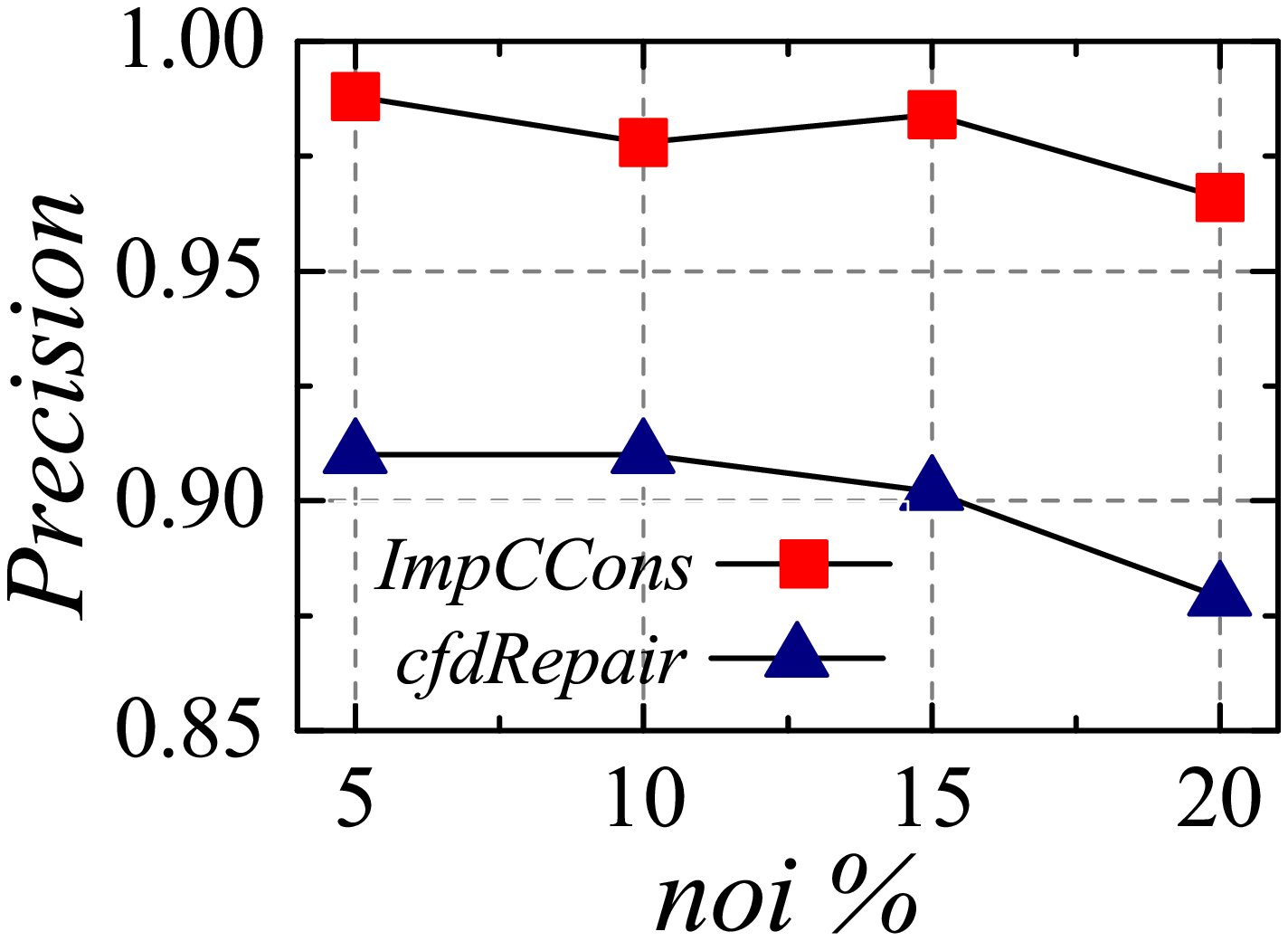}}
\subfigure[NBA, \#R=15K ]{
\includegraphics[scale=0.2]{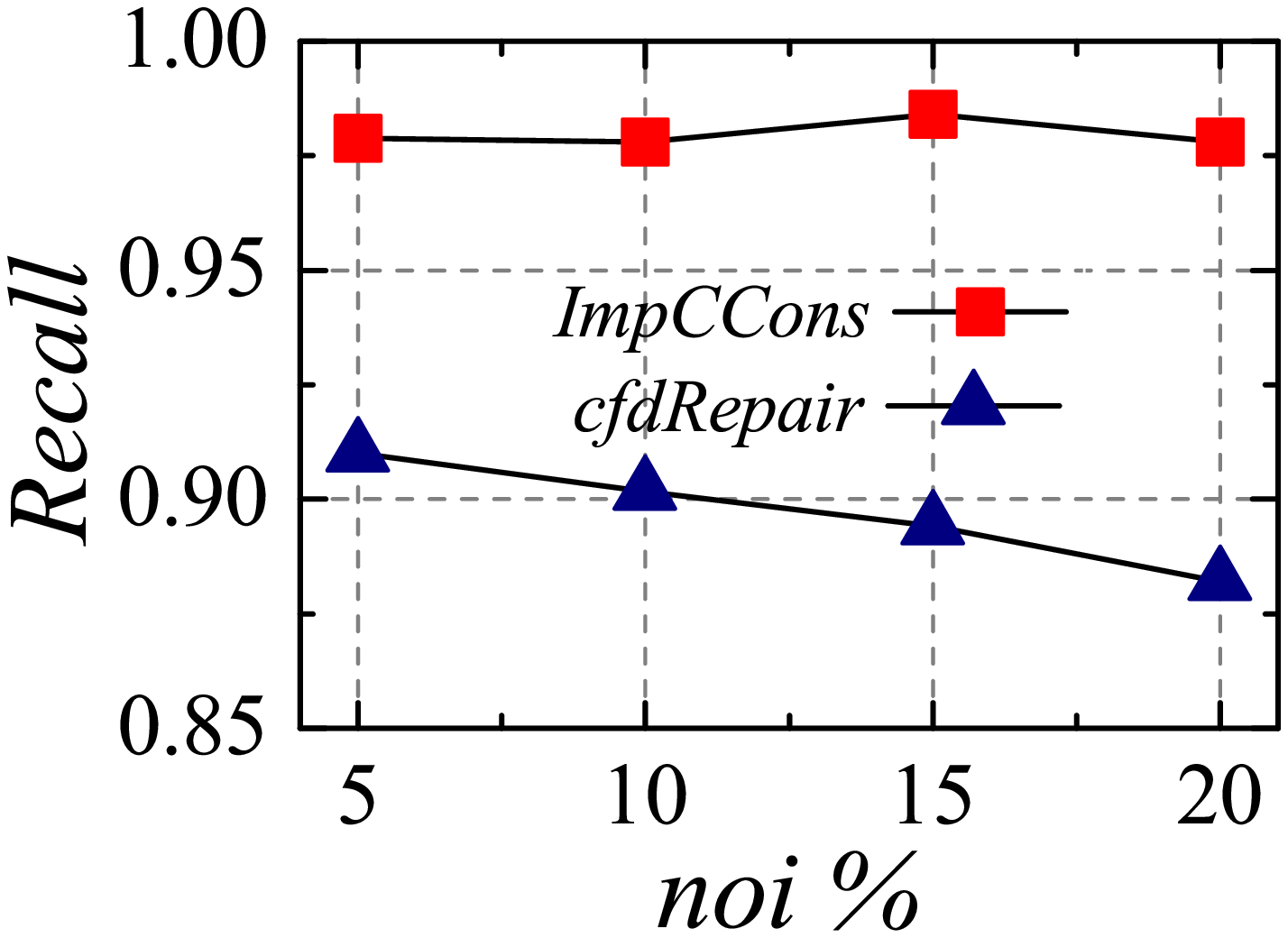}}
\subfigure[PCI, \#R=30K]{
  \includegraphics[scale=0.2]{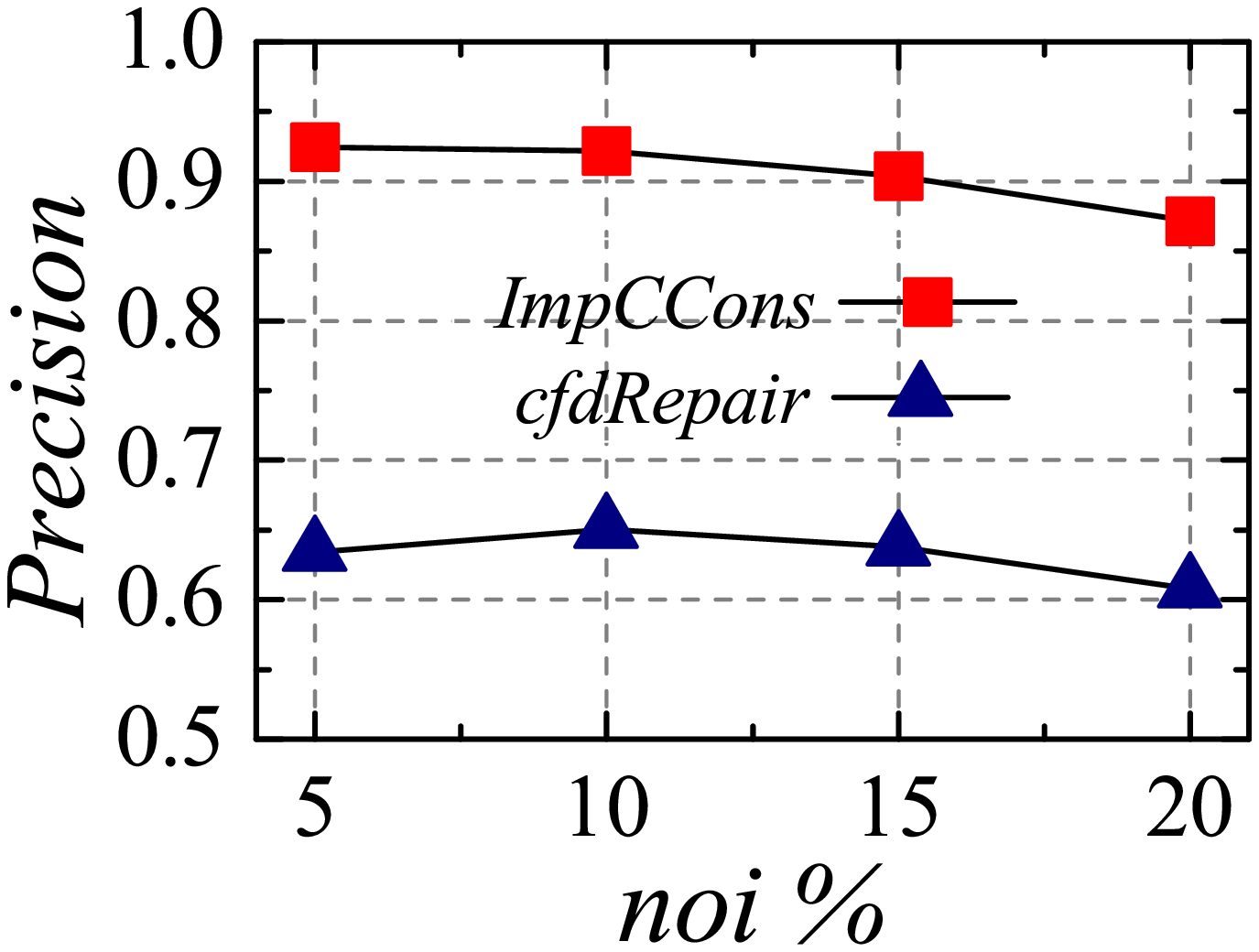}}
\subfigure[PCI, \#R=30K]{
\includegraphics[scale=0.2]{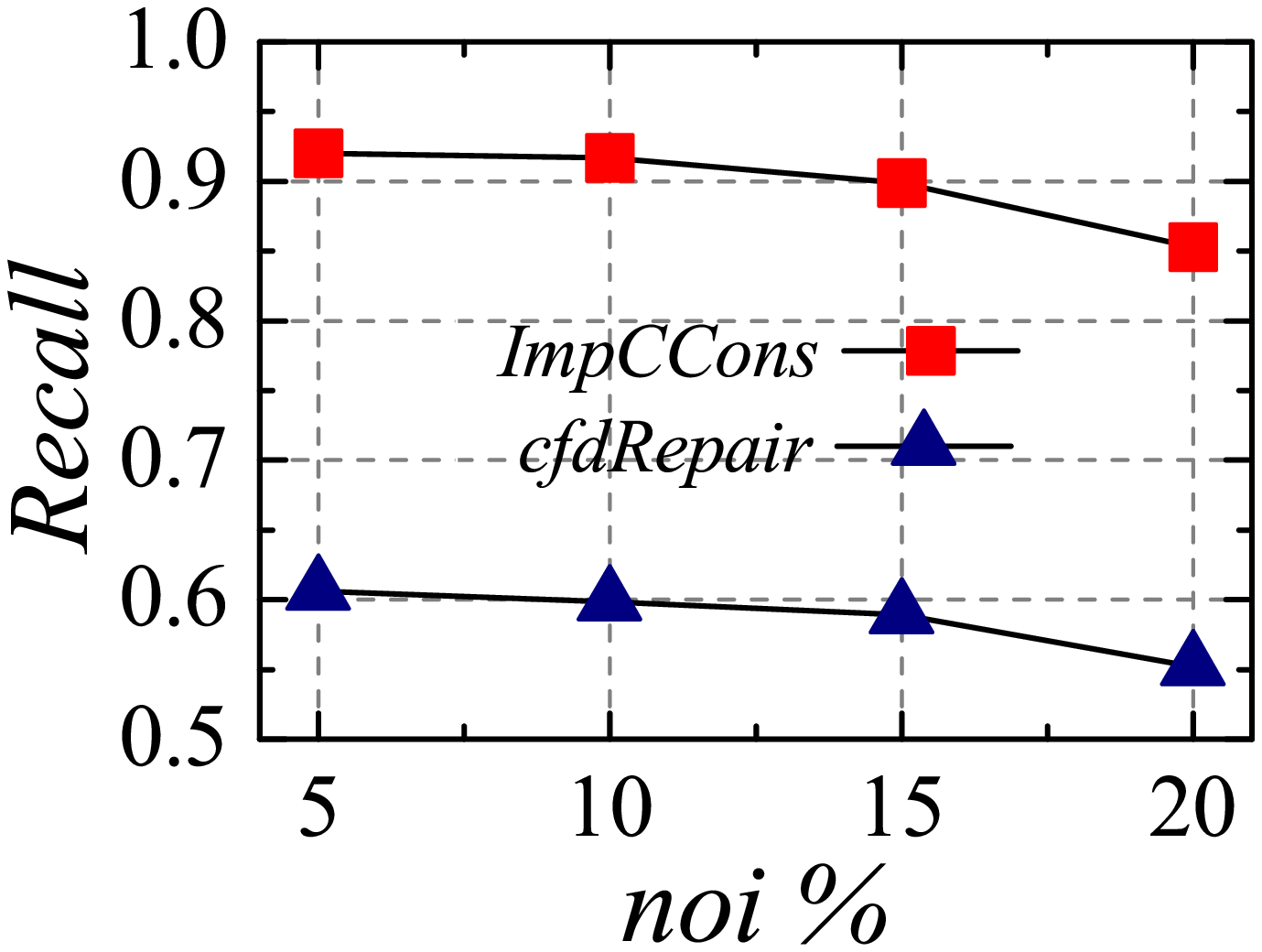}}
\\
\subfigure[NBA, \#R=15K]{
  \includegraphics[scale=0.2]{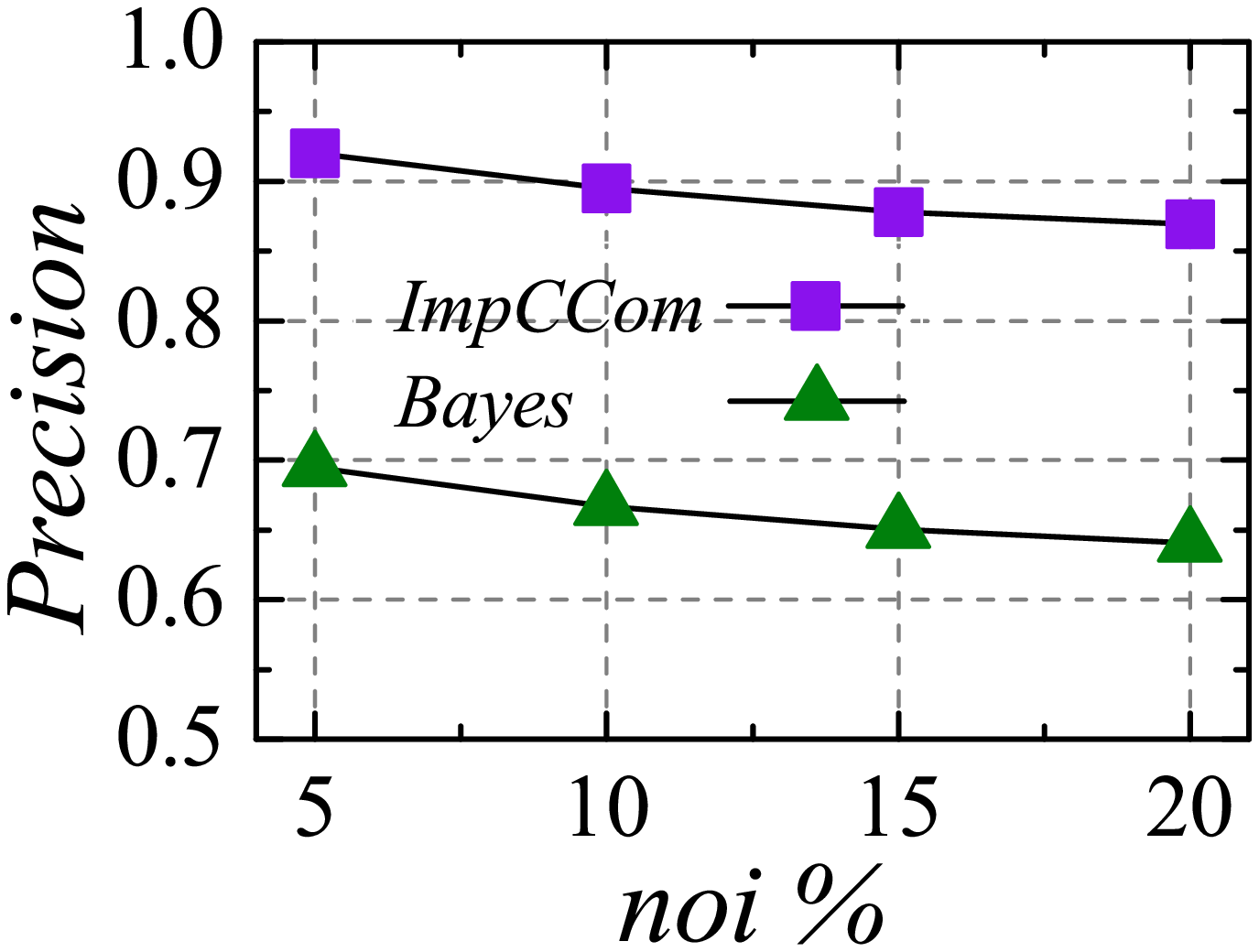}}
\subfigure[NBA, \#R=15K]{
\includegraphics[scale=0.2]{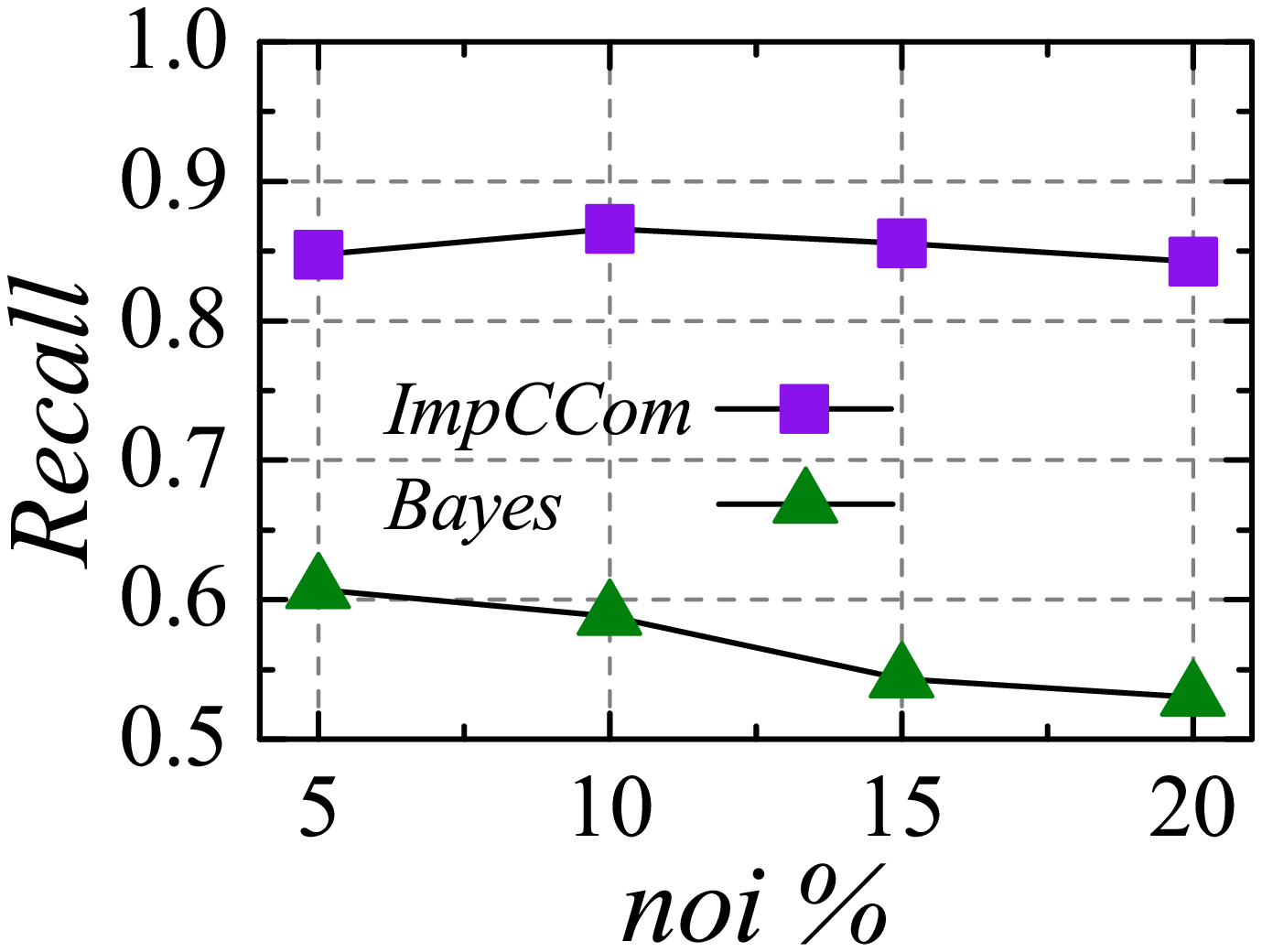}}
\subfigure[PCI, \#R=30K]{
  \includegraphics[scale=0.2]{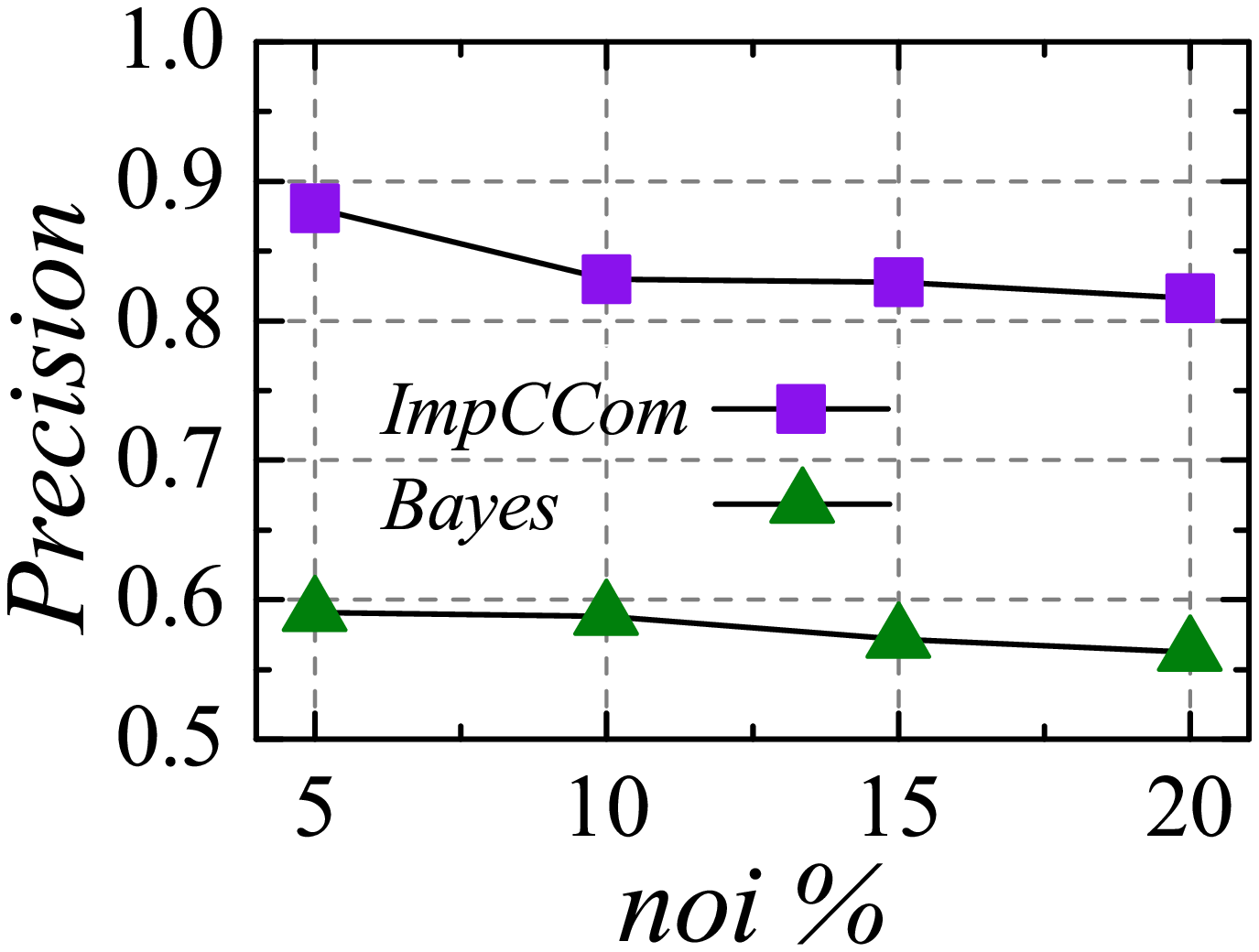}}
\subfigure[PCI, \#R=30K]{
\includegraphics[scale=0.2]{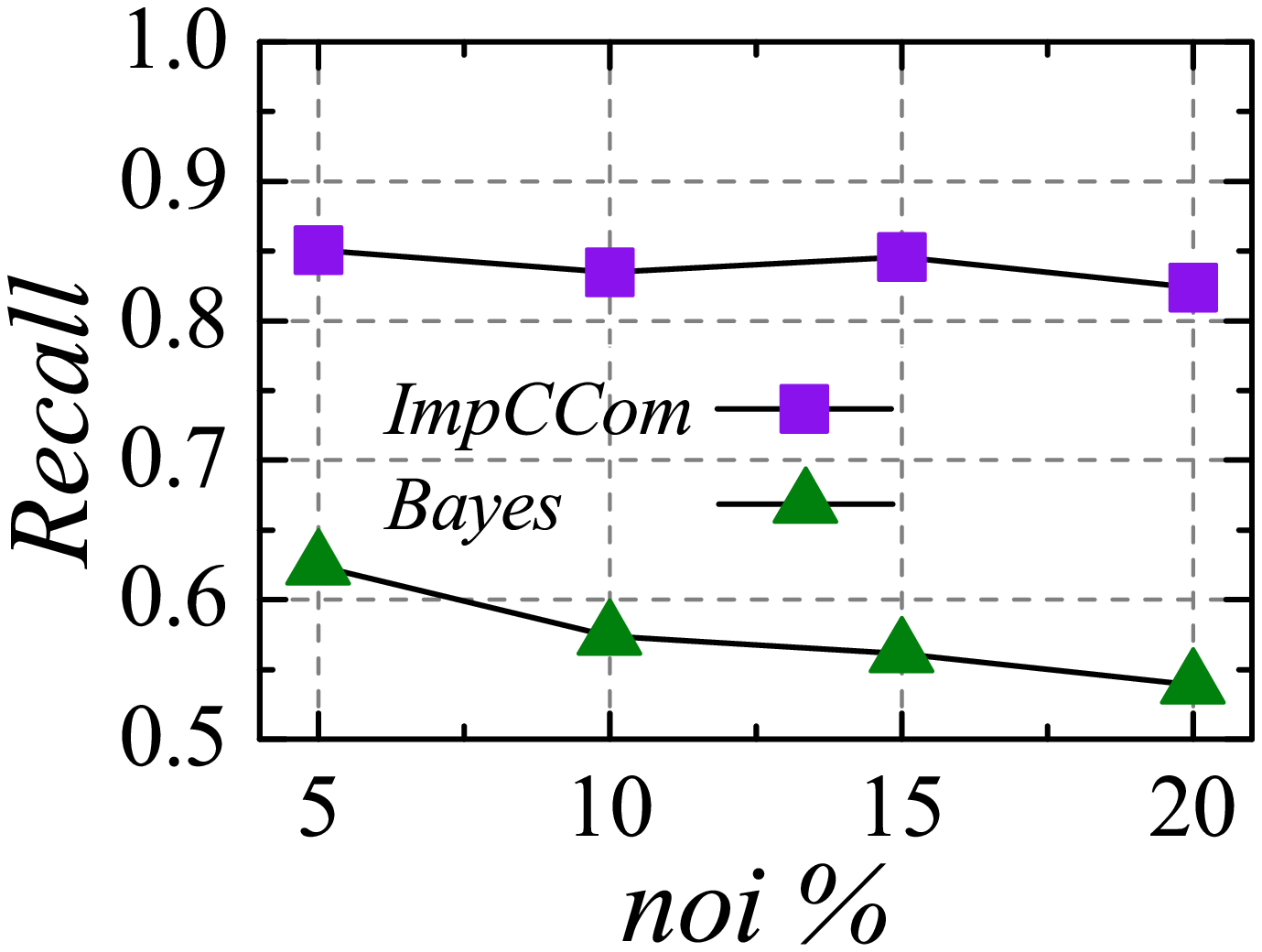}}
\caption{Noise tolerance results on two data sets}
\label{7}       % Give a unique label
\end{figure*}
\subsection{Experimental Results}
\label{5.3}
We discuss the effectiveness and efficiency of two main parameters, namely total number of records \emph{i.e.}, \#Records and the noise rate \emph{noi}\% in each data set.
\subsubsection{Effectiveness Comparison}
\label{5.3.1}
The three experiments below are ran under the conditions, \#Records varies from 5K to 25K in \emph{NBA}, and varies from 5K to 40K in \emph{PCI}, with the same noise rate \emph{noi}=10\%.
\newline \indent \underline{\emph{Exp1-1}:\textsf{ Improve3C} vs \textsf{baseRepair}}. We evaluate \textsf{P} and \textsf{R} in Figure \ref{6}(a)-(d). The proposed \textsf{Improve3C} (\textsf{ImpCCC} for short in figures) performs quite better than \textsf{baseRepair} in the two data sets. We increase precision of the baseline repair by about 25\%, and increase recall by above 27\%. \textsf{P} of \textsf{Improve3C} reaches above 0.9 with 15K records in \emph{NBA} (resp. 30K records in \emph{PCI}), while \textsf{baseRepair} merely reaches 0.7 in \emph{NBA} and around 0.62 in \emph{PCI}. It verifies \textsf{Improve3C} repair multi-errors more accurately with the proposed steps. It shows a little drop when \#Records goes over 15K in \emph{NBA} (resp. 30K records in \emph{PCI}). But it totally maintain above 0.84. \textsf{R} of \textsf{Improve3C} in both data sets shows more steady and a little higher than \textsf{P}. It indicates \textsf{Improve3C} captures more dirty data and repair them effectively than \textsf{baseRepair} does.
%\textsf{R} of \textsf{baseRepair} is poor with \#Records less than 30K in \emph{PCI}, it because \textsf{baseRepair}
\newline \indent\underline{\emph{Exp1-2}:\textsf{ ImpCCons} vs \textsf{cfdRepair}}. We discuss the two methods in Figure \ref{6}(e)-(h). Measures of the methods both show high performance in \emph{NBA}. It illustrates CFD provides a reliable and effective solution in inconsistent repair. \textsf{P} of \textsf{ImpCCons} reaches over 0.96 (resp. 0.9) in \emph{NBA} (resp. \emph{PCI}) in Figure \ref{6}(e),(f). The difference is, measures of \textsf{ImpCCons} has a little rising trend when \#Records gets larger, while measures of \textsf{cfdRepair} drop about 5\% when \#Records gets over 15K. In Figure \ref{6}(g),(h), both measures of \textsf{ImpCCons} are steady around 0.91. We increase \textsf{P} by 28\% and \textsf{R} by around 56\% from the baseline approach.
\newline \indent The preformance difference bewteen two algorithms in \emph{PCI} is quite lager than that in \emph{NBA}. It is because the team of a player is always steady and is unlikely to be changed frequently within a season, and \textsf{Score} accumulated over time is obvious for computing similarity. Thus, the advantages of \textsf{ImpCCons} measuring currency among records is not quite obvious with a small amount of records. However, the average number of records referring to an entity in \emph{PCI} is larger than in \emph{NBA}. The career changes among individuals are complex and frequent. \textsf{ImpCCons} presents a better and steady performance in \emph{PCI}. It verifies \emph{Diffcc} defined in Section 3.3 contributes to repairing the dirty data with clean ones more effectively and appropriately. It also shows the importance of currency evaluation even though timestamps is missing.
\newline \indent\underline{\emph{Exp1-3}:\textsf{ ImpCCom} vs \textsf{Bayes}}. We report the performance of \textsf{ImpCCom} compared with \textsf{Bayes} in Figure \ref{6}(i)-(l). Measures of \textsf{ImpCCom} are steady with the growth of \#Records on both data. \textsf{ImpCCom} outperforms \textsf{Bayes} on \textsf{P}. It shows the currency order is really an important factor for filling missing values with more accurate ones. In Figure \ref{6}(l), \textsf{Bayes} beats \textsf{ImpCCom} on \textsf{R} with \#Records=5K in \emph{PCI}, but it fails to 0.8 with \#Records=40K. It is because when some random noises happen to gather in some records with close currency order, \textsf{Bayes} which do not consider computing the currency may luckily well-repair a few dirty data, while \textsf{ImpCCom} fails to provide a more accurate one. However, with the growth of \#Records, \textsf{ImpCCom} preforms better and more steady than \textsf{Bayes}.
\subsubsection{Tolerance with noise}
\label{5.3.2}
We generate random erroneous data in both data sets, which consists of half inconsistent values and half missing values. We also generate random erroneous attributes with inconsistent (resp. incomplete) problems in \emph{Exp2-2} (resp. \emph{Exp2-3}). The three pairs of noise tolerance experiments are ran under the same condition that \#Records=15K in \emph{NBA}, and  \#Records=30K in \emph{PCI}.
\newline \indent\underline{\emph{Exp2-1}:\textsf{ Improve3C} vs \textsf{baseRepair}}. The tolerance degree of \textsf{Improve3C} and \textsf{baseRepair} is shown in Figure \ref{7}(a)-(d). Measures on both data shows high performance, both \textsf{P} and \textsf{R} reach 0.9 with \emph{noi}=10\%. Both \textsf{P} and \textsf{R} drop slightly when \emph{noi} increases to 20\%, but generally they maintain above 0.84. It shows \textsf{Improve3C} well outperforms \textsf{baseRepair} when there exists quite a few dirty data among records.
\newline \indent\underline{\emph{Exp2-2}:\textsf{ ImpCCons} vs \textsf{cfdRepair}}. Figure \ref{7}(e),(f) reports both \textsf{P} and \textsf{R} show good tolerance against noise in \emph{NBA}, \emph{i.e.}, \textsf{ImpCCons} can effectively find out and repair errors even though there exists much erroneous attributes. It reveals the proposed computing process on \emph{Diffcc} assists the method maintain high effectiveness. Similarly with \emph{Exp1-2}, measures on \textsf{cfdRepair} also reaches 0.9 with \emph{noi}=10\%. However, both \textsf{P} and \textsf{R} suffer a drop and reach 0.875 with \emph{noi}=20\%, which is less than about 10\% from  \textsf{ImpCCons}.
\newline \indent\underline{\emph{Exp2-3}:\textsf{ ImpCCom} vs \textsf{Bayes}}. The  experiments between \textsf{ ImpCCom} and \textsf{Bayes} are shown in Figure \ref{7}(i)-(l). It is obviously \textsf{ImpCCom} outperforms \textsf{Bayes} on both data. Specially, recall maintains a high performance in Figure \ref{7}(j),(l). It indicates \textsf{ImpCCom} has the ability to train the clean data better with the assistance of currency orders. Precision of \textsf{ImpCCom} on both data shows a slight drop, but it beats \textsf{Bayes} with 20\% in \emph{NBA} and 22\% in \emph{PCI}.
\begin{figure}[t]
\centering
\subfigure[NBA, \emph{noi}=10\% ]{
  \includegraphics[scale=0.2]{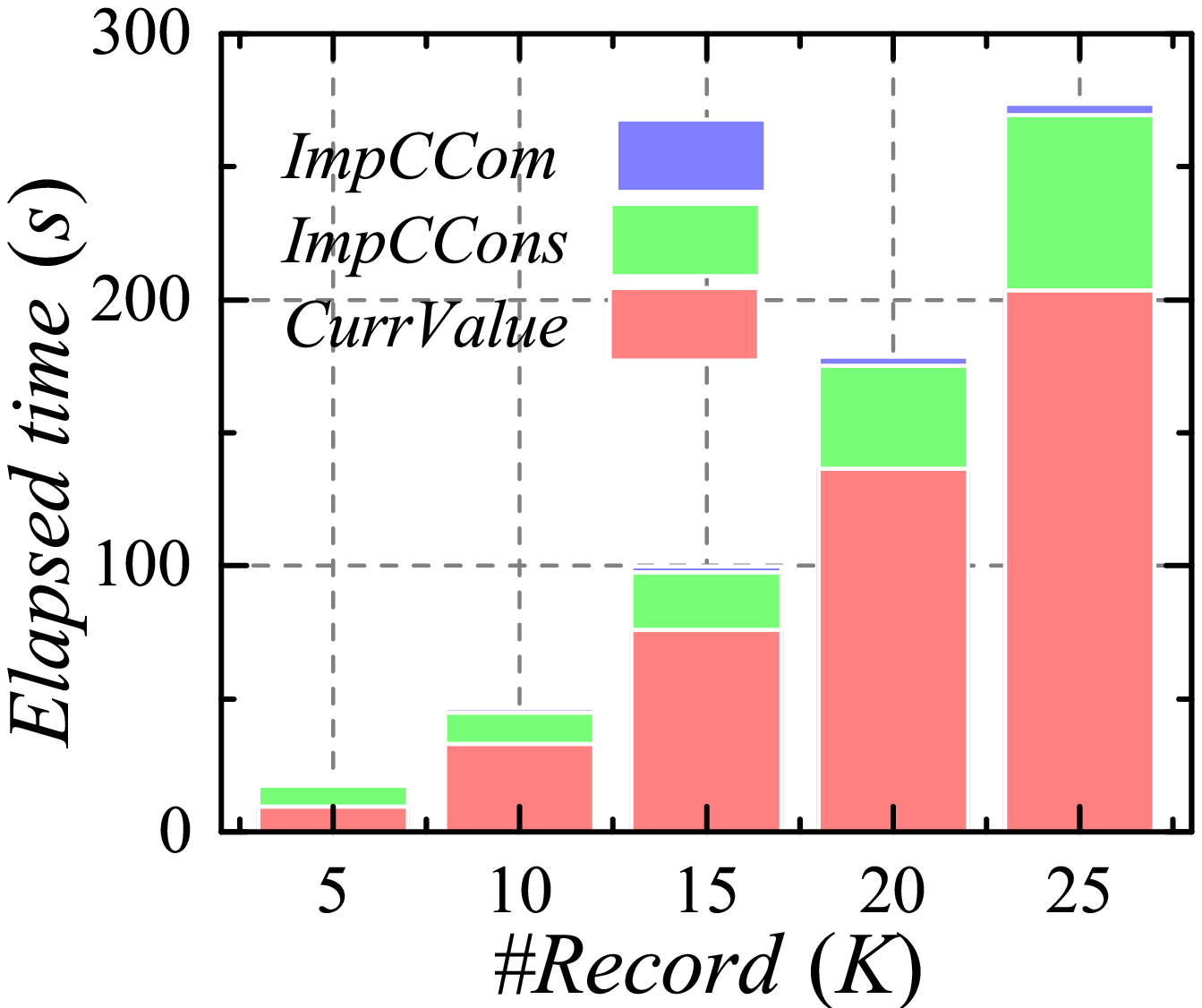}}
\subfigure[NBA, \emph{noi}=10\% ]{
\includegraphics[scale=0.2]{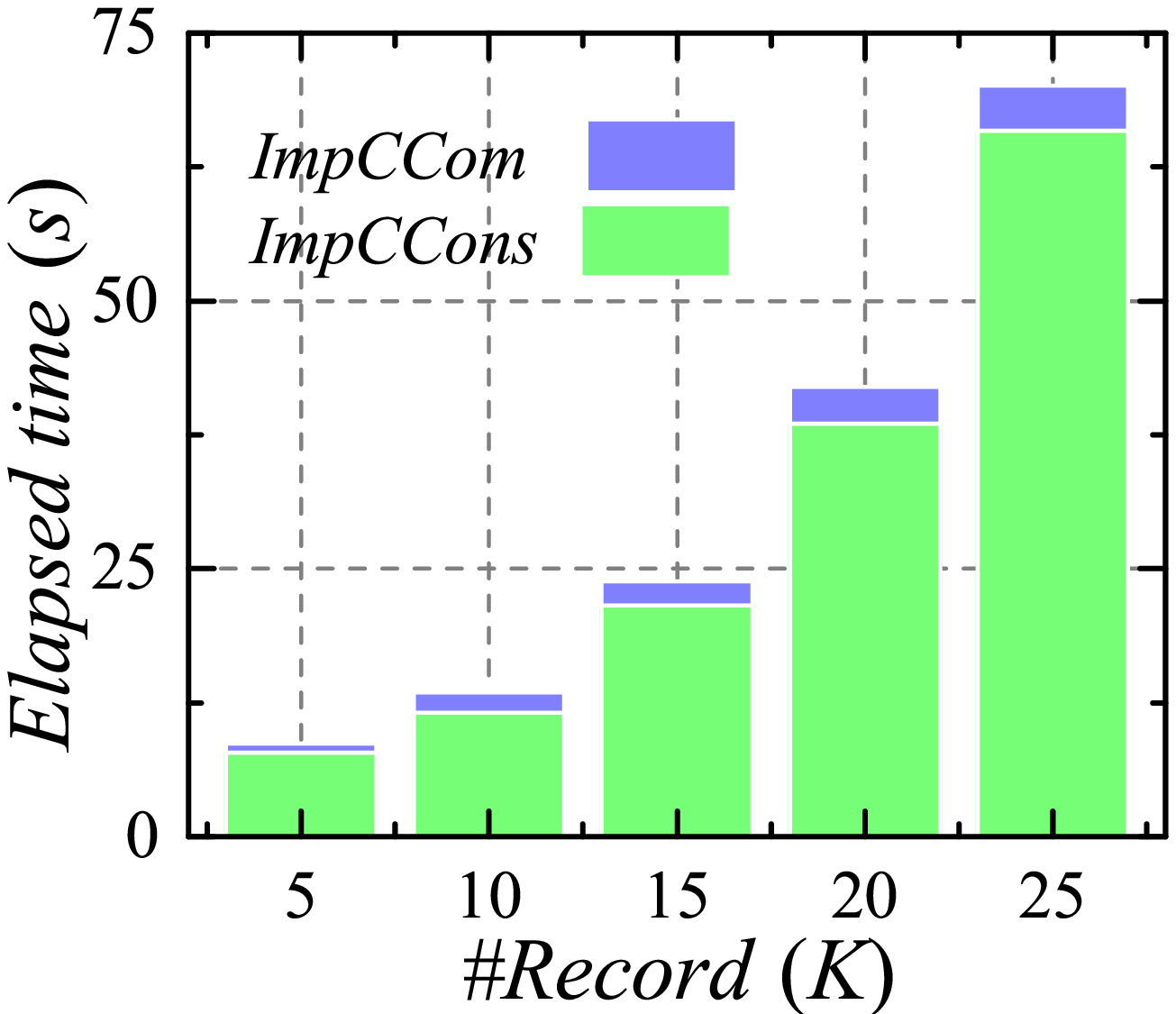}}
\\
\subfigure[PCI, \emph{noi}=10\%]{
  \includegraphics[scale=0.2]{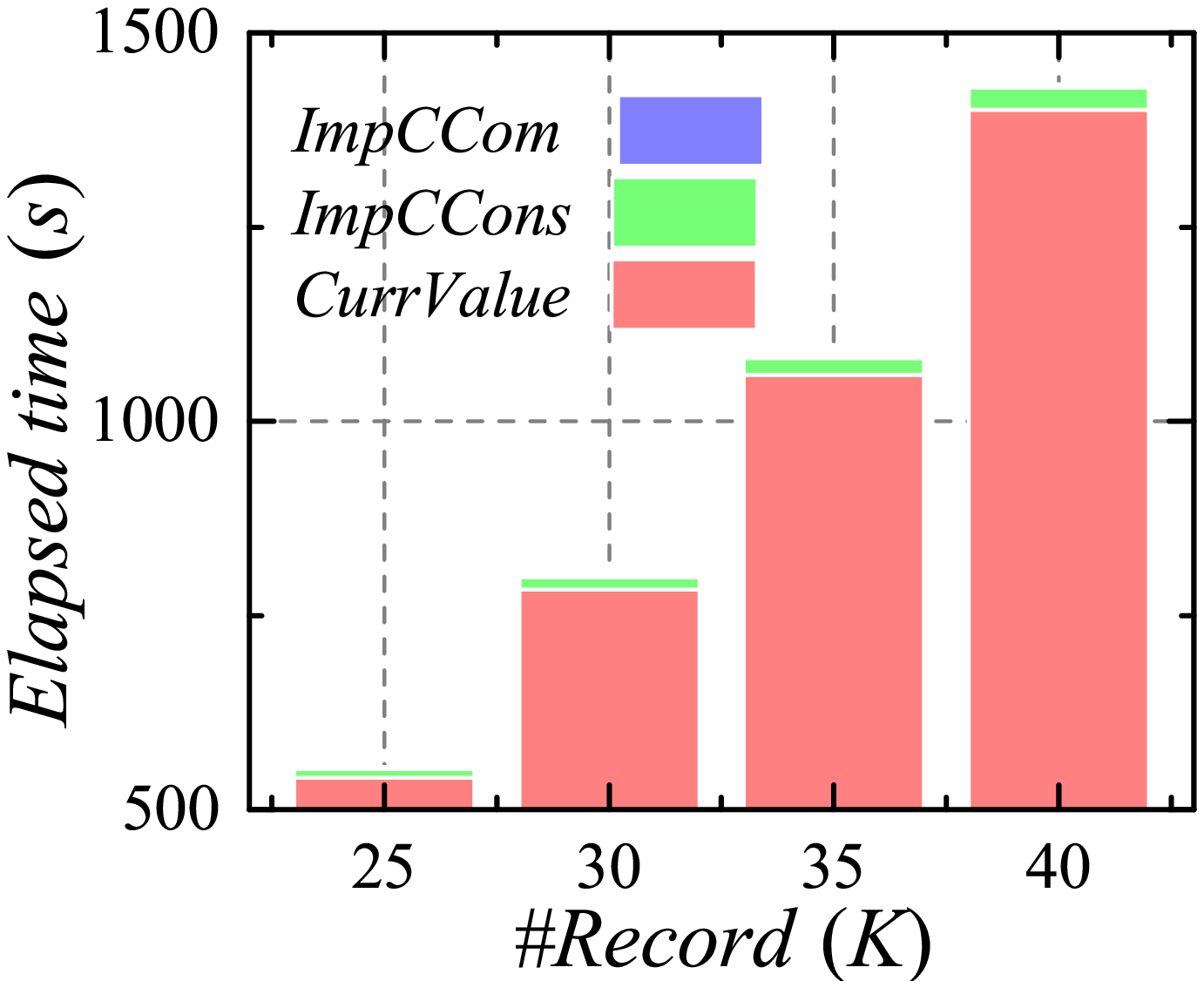}}
  \subfigure[PCI, \emph{noi}=10\%]{
  \includegraphics[scale=0.2]{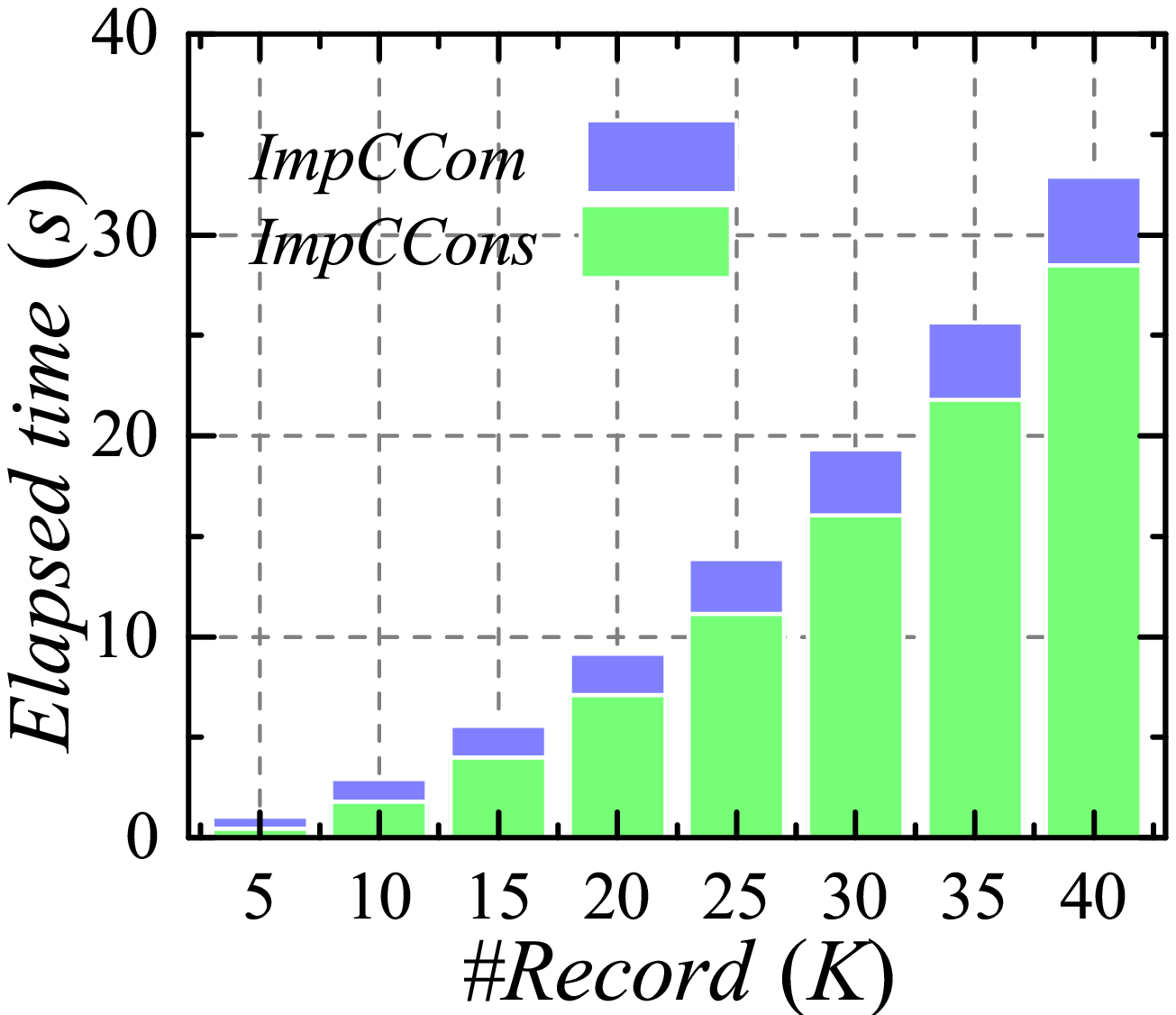}}
  \\
\subfigure[NBA, \#R=15K]{
  \includegraphics[scale=0.2]{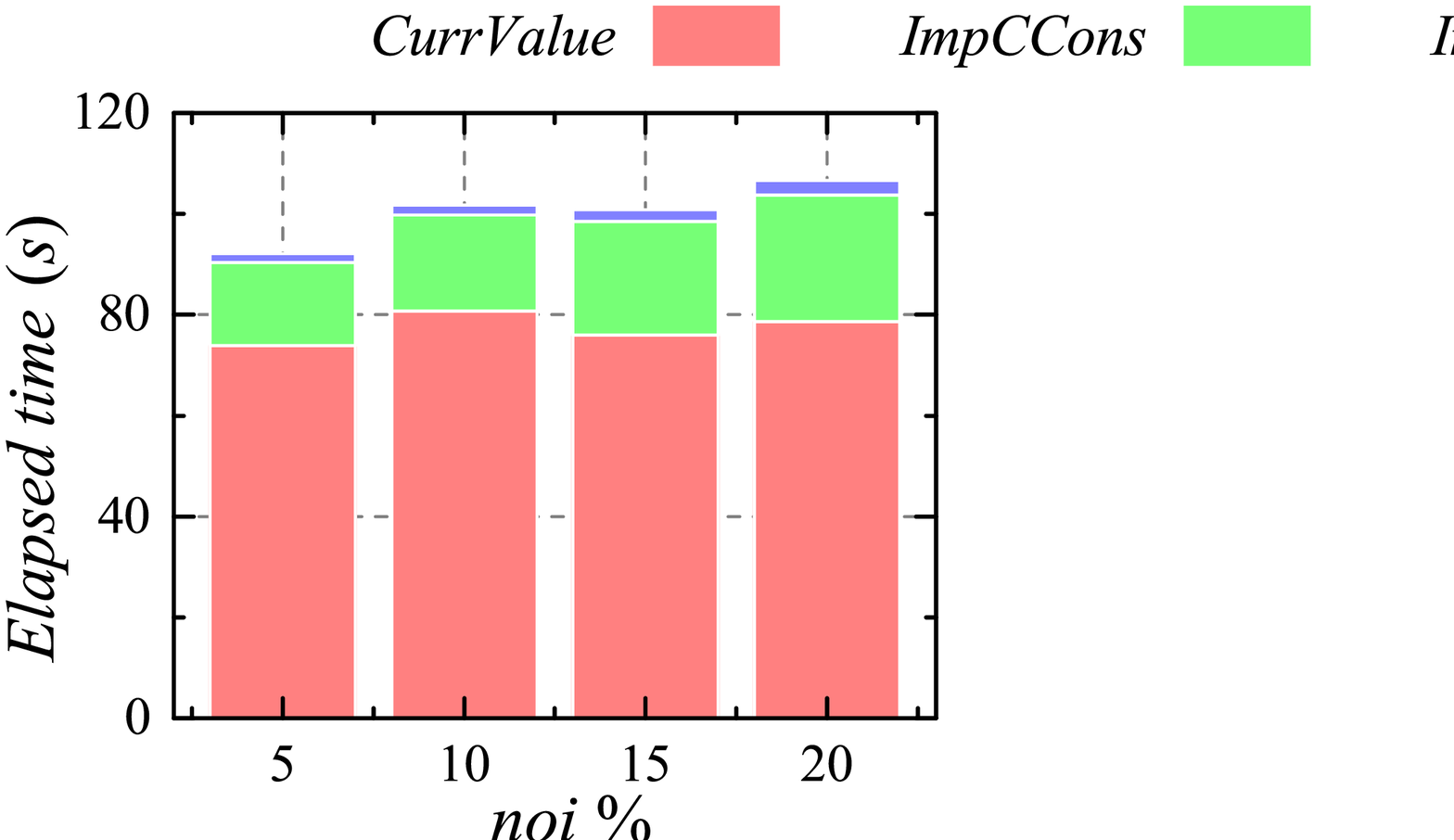}}
 % \hspace{0.5cm}
\subfigure[PCI, \#R=30K]{
\includegraphics[scale=0.2]{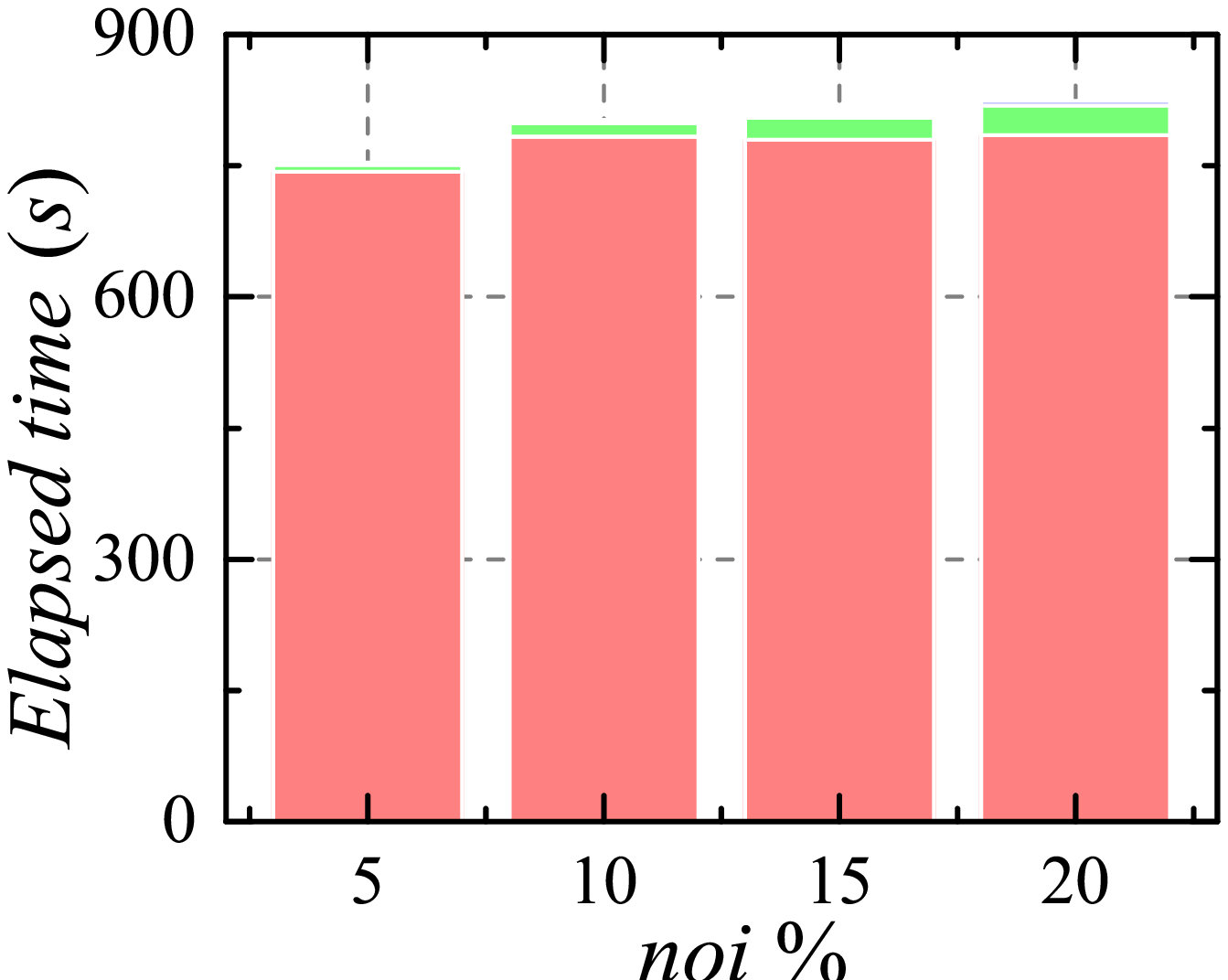}}
%\hspace{0.5cm}
\\
\subfigure[NBA, \#R=15K]{
  \includegraphics[scale=0.2]{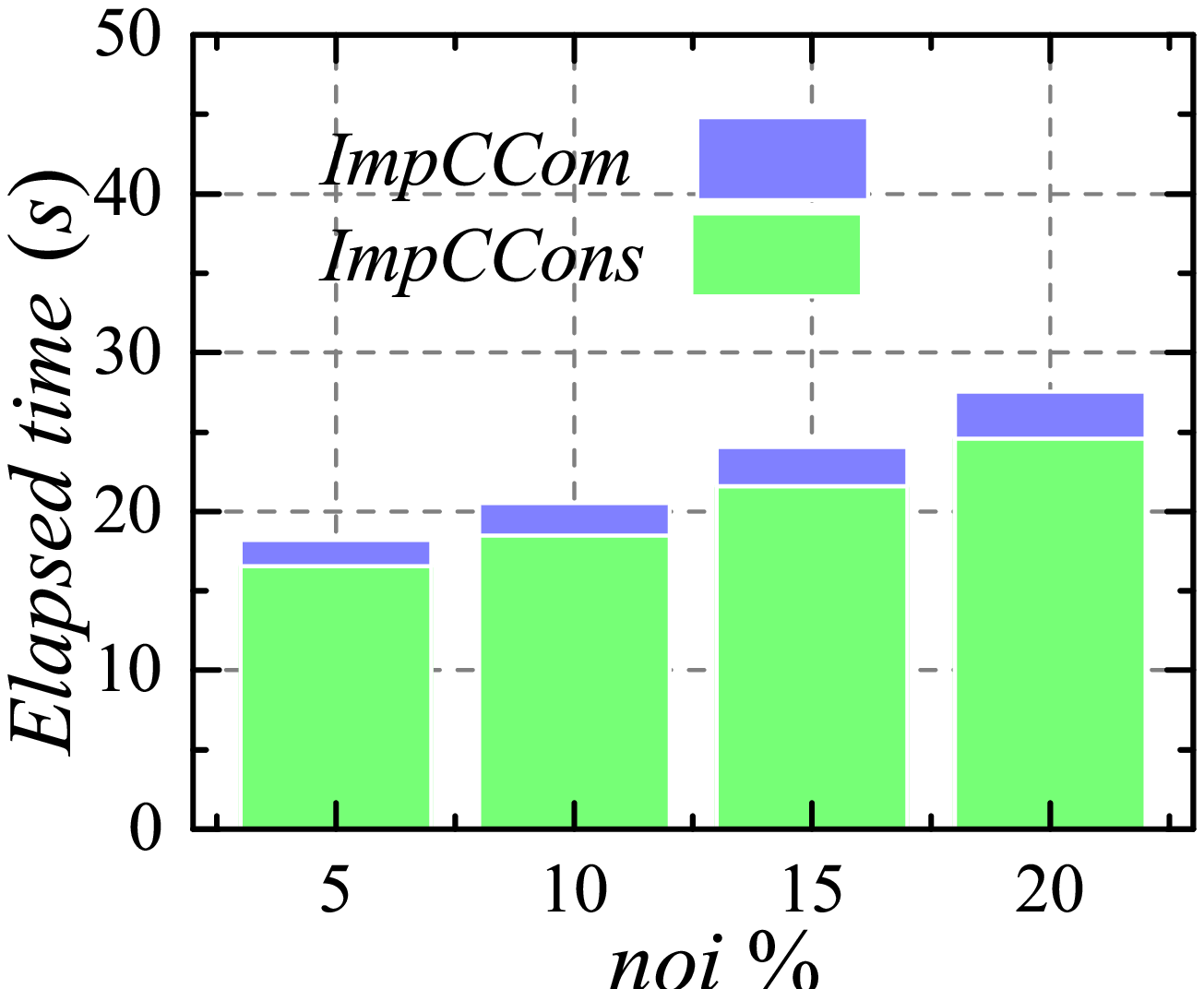}}
  %\hspace{0.5cm}
  \subfigure[PCI, \#R=30K]{
  \includegraphics[scale=0.2]{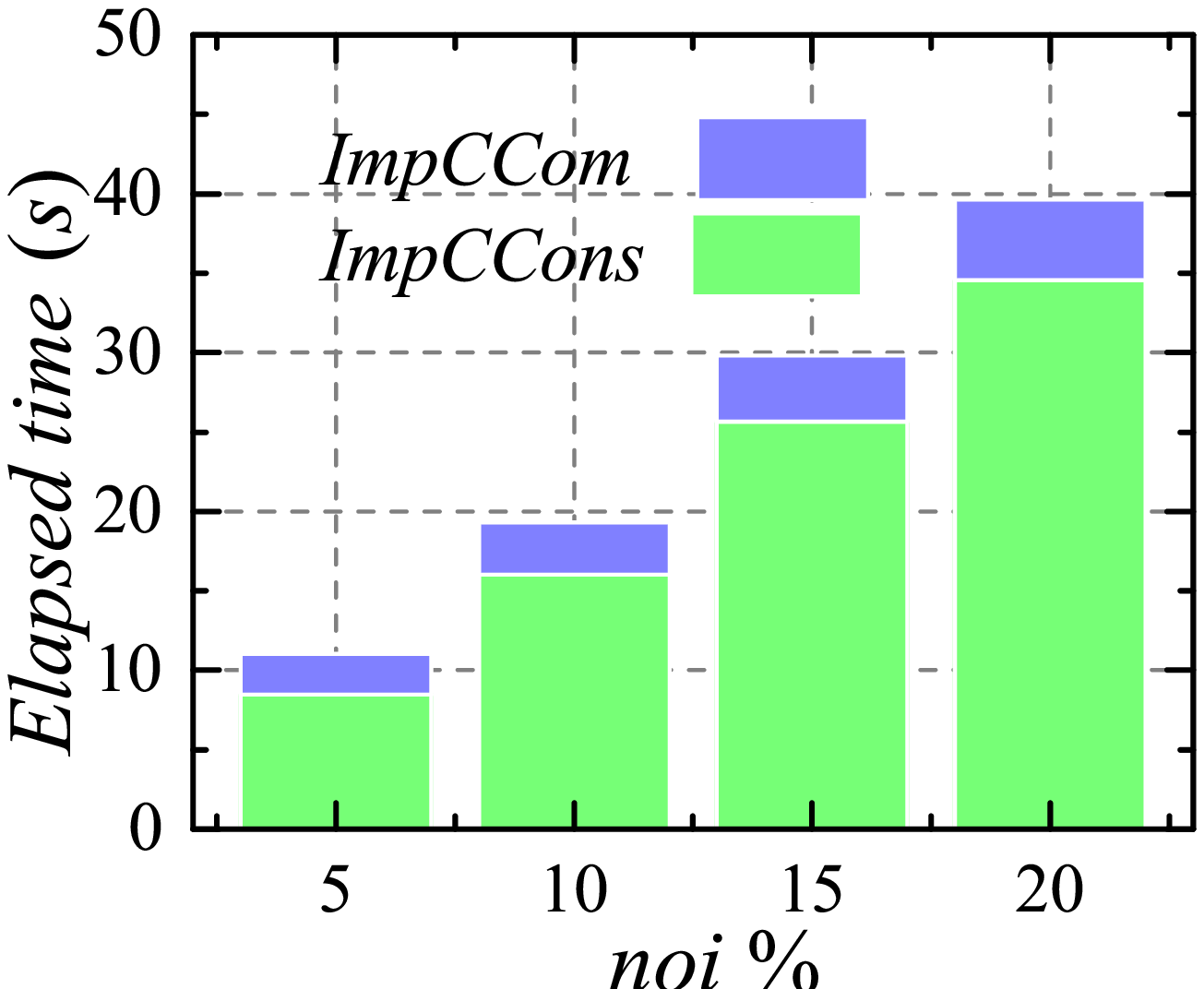}}
  \caption{Efficiency results}
\label{8}       % Give a unique label
\end{figure}
\subsubsection{Efficiency}
\label{5.3.3}
We now report the efficiency results with the time cost cumulative graphs in Figure \ref{8}. We evaluate the time cost of three critical algorithms, namely \textsf{CurrValue}, \textsf{ImpCCons} and \textsf{ImpCCom}. Efficiency of algorithms varying with the growth of \#Records under the condition \emph{noi}=10\% in Figure \ref{8}(a)-(d). It totally costs about 4.5 minutes to process the proposed methods in 25k records of \emph{NBA} and 25 minutes in 40K records of \emph{PCI}, which is acceptable referring to the record amount. From the cumulative graphs can we see algorithm \textsf{CurrValue} stands the most time of the whole method. It is because the currency graphs construction and computing currency orders takes some time. The elapsed time of \textsf{CurrValue} shows a square growth, the trend of which verifies the complexity we reports in Section 3.2. In Figure \ref{8}(b),(d), we can conclude the time costs of \textsf{ImpCCom} shows a linear growth while \textsf{ImpCCons} reports a square growth. Even though, the time costs in consistency and completeness repair is not large. Especially, \textsf{ImpCCons} based on Bayes learning is quite efficient as there are advanced training models in practical which can be easily adopted in our method. We are able to finish the repair in 0.55 minutes for 40K records in \emph{PCI}.
\newline \indent Figure \ref{8}(e)-(h) shows the time costs of these algorithms under different noise rate. It presents no impact on time costs of \textsf{CurrValue} when the noise increases as shown in Figure \ref{8}(e),(f). We make it clear in \ref{8}(g),(h) both \textsf{ImpCCons} and \textsf{ImpCCom} report a sight linear growth with the increasing \emph{noi}\%. We are able to finish repairing 20\% noises in 30K records of \emph{PCI} in 0.67 minutes, which shows a potential in scalability for large amount data quality repair issues.
\section{Related Work}
\label{Related Works}
Study on data quality is extensive for decades. Various standards and metrics are proposed to describe the quality of data in both theory and practice.
\newline \indent \textbf{Data quality dimensions}. With the demand for high-quality data, many metrics beside accuracy are necessary for quality improvement \cite{Wang1996Beyond}. \cite{Batini2009Methodologies} provides a systematic introduction of data quality methodologies. Data quality \emph{dimension} is a characteristic for data requirements, among which completeness, accuracy, consistency and currency are four important dimensions. %\emph{Accuracy} is regarded as the fundamental metric of data quality widely studied \cite{Wang1996Beyond,Yang2006Journey}. It measures the closeness of an attribute in records with its real-life value. Statistics-based and constraints-based approaches are always associated to query the accurate values \cite{Cong2007Improving}.
\emph{Completeness} measures to which degree a data set has complete attribute values to describe the corresponding real-world information \cite{Batini2009Methodologies}. Algorithms are proposed to fill the missing values \cite{Deng2010Capturing}. \emph{Consistency} describes the violation of integrity constraints. Different semantic constraints such as FD, CFD, and CIND, have been defined to guide data cleaning under specific circumstance \cite{Fan2012Foundations,Bertiequille2009Sailing}, where conditional functional dependency (CFD) is a general and effective consistency constraints for querying and inconsistency detection in database \cite{3Fan2009Conditional,Cong2007Improving}.
%Also, consistency constraints as well as data quality rules discovery problem is proposed and studied in works like \cite{Chu2014RuleMiner,Fei2011A}.
\newline \indent Furthermore, as the dimensions are not independent issues in data integration \cite{Fan2012Foundations}, data cleaning approaches have been developed with integrating several data quality dimensions. \cite{Fan2013Data} reports advanced study on critical dimensions and provides a logical framework for a uniform treatment of the issues. \cite{Cong2007Improving} propose a framework for quality improvement on both consistency and accuracy. \cite{Cappiello2008Time} discusses time-related measures with accuracy and completeness, and proposes functions of computing their mutual relationships.
%Though framework of data quality improvement and data cleaning on some key dimensions has been proposed, technological breakthroughs are still in demand in developing a comprehensive and efficient method of data quality improvement.
\newline \indent \textbf{Data currency}. Currency describes to which extent a data set is up-to-date \cite{Batini2009Methodologies}.
%Two popular topics on currency are volatility measure with recorded time (\emph{e.g.}, timestamps) and currency determination based on constraints when timestamps are imprecise. Timestamps carry valid and dependable time points, and the most current records can be easily identified via querying and computing. The concept of \emph{age} is defined as the valid time between the input time and the query requirements at task of the attribute values \cite{Heinrich2009A,Heinrich2011Assessing}, and thus, currency is computed from a function of \emph{age} in volatility.
When various data sources are integrated, timestamps are always neither complete nor uniform. It promotes the study on currency determination without available timestamps. \cite{4Fan2012Determining} is the first to propose a constraint-based model for data currency reasoning. And several fundamental theoretical problems are discussed in both \cite{4Fan2012Determining} and \cite{Fan2012Foundations}. In addition, considering the temporal changes and evolution of attribute values in records, works like \cite{5Pei2012Linking} also propose record linkage problems on temporal data.
\section{Conclusion}
\label{sec9}
This paper studies the repairing problem of low-quality data with incomplete and inconsistent values, which lacks for available timestamps. We propose a four-step framework to solve the problem. We first construct currency order graph with currency constraints, base on which a currency order determining method is presented. In addition, we introduce the currency order chain to repair the inconsistency and incompleteness data.
Various experiments on both real-life and synthetic data present the effectiveness of our method on data with mixed quality problems. Our method achieve high performance steadily with the increasing error noise up to 20\%. Moreover, the propose method outperforms the traditional repairing algorithms when the timestamps is imprecise. It verifies the propose method can validly improve completeness and consistency with currency.
\newline \indent Future works includes comprehensive data quality constraints design in semantic, various models applications in incompleteness imputation on different data sets and parallelization of \textsf{Improve3C} on big data.

%%%%%%%%%%%%%%%%%%%%%%%%

\bibliographystyle{IEEEtran}
\bibliography{1new}

\begin{thebibliography}{10}
\providecommand{\url}[1]{#1}
\csname url@samestyle\endcsname
\providecommand{\newblock}{\relax}
\providecommand{\bibinfo}[2]{#2}
\providecommand{\BIBentrySTDinterwordspacing}{\spaceskip=0pt\relax}
\providecommand{\BIBentryALTinterwordstretchfactor}{4}
\providecommand{\BIBentryALTinterwordspacing}{\spaceskip=\fontdimen2\font plus
\BIBentryALTinterwordstretchfactor\fontdimen3\font minus
  \fontdimen4\font\relax}
\providecommand{\BIBforeignlanguage}[2]{{%
\expandafter\ifx\csname l@#1\endcsname\relax
\typeout{** WARNING: IEEEtran.bst: No hyphenation pattern has been}%
\typeout{** loaded for the language `#1'. Using the pattern for}%
\typeout{** the default language instead.}%
\else
\language=\csname l@#1\endcsname
\fi
#2}}
\providecommand{\BIBdecl}{\relax}
\BIBdecl

\bibitem{Sidi2012Data}
F.~Sidi, P.~H.~S. Panahy, L.~S. Affendey, M.~A. Jabar, H.~Ibrahim, and
  A.~Mustapha, ``Data quality: A survey of data quality dimensions,'' in
  \emph{International Conference on Information Retrieval and Knowledge
  Management}, 2012, pp. 300--304.

\bibitem{Fan2013Data}
W.~Fan, F.~Geerts, S.~Ma, N.~Tang, and W.~Yu, \emph{Data Quality Problems
  beyond Consistency and Deduplication}.\hskip 1em plus 0.5em minus 0.4em\relax
  Springer Berlin Heidelberg, 2013.

\bibitem{Cappiello2008Time}
C.~Cappiello, C.~Francalanci, and B.~Pernici, ``Time related factors of data
  accuracy, completeness, and currency in multi-channel information systems,''
  in \emph{The Conference on Advanced Information Systems Engineering}, 2008,
  pp. 145--153.

\bibitem{Fan2012Foundations}
W.~Fan and F.~Geerts, \emph{Foundations of Data Quality Management}, 2012.

\bibitem{4Fan2012Determining}
W.~Fan, F.~Geerts, and J.~Wijsen, ``Determining the currency of data,''
  \emph{Acm Transactions on Database Systems}, vol.~37, no.~4, pp. 71--82,
  2012.

\bibitem{3Fan2009Conditional}
W.~Fan, F.~Geerts, and X.~Jia, ``Conditional dependencies: A principled
  approach to improving data quality,'' in \emph{British National Conference on
  Databases: Dataspace: the Final Frontier}, 2009, pp. 8--20.

\bibitem{Cong2007Improving}
G.~Cong, W.~Fan, F.~Geerts, X.~Jia, and S.~Ma, ``Improving data quality:
  consistency and accuracy,'' in \emph{International Conference on Very Large
  Data Bases}, 2007, pp. 315--326.

\bibitem{8Fan2014Conflict}
W.~Fan, F.~Geerts, N.~Tang, and W.~Yu, ``Conflict resolution with data currency
  and consistency,'' \emph{Journal of Data and Information Quality}, vol.~5,
  no. 1-2, pp. 1--37, 2014.

\bibitem{Li2017Crowdsourced}
G.~Li, J.~Fan, J.~Fan, J.~Wang, and R.~Cheng, ``Crowdsourced data management:
  Overview and challenges,'' in \emph{ACM International Conference on
  Management of Data}, 2017, pp. 1711--1716.

\bibitem{Zheng2016DOCS}
Y.~Zheng, G.~Li, and R.~Cheng, \emph{DOCS: a domain-aware crowdsourcing system
  using knowledge bases}.\hskip 1em plus 0.5em minus 0.4em\relax VLDB
  Endowment, 2016.

\bibitem{DBLP:books/daglib/0023376}
\BIBentryALTinterwordspacing
T.~H. Cormen, C.~E. Leiserson, R.~L. Rivest, and C.~Stein, \emph{Introduction
  to Algorithms, 3rd Edition}.\hskip 1em plus 0.5em minus 0.4em\relax {MIT}
  Press, 2009. [Online]. Available:
  \url{http://mitpress.mit.edu/books/introduction-algorithms}
\BIBentrySTDinterwordspacing

\bibitem{Fan2011Discovering}
W.~Fan, F.~Geerts, J.~Li, and M.~Xiong, ``Discovering conditional functional
  dependencies,'' \emph{IEEE Transactions on Knowledge \& Data Engineering},
  vol.~23, no.~5, pp. 683--698, 2011.

\bibitem{Papenbrock2015Functional}
T.~Papenbrock, J.~Ehrlich, J.~Marten, T.~Neubert, J.~P. Rudolph, J.~Zwiener,
  and F.~Naumann, ``Functional dependency discovery: an experimental evaluation
  of seven algorithms,'' \emph{Proceedings of the Vldb Endowment}, vol.~8,
  no.~10, pp. 1082--1093, 2015.

\bibitem{13Wang1996Beyond}
R.~Y. Wang and D.~M. Strong, ``Beyond accuracy: What data quality means to data
  consumers,'' \emph{Journal of Management Information Systems}, vol.~12,
  no.~4, pp. 5--33, 1996.

\bibitem{Batini2009Methodologies}
C.~Batini, C.~Cappiello, C.~Francalanci, and A.~Maurino, ``Methodologies for
  data quality assessment and improvement,'' \emph{Acm Computing Surveys},
  vol.~41, no.~3, p.~16, 2009.

\bibitem{DBLP:books/daglib/0087929}
T.~M. Mitchell, \emph{Machine learning}, ser. McGraw Hill series in computer
  science.\hskip 1em plus 0.5em minus 0.4em\relax McGraw-Hill, 1997.

\bibitem{Chu2014RuleMiner}
X.~Chu, I.~F. Ilyas, P.~Papotti, and Y.~Ye, ``Ruleminer: Data quality rules
  discovery,'' in \emph{IEEE International Conference on Data Engineering},
  2014, pp. 1222--1225.

\bibitem{Wang1996Beyond}
R.~Y. Wang and D.~M. Strong, ``Beyond accuracy: What data quality means to data
  consumers,'' \emph{Journal of Management Information Systems}, vol.~12,
  no.~4, pp. 5--33, 1996.

\bibitem{Deng2010Capturing}
T.~Deng, W.~Fan, and F.~Geerts, ``Capturing missing tuples and missing
  values,'' in \emph{Twenty-Ninth ACM Sigmod-Sigact-Sigart Symposium on
  Principles of Database Systems, PODS 2010, June 6-11, 2010, Indianapolis,
  Indiana, Usa}, 2010, pp. 169--178.

\bibitem{Bertiequille2009Sailing}
L.~Bertiequille, A.~D. Sarma, Dong, A.~Marian, and D.~Srivastava, ``Sailing the
  information ocean with awareness of currents: Discovery and application of
  source dependence,'' \emph{Computer Science}, vol.~26, no.~8, pp. 1881--3,
  2009.

\bibitem{5Pei2012Linking}
L.~I. Pei, X.~L. Dong, A.~Maurino, and D.~Srivastava, ``Linking temporal
  records,'' \emph{{PVLDB}}, vol.~4, no.~11, pp. 956--967, 2011.

\end{thebibliography}

\end{document}